\documentclass[fleqn,usenatbib]{mnras}

\usepackage[T1]{fontenc}

\DeclareRobustCommand{\VAN}[3]{#2}
\let\VANthebibliography\thebibliography
\def\thebibliography{\DeclareRobustCommand{\VAN}[3]{##3}\VANthebibliography}

\usepackage{graphicx}	
\usepackage{amsmath}	
\usepackage{amssymb}	

\usepackage{hyperref}
\usepackage{booktabs}
\usepackage{multirow}
\usepackage{multicol}
\usepackage[dvipsnames]{xcolor}
\usepackage{microtype}
\usepackage[separate-uncertainty=true,detect-weight]{siunitx}
\usepackage{cleveref}
\usepackage[figuresright]{rotating}

\usepackage{threeparttable}

\usepackage{newtxtext,newtxmath}

\newcommand {\nucfrac} {$f_{\mathrm{n}}$}
\newcommand {\cta}[1] {\citetalias{#1}}
\newcommand {\cpa}[1] {\citepalias{#1}}

\newcommand {\MLR} {$M_{\star} / L$}
\newcommand {\HST} {\textit{HST}}
\newcommand {\SDSS} {\textit{SDSS}}
\newcommand {\HLA} {\textit{HLA}}
\newcommand {\Tten} {$\Theta_{10}$}
\newcommand {\Tpeak} {$\Theta_{10}^{\mathrm{peak}}$}

\defcitealias {Cote2006} {C+06}
\defcitealias {denBrok2014} {dB+14}
\defcitealias {Munoz2015} {M+15}
\defcitealias {Turner2012} {T+12}
\defcitealias {Eigenthaler2018} {E+18}
\defcitealias {Sanchez-Janssen2019a} {SJ+19}
\defcitealias {Bell2003} {B+03}
\defcitealias {Portinari2004} {P+04}
\defcitealias {Zibetti2009} {Z+09}
\defcitealias {Into2013} {I+13}
\defcitealias {Karachentsev2013} {KMK+13}

\DeclareSIUnit \Mpc {Mpc}
\DeclareSIUnit \mag {mag}
\DeclareSIUnit \pc {pc}
\DeclareSIUnit \Msun {M_{\odot}}
\DeclareSIUnit \dex {dex}
\DeclareSIUnit \amin {arcmin}

\graphicspath{{./}}

\creflabelformat{equation}{#2#1#3}

\title[The nucleation fraction of Local Volume galaxies]{The nucleation fraction of Local Volume galaxies}

\author[N. Hoyer et al.]{
\href{https://orcid.org/0000-0001-8040-4088}{Nils Hoyer}$^{1,2}$\thanks{Contact e-mail: \href{mailto:hoyer@mpia-hd.mpg.de}{hoyer@mpia-hd.mpg.de}},
\href{https://orcid.org/0000-0002-6922-2598}{Nadine Neumayer}$^{1}$,
\href{https://orcid.org/0000-0001-8471-6679}{Iskren Y.\ Georgiev}$^{1}$,
\href{https://orcid.org/0000-0003-0248-5470}{Anil C.\ Seth}$^{3}$, and
\newauthor{
\href{https://orcid.org/0000-0002-5612-3427}{Jenny E.\ Greene}$^{4}$}
\medskip\\
$^{1}$Max-Planck-Institut f{\"{u}}r Astronomie, K{\"{o}}nigstuhl 17, D-69117 Heidelberg, Germany\\
$^{2}$Ruprecht-Karls-Universit{\"{a}}t Heidelberg, Seminarstraße 2, D-69117 Heidelberg, Germany\\
$^{3}$Department of Physics and Astronomy, University of Utah, 115 South 1400 East, Salt Lake City, UT 84112, USA\\
$^{4}$Department of Astrophysical Sciences, Princeton University, Princeton, NJ 08544, USA\\
}

\date{Accepted XXX. Received YYY; in original form ZZZ}

\pubyear{2021}

\begin{document}
\label{firstpage}
\pagerange{\pageref{firstpage}--\pageref{lastpage}}
\maketitle

\begin{abstract}
  Nuclear star clusters (NSCs) are a common phenomenon in galaxy centres and are found in a vast majority of galaxies of intermediate stellar mass \SI{\approx e9}{\Msun}.
  Recent investigations suggest that they are rarely found in the least and most massive galaxies and that the nucleation fraction increases in dense environments.
  It is unclear whether this trend holds true for field galaxies due to the limited data currently available.
  Here we present our results on the nucleation fraction for \num{601} galaxies in the Local Volume ($\lesssim \SI{12}{\Mpc}$).
  Covering more than eight orders of magnitude in stellar mass, this is the largest sample of galaxies analysed in a low-density environment.
  Within the Local Volume sample we find a strong dependence of the nucleation fraction on galaxy stellar mass, in agreement with previous work.
  We also find that for galaxies with $M_{\star} < \SI{e9}{\Msun}$, early-type galaxies have a higher nucleation fraction than late-types.
  The nucleation fraction in the Local Volume correlates independently with stellar mass, Hubble type, and local environmental density.
  We compare our data to those in galaxy cluster environments (Coma, Fornax, and Virgo) by compiling previous results and calculating stellar masses in a homogeneous way.
  We find significantly lower nucleation fractions (up to \SI{40}{\percent}) in galaxies with $M_{\star} \lesssim \SI{e9.5}{\Msun}$, in agreement with previous work.
  Our results reinforce the connection between globular clusters and NSCs, but it remains unclear if it can explain the observed trends with Hubble type and local environment.
  We speculate that correlation between the nucleation fraction and cluster environment weakens for the densest clusters like Coma and Virgo.
\end{abstract}

\begin{keywords}
galaxies: general -- galaxies: nuclei -- galaxies: star clusters: general -- galaxies: clusters: general -- galaxies: groups: general
\end{keywords}



\section{Introduction}
\label{sec:introduction}

Nuclear star clusters (NSCs) are a common phenomenon in the centres of galaxies \citep[e.g.][]{Phillips1996,Boeker2002,Scarlata2004,Seth2006,Georgiev2009,Neumayer2011,Georgiev2014}, but are not always present.
The nucleation fraction of galaxies (denoted as {\nucfrac}) measures the fraction of galaxies with an NSC.
It increases as a function of galaxy stellar mass (denoted as $M_{\star}$), but seems to hit a peak at \SI{\approx e9}{\Msun} \citep{denBrok2014,Munoz2015,Sanchez-Janssen2019a,Neumayer2020}.

Early studies considered {\nucfrac} in samples of galaxies over a limited range in Hubble type and / or stellar mass in both field \citep[e.g.][]{Balcells2007,Baldassare2014,Carollo1997,Carollo1998,Carollo2002,Lauer2005} and cluster \citep[][]{Cote2006,denBrok2014,Munoz2015,Sanchez-Janssen2019a} environments.
Recently, \citet[][hereafter \cta{Sanchez-Janssen2019a}]{Sanchez-Janssen2019a} investigated NSCs in the core of the Virgo galaxy cluster and combined their results with ancillary data for high-mass galaxies from \citet{Cote2006}.
To compare their results, they added the data of \citet{denBrok2014} for the Coma and \citet{Munoz2015} for the Fornax galaxy clusters.
They found that {\nucfrac} coincides between the Fornax and Virgo galaxy clusters and that it is elevated for the Coma galaxy cluster, and reason that this difference is due to differences in host halo mass.
Further evidence for the role of the environment in determining {\nucfrac} comes from studies that show that nucleated galaxies are preferentially located at the centres of galaxy clusters \citep[e.g.][]{Binggeli1987,Ferguson1989,Lisker2007,Lim2018,Ordenes-Briceno2018}.

To compare {\nucfrac} to a low-density environment, \cta{Sanchez-Janssen2019a} visually inspected and assigned a nuclear classification to \num{55} low-mass satellite galaxies of the Milky Way, M$\,$31, and M$\,$81 using high-resolution \textit{Hubble Space Telescope} ({\HST}) data.
The resulting {\nucfrac} is smaller than for all cluster environments, but the uncertainties on their data points are large and still compatible with the values for both the Fornax and Virgo galaxy clusters (see their Figure~\num{2}).
At the high-mass end ($M_{\star} \geq \SI{e9}{\Msun}$) \citet{Baldassare2014} analysed the central regions of \num{28} galaxies and report a consistency of {\nucfrac} between early-type field galaxies and Virgo galaxy cluster members of similar stellar mass.
However, given the limited sample sizes and coverage in stellar mass and Hubble type, the significance of the offset between {\nucfrac} for field and cluster environments remained unclear.

\citet{Carlsten2021b} investigated the globular cluster and NSC populations of \num{177} low-mass early-type galaxies ($M_{\star} \lesssim \SI{e8.5}{\Msun}$) around \num{27} massive hosts in the Local Volume.
They confirmed the difference in {\nucfrac} between the field and cluster environments at the low-mass end.
Furthermore, they were able to show that galaxies in close proximity to their host galaxy show an elevated {\nucfrac} compared to galaxies further away for all stellar masses.
This led the authors to question whether the parent halo mass or distance from the centre of the halo is the stronger influence on the nucleation fraction.

The Local Volume (LV; $d \lesssim \SI{12}{\Mpc}$) is a natural place to study NSC demographics since the sample of galaxies is nearly complete down to the lowest galaxy masses \citep{Karachentsev2013} and nuclei are at least partially resolved \citep[e.g.][]{Pechetti2019}.
Despite being an ideal laboratory for studying {\nucfrac}, thus far there have been no studies of the nucleation fraction for the complete population of LV galaxies.
In this paper we present our analysis of {\nucfrac} of galaxies in the LV.
The sample of galaxies and volume is large enough that it provides a good proxy for the majority of galaxies that live in field and group environments.
Based on the most complete catalogue of galaxies in the LV available today, we both collect literature nuclear classifications where they are available and classify other galaxies ourselves based on high-resolution {\HST} data.
We use self-consistent mass-to-light ratios from the literature to calculate galaxy stellar masses for all environments in a homogeneous way.
Our final sample contains galaxies of all Hubble types over a wide range of stellar mass ($\SI{e2.5}{\Msun} \lesssim M_{\star} \lesssim \SI{e11.5}{\Msun}$) and environments, from isolated field galaxies to rich groups.
We confirm that the stellar mass is the primary indicator for the nucleation of a galaxy and show that the Hubble type also correlates with {\nucfrac}.
Likewise, we find a correlation between {\nucfrac} and the local environment of a galaxy.
Furthermore, we study {\nucfrac} as a function of galactocentric distance to find that it increases for the LV in the central regions, slightly increased for the Fornax galaxy cluster, and decreases for the Virgo galaxy cluster.
The latter observation seems to disagree with the literature.

The paper is structured as follows:
\S\ref{sec:data} introduces the galaxy catalogue of the LV and briefly describes the nuclear classification scheme of its members without a decisive nuclear classification.
Here we also introduce the data used to calculate {\nucfrac} for galaxy clusters.
In \S\ref{sec:analysis} we highlight the mass determination scheme and show {\nucfrac} for different environments as a function of stellar mass, Hubble type, tidal index, and galactocentric distance.
In this section we also fit {\nucfrac} with a logistic function.
Based on our analysis, we discuss potential consequences for NSC formation and evolution theories in \S\ref{sec:discussion}.
Finally, we conclude in \S\ref{sec:conclusions} and present an outlook for further investigations.

\section{Data}
\label{sec:data}

\subsection{The Local Volume galaxy sample}
\label{subsec:galaxy_sample}

One of the earliest catalogues of LV galaxies contained \num{179} objects within $d\lesssim \SI{10}{\Mpc}$ \citep{Kraan-Korteweg1979}.
Over the years, the number of objects steadily increased and reached \num{869} in \num{2013} following the release of the `Updated Nearby Galaxy Catalog' \citep[UNGC;][hereafter \cta{Karachentsev2013}]{Karachentsev2013}.
It is updated regularly\footnote{The most recent version can be obtained via \url{https://www.sao.ru/lv/lvgdb/tables.php}.} and contains \num{1246} objects as of late February \num{2021}.

The UNGC is the most complete catalogue of the LV which is complete down to $M_{B} = \SI{-11}{mag}$.
For fainter magnitudes the estimated completeness was between \SI{40}{\percent} and \SI{60}{\percent} in \num{2013} (see discussion in \cta{Karachentsev2013}), where half of the ultra-faint dwarf companions around massive galaxies would be missing.
Indeed, recently many new ultra-faint dwarfs were discovered in the LV \citep[e.g.][]{Carlsten2020a,Habas2020} which are currently not part of the UNGC.
Although these studies include a nuclear classification, we do not add them to the UNGC as many distance estimates and apparent magnitudes are missing.
However, their results will be used to investigate potential completeness biases in the UNGC (\textit{cf}.\ \S\ref{subsec:missing_galaxies}).

\cta{Karachentsev2013} constructed the UNGC by setting two limitations to potential members:
(1) the distance estimate of a galaxy must be smaller than \SI{11}{\Mpc} or (2) the radial velocity component of its velocity vector with respect to the Local Group centroid is \SI{\leq 600}{\kilo\metre\per\second}.
The latter choice was motivated by the uncertainty on today's value of the Hubble parameter and perturbations by both the Local Void \citep{Tully1988} and the nearby Virgo galaxy cluster.
As a result, some galaxies have reliable distance estimates larger than \SI{11}{\Mpc} but are only included because of (2).
Since we use LV galaxies as a proxy for the field environment, we keep these galaxies in the catalogue if they do not belong to any galaxy cluster.

We removed twelve objects from the UNGC because they are classified as globular clusters of M$\,$31 \citep[e.g.][]{Huxor2014,McConnachie2018}.
An additional \num{13} galaxies are removed because they contain a `Virgo Cluster Catalog' \citep{Binggeli1985} identification number and are, therefore, classified as Virgo galaxy cluster members.
Finally, we also removed `Kim 2' which has been identified as a globular cluster associated to the Milky Way \citep{Kim2015}.

Consequentially, our updated LV catalogue contains \num{1220} galaxies and is presented in \Cref{tab:data_table_local_volume} (its full version is available online only).
It contains the main identifier, angular coordinates, a distance estimate, apparent magnitudes in the $B$-, $g$-, $V$-, $r$-, and $K$-bands corrected for Galactic extinction, $(B - V)_{0}$ and $(g - r)_{0}$ colours, $E(B - V)$, the Hubble type, three tidal indices and a list of the ten most influential neighbouring galaxies (\textit{cf}.\ \S\ref{subsec:nucleation_fraction_as_a_function_of_tidal_index}), a mass estimate (\textit{cf}.\ \S\ref{subsec:stellar_mass_estimates}), and a nuclear classification with literature references, if available.

\subsection{Galaxy parameters}
\label{subsec:galaxy_parameters}

The UNGC contains some basic galaxy properties including physical coordinates, a Hubble type value (denoted as $T$), and apparent magnitudes in the $B$- and $K_{\mathrm{s}}$-bands.
We use the HyperLEDA\footnote{\url{http://leda.univ-lyon1.fr/}} and SIMBAD\footnote{\url{https://simbad.harvard.edu/simbad/}} data bases to supplement the information from the UNGC for the following reasons:
\begin{enumerate}
  \item[$\bullet$] to calculate stellar masses by adding magnitudes in the Johnson-Cousins and {\SDSS} filter bands ($B$, $g$, $V$, $r$, and $K$);

  \item[$\bullet$] to provide uncertainties on apparent magnitudes and the Hubble type, which are currently lacking;

  \item[$\bullet$] to update $B$-band magnitudes for galaxies for which those value was estimated by eye from a comparison to a `similar looking' galaxy (\textit{cf}.\ \cta{Karachentsev2013});

  \item[$\bullet$] to improve the sampling and range of the Hubble Type, currently limited to $T \in [-3, \, 10]$.
\end{enumerate}

We search both data bases using the angular coordinates from the UNGC and apply a search box and diameter of \SI{10}{\arcsec}, respectively.
The angular coordinates of five galaxies (Draco, Fornax, ESO$\,$174-1, NGC$\,$55, NGC$\,$5194) differ by more than \SI{100}{\arcsec} between the UNGC and the HyperLEDA data base.
For Draco, Fornax, and NGC$\,$55 this difference is due to their large spatial extent.
For NGC$\,$5194 it seems likely that the angular coordinates of M$\,$51 were used in the UNGC.
We are unable to explain the difference for ESO$\,$174-1.
For all five galaxies we adopt the angular coordinates presented in the HyperLEDA data base.

Parameters from both online data bases and the UNGC are combined in the following way:
\begin{enumerate}
  \item[$\bullet$] Main identifier: the main identifier is taken from the UNGC.

  \item[$\bullet$] Distance: we adopt the value given in the UNGC if it is reliable, i.e.\ based on the TRGB, the luminosity of Cepheids or RR Lyrae stars, or surface brightness fluctuations.
  Otherwise, we adopt the \texttt{modbest} value\footnote{The \texttt{modbest} parameter is a weighted average of \texttt{mod0} taken from a distance catalogue and the luminosity distance inferred from the redshift.} from HyperLEDA and, if unavailable, adopt the value from SIMBAD.
  Finally, if no distance estimate is available in either online data base, we use the unreliable value in the UNGC.

  \item[$\bullet$] Magnitudes: From HyperLEDA we adopt total apparent magnitudes in the $B$-, $V$-, and $K$-bands.
  From SIMBAD we take the same Johnson-Cousins magnitudes and additional {\SDSS} magnitudes in the $g$- and $r$- bands, if available\footnote{Most of the $g$- and $r$-band magnitudes stem from the {\SDSS} DR7 \citep{Abazajian2009} and those are model-fit galaxy magnitudes. See \url{http://classic.sdss.org/dr7/algorithms/photometry.html} for details.}.
  The addition of magnitudes to the UNGC values is similar as for the distance values:
  first we consider HyperLEDA, then SIMBAD, and finally the values presented in the UNGC.
\end{enumerate}
From both online data bases we include apparent magnitudes and a distance estimate.
In addition, we search for the corrected asymptotic $(B - V)_{0}$ colour in the HyperLEDA data base (catalogued as \texttt{bvtc}) as it is corrected for internal extinction.
Apparent magnitudes are corrected for foreground Galactic extinction based on the \citet{Schlafly2011} re-calibration of the \citet{Schlegel1998} dust maps, assuming the \citet{Fitzpatrick1999} reddening law of $R_{V} = 3.1$.
Our final $(B - V)_{0}$ magnitudes are calculated as follows:
(1) where available, we use the \texttt{bvtc} parameter from HyperLEDA.
(2) where \texttt{bvtc} is unavailable, we calculate $B - V$ using apparent magnitudes from HyperLEDA, SIMBAD, or the UNGC (see above) and correct the colour for Galactic extinction.
Similarly, the final $(g - r)_{0}$ is calculated using the apparent magnitudes from SIMBAD and correcting for Galactic extinction.
All final apparent magnitudes and $E(B - V)$ are given in \Cref{tab:data_table_local_volume}.

\begin{figure}
  \centering
  \includegraphics[width=\columnwidth]{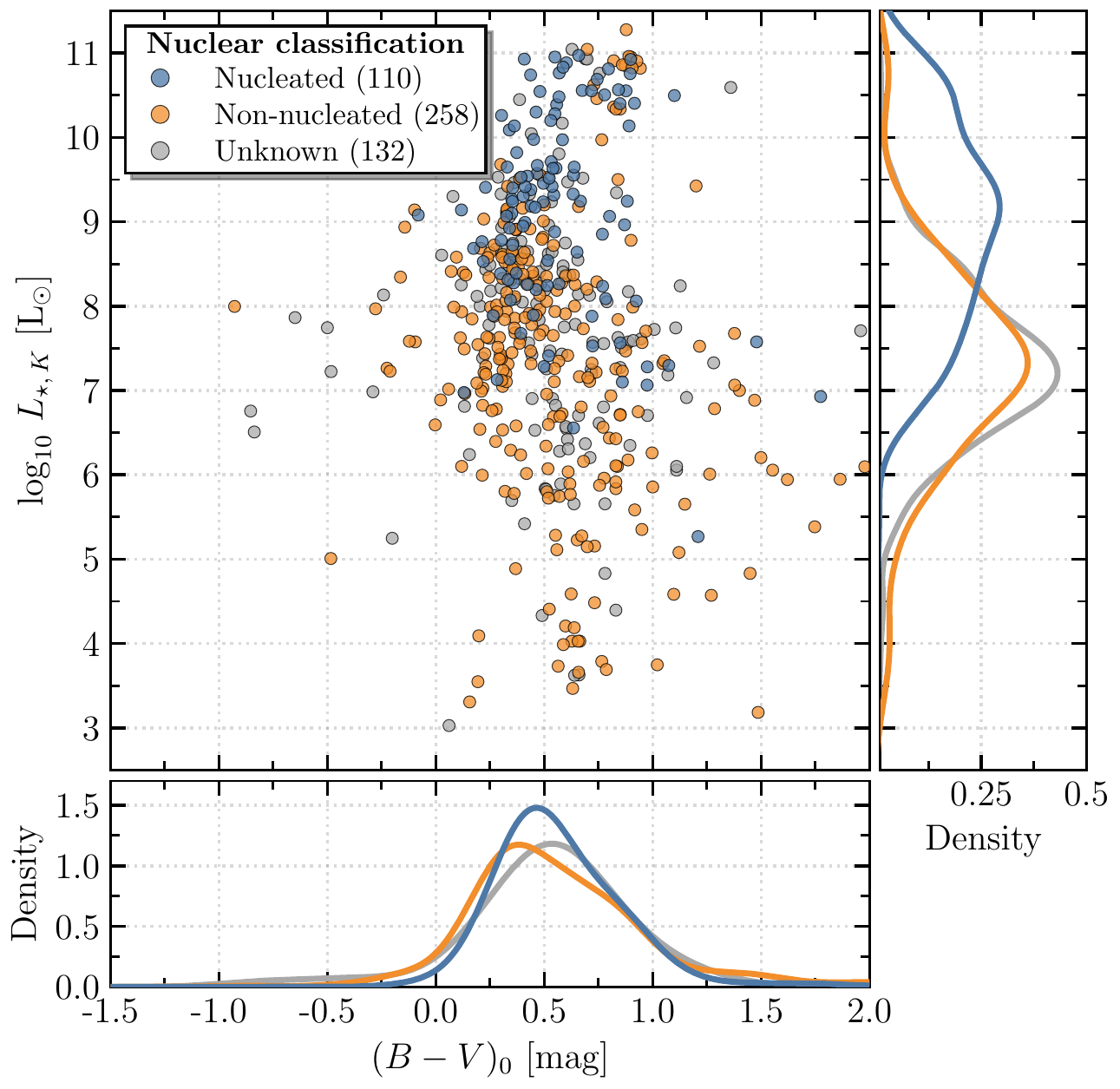}
  \caption{%
    Logarithmic $K$-band luminosity as a function of $(B - V)_{0}$ colour.
    Galaxies are colour-coded based on their nuclear classification into nucleated (blue), non-nucleated (orange) and unknown (gray).
    Uncertainties are omitted for clarity.
    The \textit{right} and \textit{bottom} panels show the Gaussian kernel density estimates along the projected axes.
    As indicated by the numbers in the legend, only part of the Local Volume sample has a $(B - V)_{0}$ colour estimate.
  }
  \label{fig:klum_bmv}
\end{figure}

We show the $K$-band luminosity as a function of optical $(B - V)_{0}$ colour in \Cref{fig:klum_bmv} where galaxies are already split based on their nuclear classification (\textit{cf}.\ \S\ref{subsubsec:galaxy_classification_local_volume_galaxies}).
For \num{842} galaxies (\SI{\approx67.6}{\percent}) a direct measurement of the $K$-band magnitude of a galaxy is unavailable.
The $K$-band magnitude is a good proxy to estimate galaxy stellar masses if optical $(B - V)_{0}$ or $(g - r)_{0}$ colours are unavailable.
Therefore, we calculate missing apparent $K$-band magnitudes $m_{K}$ by using the apparent $B$-band magnitude $m_{B}$ and the Hubble type value $T$ via
\begin{equation}
  \label{equ:b_k_mag}
  \langle m_{B} - m_{K} \rangle_{0} = 
  \begin{cases}
    \num{4.10}                       & \mathrm{for} \hspace{1ex} T < \num{3} \, , \\
    \num{2.35}                       & \mathrm{for} \hspace{1ex} T > \num{8} \, , \\
    \num{4.60} - \num{0.25} \times T & \mathrm{otherwise}                    \, ,
  \end{cases}
\end{equation}
which is taken from the 2MASS Large Galaxy Atlas \citep{Jarrett2003} and was already used by \cta{Karachentsev2013} for the UNGC\footnote{Due to the updated $m_{B}$ and $T$ values, we do not adopt $m_{K}$ from the UNGC, if calculated via \Cref{equ:b_k_mag}.}.
The relation is based on the observation that $B - K$ depends on Hubble type ranging between $\approx 4$ and $\approx 2$ for early- and late-type galaxies, respectively \citep[Figure 19 of][]{Jarrett2003}.
Note that the scatter of $B - K$ increases with Hubble type to \num{\approx 2} for $T = 9$, resulting in large uncertainties.

The uncertainties for the galaxy parameters in our catalogue are obtained as follows:
depending on the filter band, up to one third of all galaxies have no uncertainty tabulated on their apparent magnitude.
For those cases we adopt an uncertainty of \SI{0.5}{\mag} which equals the value adopted by \cta{Karachentsev2013} for their `visually' derived $B$-band magnitudes.
In addition, we consider uncertainties on $g$- and $r$-band magnitudes in SIMBAD below \SI{0.1}{\mag} to be unreliable\footnote{For high-mass galaxies the uncertainty on the apparent magnitude often is $\delta m = \SI{0.001}{\mag}$ which likely does not take into account systematic contributions.} and adopt a value of \SI{0.1}{\mag}.
For galaxies without an uncertainty on their Hubble type (\num{358} galaxies) we assume an uncertainty of \num{1.0} which seems to be a typical assumption made by other works like \cta{Karachentsev2013} and \citet{Jarrett2003}.

To compare the nucleation fraction as a function of stellar mass for different environments, the galaxy parameters given in the reference papers (see next sections) are insufficient.
For both the Fornax and Virgo galaxy clusters we consider all galaxies flagged as members by the HyperLEDA data base (\num{337} and \num{2093} galaxies, respectively).
For the Coma galaxy cluster we use the data set of \citet{denBrok2014}.
We search the HyperLEDA and SIMBAD data bases for apparent magnitudes and Hubble types in the same way as for LV galaxies, including an estimate of their $K$-band magnitudes via \Cref{equ:b_k_mag}, if missing.
For the NGVS sample of \cta{Sanchez-Janssen2019a}, we use the absolute $g$- and $r$- magnitudes from their Table 4.
We assume the following distances if no measurement is available:
$d_{\mathrm{Coma}} = \SI{35.10 \pm 0.55}{\mag}$ \citep{Ferrarese2000b}, $d_{\mathrm{Fornax}} = \SI{31.51 \pm 0.15}{\mag}$ \citep{Blakeslee2009}, and $d_{\mathrm{Virgo}} = \SI{31.09 \pm 0.01}{\mag}$ \citep{Mei2007}.

\subsection{Galaxy nuclear classification}
\label{subsec:galaxy_classification}

\subsubsection{Local Volume}
\label{subsubsec:galaxy_classification_local_volume_galaxies}

Roughly a third of the galaxies (451) in the predecessor of the UNGC \citep[called `Catalog of Neighboring Galaxies'; ][]{Karachentsev2004} have been broadly classified as containing a `star-like nucleus' \citep{Karachentsev2002}.
The large amount of recent literature and available high-resolution imaging data in the {\HST} archive prompted us to make a comprehensive update to the galaxy nucleation classification in the LV as follows:
we performed an \textit{ADS}\footnote{\url{https://ui.adsabs.harvard.edu/}} search based on galaxy (object) name taken from the UNGC and HyperLEDA.
The abstract keyword search contained `nuclear star cluster', `NSC', `nuclear cluster', and `nucleus'.
To mostly include reliable classifications which are primarily based on {\HST} data, we limited the publication date to be more recent than January \num{1990}.
For \num{310} galaxies we detected literature sources which are given in \Cref{tab:data_table_local_volume}.

We search through the \textit{Hubble Legacy Archive}\footnote{\label{note1}\url{https://hla.stsci.edu}} ({\HLA}) to classify galaxies whose literature classification is ambiguous or non-existent.
Given the small spatial extent of NSCs (effective radius $\lesssim \SI{50}{\pc}$, \citealp[e.g.][]{Neumayer2020}), we limit our search to the high-resolution {\HST} cameras ACS, WFC3, and WFPC2, and search within a radius of \SI{36}{\arcsecond} around the galaxy coordinates in our catalogue.
No restrictions are set on the filter, but we focused our analysis on red / NIR filters to minimise effects due to dust extinction.
This eases the confirmation of extended sources and thus allows for a reliable nuclear classification.
Including the previously removed globular clusters of M$\,$31 and Virgo galaxy cluster members, we find available {\HLA} data for \num{633} objects ($\approx \SI{50.8}{\percent}$) in at least one filter.
With the above constraints, no {\HST} data is found for the remaining \num{613} ($\approx \SI{49.2}{\percent}$) galaxies.

Using the available {\HLA} data, we classify galaxies with ambiguous or without a nuclear classification into nucleated (`1'), non-nucleated (`0'), or `remains unknown' (`?').
The latter category is adopted if all available data is either significantly affected by dust extinction or if a galaxy centre is not within the {\HST} camera field of view.
We do \emph{not} re-analyse galaxies previously classified as nucleated, but we did look at all galaxies without an NSC and with available {\HST} data.
The classification process includes a visual inspection of all available {\HST} data, a determination of the position of the potential NSC with respect to the photometric centre of the host galaxy at different isophotal radii, and a verification that the potential NSC is extended by using an oversampled PSF and two-dimensional fitting techniques.
This scheme is comparable to the one described in \citet{Georgiev2014} and will be detailed further in an upcoming paper (Hoyer et al.~in prep.).
In that paper we will also discuss the properties of the newly discovered NSCs in the LV while this paper solely focuses on the nucleation fraction.

In total, using the {\HLA} data for 633 objects we identify 22 new detections of NSCs.
In combination with the compilation of literature classifications, this results in \num{145} nucleated and \num{456} non-nucleated galaxies.
Currently, \num{46} galaxies with available {\HST} data are part of the `remains unknown' category (`?' in the data table) where no nuclear classification could be assigned.
All galaxies and their parameters used in this work are given in \Cref{tab:data_table_local_volume}.
Their nuclear classifications are grouped into (`1'), (`0'), and (`?').
For simplicity, the latter category combines galaxies with insufficient and without available {\HST} data (the `unknown' distribution of galaxies in \Cref{fig:klum_bmv}).

The nuclear classification of galaxy cluster members from the literature sources is described but \emph{not} verified in the following sections.

\subsubsection{Coma galaxy cluster}
\label{subsubsec:galaxy_classification_coma_galaxy_cluster}

As part of the Advanced Camera for Surveys Coma Cluster Survey (ACSCCS), \citet{denBrok2014} used {\HST} ACS F814W images and PSFs created with \textsc{TinyTim} \citep{Krist1993,Krist1995} to assign a nuclear classification to their data.
If a fit with a central Gaussian source yielded a higher evidence than a fit without a central source, the galaxy was classified as nucleated.
Therefore, we classify all galaxies as nucleated if an apparent magnitude for the central source of a galaxy is available in their Table A1.
If no apparent magnitude is presented, we assume that the galaxy is not nucleated.
Note that the original data table contains five duplicates and bad identifiers for some galaxies which have been removed (Mark den Brok, priv. comm.).

Recently, \citet{Zanatta2021} investigated the core of the Coma galaxy cluster using {\HST} ACS imaging to classify \num{66} elliptical dwarf galaxy candidates.
However, due to a lack of photometric parameters needed to determine stellar masses, we do not use their data set.

\subsubsection{Fornax galaxy cluster}
\label{subsubsec:galaxy_classification_fornax_galaxy_cluster}

\citet{Munoz2015} used the Next Generation Fornax Cluster Survey (NGFS) to classify dwarf galaxies in the central regions of the Fornax galaxy cluster.
They used \textsc{galfit} \citep{Peng2010} to fit the surface brightness of a galaxy with a S{\'{e}}rsic profile \citep{Sersic1968}, and used an additional profile if needed to fit the NSC.

\citet{Turner2012} analysed \num{43} high-luminosity galaxies in the Fornax galaxy cluster as part of the Advanced Camera for Surveys Fornax Cluster Survey (ACSFCS).
Both studies classify galaxies as either nucleated or non-nucleated.
Two galaxies are part of both surveys and were assigned the same nuclear classification.

Furthermore, the Fornax Deep Survey \citep[FDS; ][]{Iodice2016} covers the whole Fornax galaxy cluster and the Fornax A group.
While both \citet{Venhola2018} and \citet{Su2021} present nuclear classifications for the dwarf elliptical cluster members, their classifications differ for \num{55} galaxies despite using the same photometric data products.
In addition, their nuclear classifications do not agree with \citet{Munoz2015} for \num{175} and \num{26} galaxies, respectively.
As a result, we do not use the data of \citet{Venhola2018} at all and only consider the data of \citet{Su2021} for our analysis of the nucleation fraction versus galactocentric radius (\textit{cf}.\ \Cref{subsec:nucleation_fraction_as_a_function_of_galactocentric_distance}).
Finally, four duplicates have been removed from the FDS: FDS12\_0241, FDS25\_0232, FDS25\_0296, and FDS4\_0098.

\subsubsection{Virgo galaxy cluster}
\label{subsubsec:galaxy_classification_virgo}

\cta{Sanchez-Janssen2019a} investigated NSCs in mainly low-mass systems in the core of the Virgo galaxy cluster in the Next Generation Virgo Cluster Survey (NGVS).
For NSC identification, the authors modelled the galaxy body with a single S{\'{e}}rsic profile and added a second component for an NSC, if central light excess was found.

Based on the Advanced Camera for Surveys Virgo Cluster Survey (ACSVCS), \citet{Cote2006} classified mainly high-luminosity early-type galaxies.
Their classification is split into three classes which contain further subcategories (their Table 2):
nucleated (`I'), non-nucleated (`II'), and unknown (`0').
We only consider galaxies classified as (`Ia') and (`Ib') as nucleated as the authors considered (`Ic'), (`Id'), and (`Ie') as uncertain.

\citet{Lisker2007} assigned a nuclear classification to \num{413} early-type dwarf galaxies using \textit{SDSS} imaging data.
The authors note that it seems likely that some galaxies classified as non-nucleated may host a faint NSC but escaped detection due to high-surface brightness.
Therefore, we remove the classification of galaxies flagged as non-nucleated and `bright' in their Table 1.
Because this choice will artificially elevate (i.e.\ bias) the nucleation fraction towards higher values, we only use their data in \Cref{subsec:nucleation_fraction_as_a_function_of_galactocentric_distance} where we investigate the nucleation fraction versus galactocentric distance.
In that section we are only interested in general trends of {\nucfrac} and not in exact values.

Data tables for the Coma, Fornax, and Virgo galaxy clusters are presented in \Cref{tab:data_table_coma}, \Cref{tab:data_table_fornax}, and \Cref{tab:data_table_virgo}, respectively.
Their structure is the same as for the LV (\textit{cf}.\ \Cref{tab:data_table_local_volume}), however, in addition we include the nuclear classifications of each survey.
The full data tables are available in the online supplementary material.

\section{Analysis}
\label{sec:analysis}

\subsection{Galaxy stellar mass estimate}
\label{subsec:stellar_mass_estimates}

The stellar photometric mass of a galaxy is often calculated by fitting its spectral energy distribution (SED) or by using (colour-dependent) mass-to-light ratios (denoted as {\MLR}).
SED fitting gives a more accurate stellar mass, as exemplified by the analysis of \cta{Sanchez-Janssen2019a} of more than \num{400} galaxies finding a typical uncertainty of \SI{0.15}{\dex}.
This is considerably smaller than the scatter derived from single colour-based {\MLR} \citep[approximately \SI{0.5}{\dex}, e.g.][]{McGaugh2014}, nevertheless, it has been shown to yield very comparable and unbiased results compared to SED fitting \citep[e.g.][]{Roediger2015}.

Therefore, we calculate the galaxy photometric stellar mass using the $B$-, $g$-, $V$-, and $r$--band magnitudes from our catalogue via the {\MLR}-colour relation $\log_{10} (M_{\star} / L)_{\lambda} = \alpha_{\lambda} + \beta_{\lambda} \times (\mathrm{colour})$, where $\lambda \in [V, \, r]$ and $\mathrm{colour} \in [(B - V)_{0}, \, (g - r)_{0}]$.
We chose these two colours because (1) they are most readily available for the majority of LV galaxies, and (2) although rigorously investigated and debated in the literature, which colour is the ideal proxy for the stellar mass of a galaxy, \citet{McGaugh2014} and \citet{Du2020b} find that $(B - V)_{0}$ and $(g - r)_{0}$, respectively, are among the best proxies and are less sensitive to the uncertain contribution of thermally pulsating AGB stars.
This is not surprising because both $(B, \, V)$ and $(g, \, r)$ filter transmission curves are very close to each other.

The works of \citet{McGaugh2014} and \citet{Du2020b} build upon earlier studies by \citet{Bell2003}, \citet{Portinari2004}, \citet{Zibetti2009}, \citet{Into2013}, and \citet{Roediger2015} and re-calibrate their relations to ensure self-consistency within the model:
as pointed out by \citet{McGaugh2014}, NIR luminosities are over-predicted compared to optical luminosities leading to different {\MLR} and, therefore, to different stellar mass estimates.
All coefficients $(\alpha_{V}, \, \beta_{V})$ and $(\alpha_{r}, \, \beta_{r})$ that we use for the calculation of stellar masses are listed in \Cref{tab:mass_to_light_ratios}.
We also give $M_{\star} / L_{K}$ from \citet{McGaugh2014} for an assumed $(B - V)_{0} = \SI{0.6}{\mag}$ as they show that a stellar mass estimate from the NIR luminosity is only weakly dependent on colour.
Also, the scatter of {\MLR} is expected to be smaller in the NIR compared to that in the optical \citep{Bell2001,Portinari2004}.
The assumption of $(B - V)_{0} = \SI{0.6}{\mag}$ seems to be justified for the LV (\textit{cf}.\ \Cref{fig:klum_bmv}).

All five publications differ in their model by either using a different initial mass function or stellar population models, both of which influence {\MLR} and contribute to the systematic bias and uncertainty.
\citet{Bell2003} and \citet{Roediger2015} test their models on galaxies of different Hubble type whereas the other three studies mainly focus on late-type galaxies.
However, the difference between calculated stellar masses from different models (typically \SI{\leq 0.3}{\dex}) is often below the uncertainty of each value (typically \SI{\approx 0.4}{\dex}).
To minimise the bias from adopting a specific {\MLR}, we use the relations from all publications, i.e.\ the coefficients in \Cref{tab:mass_to_light_ratios} and depending on the filter availability, we calculate up to nine stellar masses per galaxy from the two colours\footnote{Four masses estimates stem from $(B - V)_{0}$ and five from $(g - r)_{0}$, respectively.}.
We remind again that if the \texttt{bvtc} colour is unavailable, only then we use the calculated optical $(B - V)_{0}$ colour.

For each calculated stellar mass we determine its formal uncertainty by error propagation of the uncertainty on the {\MLR}, its photometry, the absolute magnitude of the Sun\footnote{Assumed to be \SI{0.04}{\mag}; \url{http://mips.as.arizona.edu/~cnaw/sun.html}}, and the distance measurement.
The uncertainty on {\MLR} is assumed to be \SI{0.3}{\dex} for all mass-to-light ratios.
This value roughly corresponds to the dispersion in stellar mass-to-light ratio between \citet{McGaugh2014} and \citet{Du2020b} for the relation of \citet{Zibetti2009} caused by different reference stellar masses.
We note that this uncertainty of \SI{0.3}{\dex} dominates the stellar mass error-budget for most galaxies.
The final stellar mass, as reported in our data tables, are calculated as the uncertainty weighted mean from the masses computed from each {\MLR}.
The final uncertainty equals the standard error of the weighted mean.
To avoid systematic differences between environments due to different calculations of stellar mass performed by other studies, we re-calculate stellar masses for all Coma, Fornax, and Virgo galaxies using the same approach (\textit{cf}.\ \Cref{subsubsec:consistency_of_stellar_masses_with_previous_studies}).
Depending on the data set, the mean standard deviations between individual mass estimates ranges between \SI{0.09}{\dex} and \SI{0.12}{\dex}.

Note that there are exceptions to this scheme.
Some galaxies have unreliable apparent magnitudes that result in poor stellar mass estimates.
For these galaxies we adopt the stellar mass based on $M_{\star} / L_{K}$ given in \Cref{tab:mass_to_light_ratios}.
The masses for the Milky Way and the Sag dSph galaxy are assumed to be the following: $M_{\star, \, \mathrm{MW}} = \SI{6.43e10}{\Msun}$ \citep{McMillan2011} and $M_{\star, \, \text{Sag dSph}} = \SI{1e8}{\Msun}$ \citep{Vasiliev2020}.
Although \cta{Sanchez-Janssen2019a} calculate stellar masses using the more accurate SED fitting technique, we will use our mass estimates for further analysis to avoid a systematic bias in mass estimates between different environments.
In the following section we show that our calculated stellar masses are consistent with their values.
\begin{table}
  \small
  \caption{%
    Parameters $(\alpha_{V}, \, \beta_{V})$ and $(\alpha_{r}, \, \beta_{r})$ are used to calculate stellar masses and are taken from \citet{McGaugh2014} and \citet{Du2020b}, respectively.
    The authors re-calibrate the initial parameters provided by \citet{Bell2003,Portinari2004,Zibetti2009,Into2013}, and \citet{Roediger2015} to ensure self-consistency.
    `BC03' and `FSPS' refer to the models of \citet{Bruzual2003} and \citet{Conroy2009}.
  }
  \begin{center}
    \begin{threeparttable}
      \begin{tabular}{%
        S[table-format=1.3]
        S[table-format=1.3]
        S[table-format=1.3]
        S[table-format=1.3]
        S[table-format=1.2]
        S[table-format=1.2]
        c
      }
        \toprule
        {$\alpha_{V}$} & {$\beta_{V}$} & {$\alpha_{r}$} & {$\beta_{r}$} & \multicolumn{2}{c}{$M_{\star}\ /\ L_{K}$} & \multirow{2}{*}{Ref.\tnote{(c)}} \\
        \cmidrule(lr){1-2}
        \cmidrule(lr){3-4}
        \cmidrule(lr){5-6}
        \multicolumn{2}{c}{Revised\tnote{(a)}} & \multicolumn{2}{c}{Revised\tnote{(b)}} & {Original} & {Revised\tnote{(a)}} & {} \\
        \midrule

        -0.628 & 1.305  & -0.306 & 1.097  & 0.73   & 0.60   & {(1)} \\
        -0.654 & 1.290  & {-{-}} & {-{-}} & 0.50   & 0.60   & {(2)} \\
        -1.075 & 1.837  & -0.840 & 1.654  & 0.21   & 0.50   & {(3)} \\
        -0.900 & 1.627  & -0.663 & 1.530  & 0.41   & 0.54   & {(4)} \\
        {-{-}} & {-{-}} & -0.792 & 1.629  & {-{-}} & {-{-}} & {(5)} \\
        {-{-}} & {-{-}} & -0.647 & 1.497  & {-{-}} & {-{-}} & {(6)} \\

        \bottomrule
      \end{tabular}
      \begin{tablenotes}
        \item[(a)] Revised by \citet{McGaugh2014} to ensure self-consistency.
        \item[(b)] Revised by \citet{Du2020b} to ensure self-consistency.
        \item[(c)] (1) \citet{Bell2003}; (2) \citet{Portinari2004}; (3) \citet{Zibetti2009}; (4) \citet{Into2013}; (5) \citet[][BC03]{Roediger2015}; (6) \citet[][FSPS]{Roediger2015}.
      \end{tablenotes}
    \end{threeparttable}
  \end{center}
  \label{tab:mass_to_light_ratios}
\end{table}

\subsubsection{Consistency of stellar masses with previous studies}
\label{subsubsec:consistency_of_stellar_masses_with_previous_studies}

\begin{figure}
  \centering
  \includegraphics[width=\columnwidth]{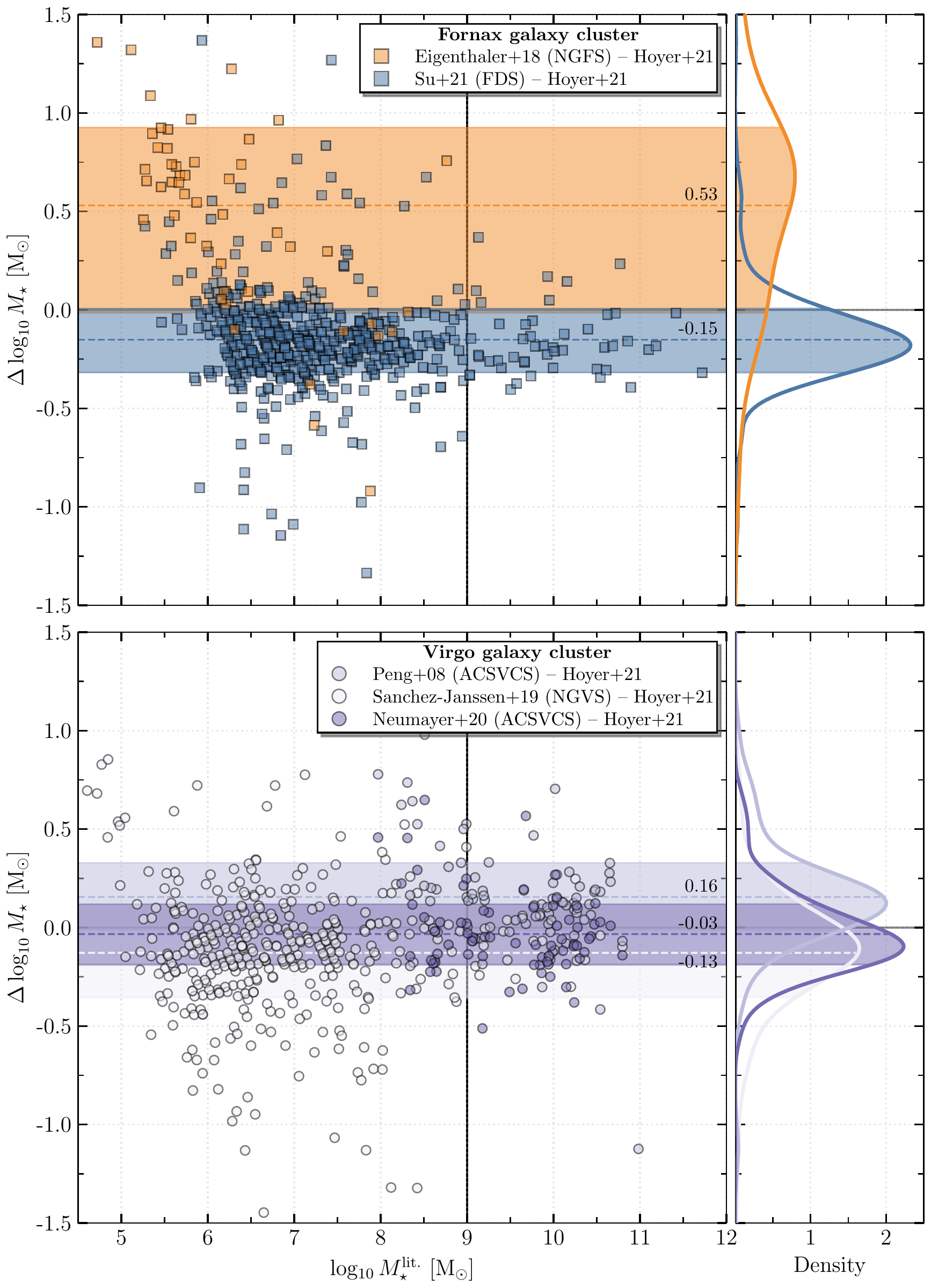}
  \caption{%
    Comparison between stellar mass estimates in the literature and our values.
    \textit{Top panel}: Comparison to \citet{Eigenthaler2018} for the NGFS and \citet{Su2021} for the FDS (both Fornax galaxy cluster).
    \textit{Bottom panel}: Comparison to \citet{Peng2008} and \citet{Neumayer2020} for the ACSVCS, and \citet{Sanchez-Janssen2019a} for the NGVS (all Virgo galaxy cluster).
    The means of the logarithmic mass differences are shown with dashed lines and the $1\sigma$ interval is highlighted with a shaded region.
    One-dimensional Gaussian kernel density estimates present the density of each distribution (\textit{right panels}).
    Typical uncertainties on the stellar masses range between \SI{0.3}{\dex} and \SI{0.5}{\dex} but are omitted for clarity.
  }
  \label{fig:mass_comparison}
\end{figure}

To test the robustness of our stellar mass estimates, we compare our values to the literature values of \citet{Eigenthaler2018} for the NGFS, \citet{Su2021} for the FDS, \citet{Peng2008} and \citet{Neumayer2020} for the ACSVCS, and \cta{Sanchez-Janssen2019a} for the NGVS.
The difference between our mass estimates and the literature values is shown in \Cref{fig:mass_comparison}.
A thick dashed line gives the mean of the difference and shaded regions give the $1\sigma$ interval.
For reference, the solid line at \SI{e9}{\Msun} highlights the peak in nucleation fraction found by \cta{Sanchez-Janssen2019a}.

A comparison with the data of \citet{Eigenthaler2018} for the NGFS (top panel; orange colour) shows the largest scatter with the $1\sigma$ interval covering a range of \SI{\approx 1}{\dex}.
The mean of the difference has a value of \SI{\approx 0.53}{\dex} and is significantly different from zero.
The authors base their mass estimates on non-revised the stellar mass-to-light ratios from \citet{Bell2003}.
In comparison, the agreement with the stellar mass estimates of \citet{Su2021} for the FDS (top panel; blue colour) is better with a mean difference of \SI{\approx -0.15}{\dex}.

Likewise, the comparison to the data of \cta{Sanchez-Janssen2019a} (bottom panel; light violet colour) gives a similar value for the mean difference (\SI{\approx -0.13}{\dex}).
The comparison for the ACSVCS mass estimates of \citet{Peng2008} (bottom panel; medium violet colour) results in a similar value but with opposite sign (\SI{\approx 0.16}{\dex}).
Finally, a comparison to the mass estimates of \citet{Neumayer2020} (bottom panel; dark violet colour) shows the best agreement with a difference of only \SI{\approx -0.03}{\dex}.

As a result, it is important to use a homogeneous approach to estimate stellar masses for galaxies in different environments, especially at the low-mass end.
In other words, adopting stellar mass estimates from different literature sources introduces different systematic biases which then affect our analysis of {\nucfrac}.
Although the overall agreement with literature estimates is below the typical uncertainty on stellar masses (\SI{\approx 0.3}{\dex} in log-space) we use our mass estimates for all environments.

\subsection{Nucleation fraction as a function of galaxy stellar mass}
\label{subsec:nucleation_fraction_as_a_function_of_stellar_mass}

The nucleation fraction is calculated as the number of nucleated galaxies divided by the total number of classified galaxies.
Here we examine its dependence on the galaxy stellar mass \citep[see also e.g.\ Figure \num{3} in][]{Neumayer2020} including our LV galaxy measurements.
We choose uniform binning across the whole mass range (unlike \cta{Sanchez-Janssen2019a}) with a bin width of \SI{0.5}{\dex}.

\begin{figure}
  \centering
  \includegraphics[width=\columnwidth]{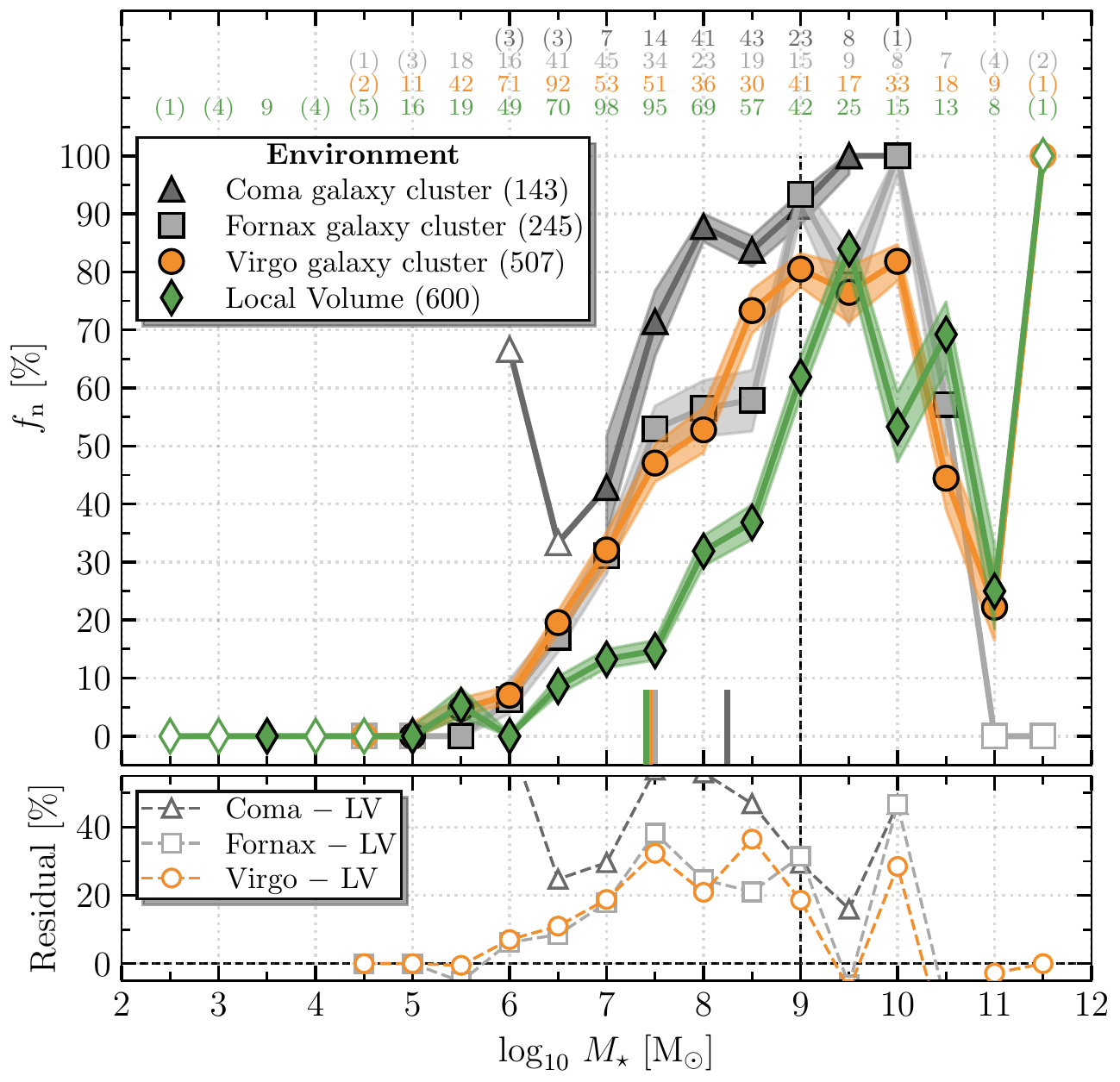}
  \caption{%
    \textit{Top panel}: Nucleation fraction ({\nucfrac}) as a function of logarithmic stellar mass ($\log_{10} \, M_{\star}$) for the Local Volume (green diamonds) and the galaxy clusters Coma (dark gray triangles), Fornax (light gray squares), and Virgo (orange circles).
    Shaded areas indicate the \SI{68}{\percent} confidence interval and the number of galaxies per bin is indicated at the top.
    Open symbols show mass bins with fewer than seven contributing galaxies.
    The vertical dashed line shows the peak of {\nucfrac} identified by \citet{Sanchez-Janssen2019a}.
    Colour-coded lines at the bottom show the mean mass for each by environment.
    The mean mass for the Local Volume and the Virgo galaxy clusters are similar which is why only the vertical line for the Local Volume is visible.
    The total number of galaxies are shown in the legend.
    For the Coma galaxy cluster, we use the data set of \citet{denBrok2014} (ACSCCS).
    For the Fornax galaxy cluster, we use the data sets of \citet{Munoz2015} (NGFS), \citet{Turner2012} (ACSFCS), and four galaxies of \citet{Georgiev2014}.
    Finally, for the Virgo galaxy cluster, we use the data sets of \citet{Sanchez-Janssen2019a} (NGVS), \citet{Cote2006} (ACSVCS), and \num{25} galaxies of \citet{Georgiev2014}.
    \textit{Bottom panel}: Difference between nucleation fractions as a function of stellar mass.
    The horizontal dashed line shows agreement between {\nucfrac} of the Local Volume and a galaxy cluster.
  }
  \label{fig:nucfrac_stellar_mass}
\end{figure}

In \Cref{fig:nucfrac_stellar_mass} we show the nucleation fraction ({\nucfrac}) as a function of logarithmic stellar mass ($\log_{10} \, M_{\star}$) for the LV (green diamonds) and the galaxy clusters Coma (dark gray triangles), Fornax (light gray squares), and Virgo (orange circles), respectively.
A \SI{68}{\percent} confidence interval\footnote{Based on the Agresti-Coull interval; \citealp{Agresti1998}.} of a binomial distribution is shown with colour-shaded areas and the number of galaxies per mass bin is given at the top.
Surveys and publications from which we obtained data for the galaxy clusters are indicated in the caption.
The total number of galaxies per environment are indicated in the legend.
Mass bins containing fewer than seven galaxies are shown with open symbols and without the shaded confidence intervals.
A vertical dashed line at \SI{e9}{\Msun} indicates the peak of the nucleation fraction as identified by \cta{Sanchez-Janssen2019a}.
Colour-coded vertical lines at the bottom of the panel show the mean mass of classified galaxies split by environment.
In addition to the previously mentioned reference papers for each galaxy cluster, we add \num{4} and \num{25} classified galaxies from \citet{Georgiev2014} for the Fornax and Virgo galaxy clusters, respectively, to include late-type galaxies in galaxy clusters and to increase the statistical significance at the high-mass end.
The data of \citet{Lisker2007} and \citet{Su2021} are not included.

We see that, regardless of environment, {\nucfrac} is a strong function of $M_{\star}$ and peaks between \SI{e9}{\Msun} and \SI{e9.5}{\Msun}.
A direct comparison between the different {\nucfrac} suggest an environmental dependence, i.e.\ at fixed $M_{\star}$, {\nucfrac} is higher for denser galaxy cluster environments (up to \SI{\approx 40}{\percent} in the mass range \SI{e7.5}{\Msun} to \SI{e9}{\Msun} for the Fornax and Virgo galaxy clusters).
Differences beyond \SI{e9.5}{\Msun} may be attributed to low statistics.
We investigate possible selection biases due to classified versus unclassified galaxies and non-detections of nuclei due to the resolution limits of {\HST} in \S\ref{sec:selection_effects}.
We note that our results may differ from those presented in \cta{Sanchez-Janssen2019a} due to different mass calculations being used to ensure consistency between the samples (see previous section).

Independent of environment, nucleation seems to start in the range \SIrange{e5}{e5.5}{\Msun} and peaks between \SI{e9}{\Msun} and \SI{e9.5}{\Msun}.
Despite poor number statistics at the highest masses, nucleation beyond \SI{\approx e10.5}{\Msun} is quite rare, regardless of environment.
This suggests that nucleation at these stellar masses may not be dependent on environment, which could be due to the emergence and influence of supermassive black holes (\textit{cf}.~\S\ref{subsec:intermediate_high_mass_galaxies}) and the fact that the gas accretion to (re-)build NSCs after mergers cannot be sustained in these high-mass galaxies \citep[e.g.][]{Antonini2015}.

\subsubsection{A generic model}
\label{subsubsec:generic_model}

To study the effects of the stellar mass of a galaxy and its environment in more detail, we use a simple model to fit the nucleation fraction.
For simplicity and the apparent shape of the nucleation fraction, we model it with a logistic function $L$ with slope $\kappa$ and midpoint (x-axis offset) $M_{50}$,
\begin{equation}
  \label{equ:model}
  L(\kappa, \, M_{50}) = \frac{1}{1 + \exp [ -\kappa (\log_{10} M_{\star} - M_{50}) ] } \quad .
\end{equation}
Note that the amplitude of the logistic function is set to one and {\nucfrac} values are only fit to galaxies with $M_{\star} < \SI{e9.5}{\Msun}$ where the behaviour of {\nucfrac} with galaxy mass is found to be monotonic in all environments.

\begin{figure*}
  \centering
  \includegraphics[width=\textwidth]{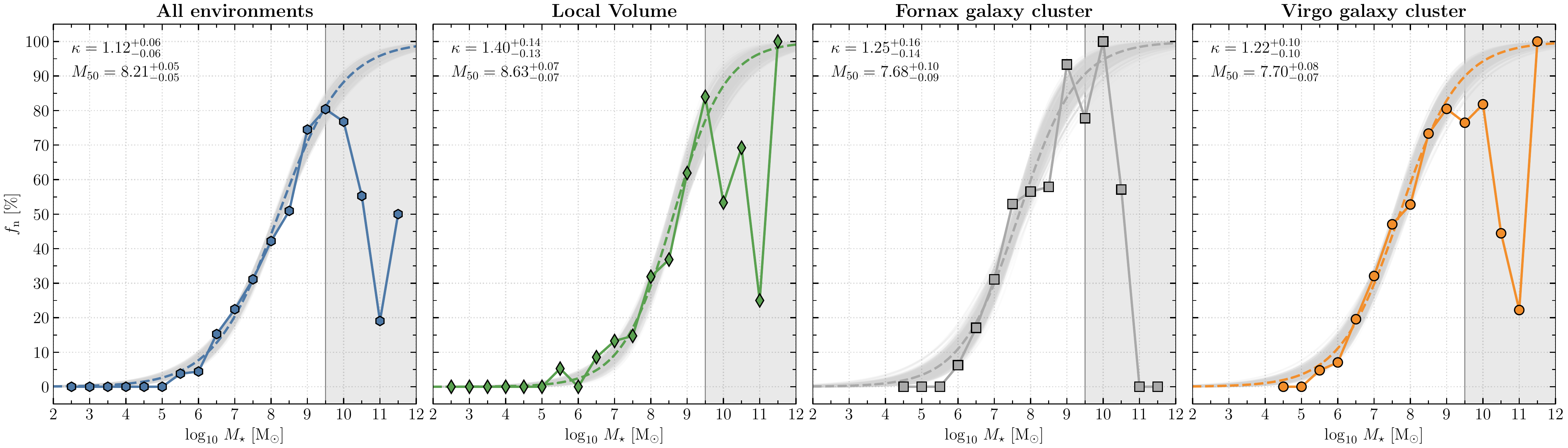}
  \caption{%
    Nucleation fraction ({\nucfrac}) as a function of logarithmic stellar mass ($\log_{10} \, M_{\star}$) for all combined environments (\textit{first}), the Local Volume (\textit{second}), and the Fornax and Virgo galaxy clusters (\textit{third} and \textit{fourth panels}).
    Data above $M_{\star} = \SI{e9.5}{\Msun}$ (gray shaded area) are excluded from the Markov-Chain Monte Carlo (MCMC) analysis.
    The best fit is shown with a dashed colour-coded line.
    Additionally, we show \num{500} randomly drawn MCMC samples for each analysis.
    The parameters for the best fitting results are indicated where $\kappa$ is the slope and $M_{50}$ the midpoint of the logistic function.
  }
  \label{fig:fit_results}
\end{figure*}

To fit this model to the unbinned data of all environments\footnote{We do not fit the data for the Coma galaxy cluster due an incompleteness of galaxies at the low-mass end. See \citet{Zanatta2021} for a recent investigation.}, we use the \texttt{emcee} \citep{Foreman-Mackey2013} package to perform a Markov-Chain Monte Carlo analysis.
Further details as well as a corner plot for the each data set are presented in \Cref{sec:mcmc_analysis}.
The resulting parameters of our model are presented in \Cref{tab:fit_results}.
\Cref{fig:fit_results} compares the best fitting model to the data.

The slope and midpoints of the logistic functions differ between a field and cluster environment:
For the Fornax and Virgo galaxy clusters the parameters are practically the same while for the LV the midpoint of the logistic function is shifted to higher masses by \SI{\approx 1}{\dex}, and the slope paramater is \SI{10}{\percent} to \SI{15}{\percent} larger (and thus steeper) than in the cluster environments.
\begin{table}
  \small
  \caption{%
    Logistic function model parameters $\kappa$, $M_{50}$, and $\tau$ obtained from the MCMC posteriors by fitting \Cref{equ:model} to the unbinned data of the respective environments.
    The uncertainties are the posterior $1\sigma$ percentiles (\textit{cf}.\ \Cref{fig:lv_corner,fig:f_corner,fig:v_corner,fig:a_corner}).
  }
  \begin{center}
    \begin{threeparttable}
      \begin{tabular}{
        l
        c
        c
      }
      \toprule
      Environment & $\kappa$ & $M_{50}$\\
      \midrule

      {All environments}      & $1.12^{+0.06}_{-0.06}$ & $8.21^{+0.05}_{-0.05}$ \\[10pt]
      {Local Volume}          & $1.40^{+0.14}_{-0.13}$ & $8.63^{+0.07}_{-0.07}$ \\[3pt]
      {Fornax galaxy cluster} & $1.25^{+0.16}_{-0.14}$ & $7.68^{+0.10}_{-0.09}$ \\[3pt]
      {Virgo galaxy cluster}  & $1.22^{+0.10}_{-0.10}$ & $7.70^{+0.08}_{-0.07}$ \\

      \bottomrule
      \end{tabular}
    \end{threeparttable}
  \end{center}
  \label{tab:fit_results}
\end{table}

\subsection{Nucleation fraction as a function of Hubble type}
\label{subsec:nucleation_fraction_as_a_function_of_hubble_type}

\begin{figure}
  \centering
  \includegraphics[width=\columnwidth]{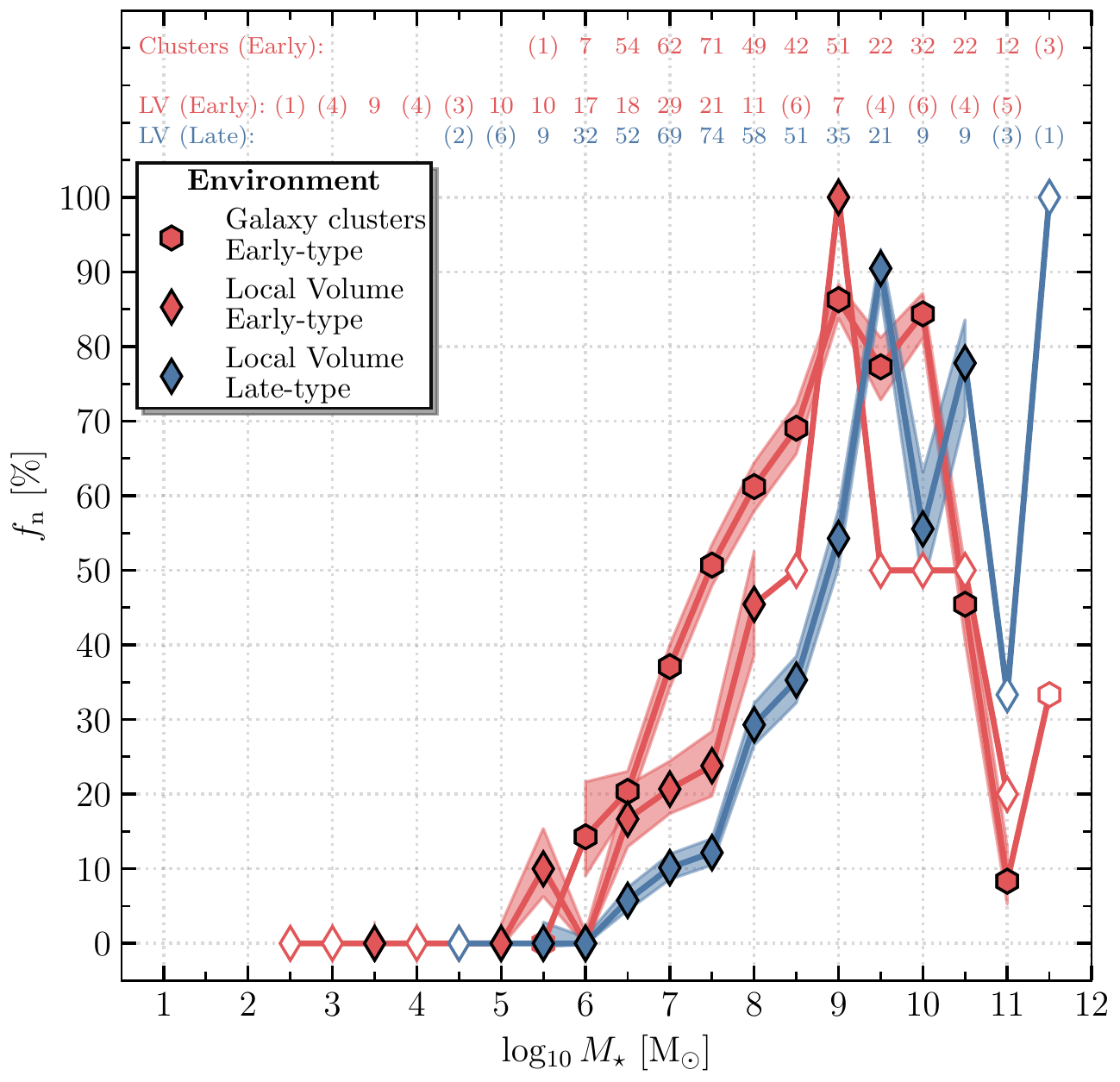}
  \caption{%
    Nucleation fraction ({\nucfrac}) as a function of logarithmic stellar mass ($\log_{10} \, M_{\star}$) for the Local Volume (diamonds) and galaxies in the Coma, Fornax, and Virgo galaxy clusters (hexagons).
    The Local Volume data set is split by Hubble type into early- (red) and late-types (blue).
    Shaded areas show the \SI{68}{\percent} confidence interval and mass bins with fewer than seven galaxies are shown with open symbols.
  }
  \label{fig:nucfrac_hubble_type}
\end{figure}

Here we examine if there exists a difference between early- and late-type morphologies.
We split the LV data into early- and late-types based on their Hubble type.
All galaxies of the ACSCCS, ACSFCS, ACSVCS, NGFS, and NGVS are of early-type.
We combine data for all galaxy clusters into one {\nucfrac} and compare it to the LV in \Cref{fig:nucfrac_hubble_type}.
Galaxy cluster data are shown with hexagons and LV data with diamonds.
Red colour indicates early- and blue colour late-type galaxies.
Similar to \Cref{fig:nucfrac_stellar_mass}, the colour-coded area shows the confidence interval and galaxies per bin are indicated at the top.

We can see that, considering only early-type galaxies, the significant difference between the LV and the galaxy clusters persists.
For the LV, it appears that there is an apparent difference in {\nucfrac} between early- and late-type galaxies in the mass range $\SI{e6.5}{\Msun} \lesssim M_{\star} \lesssim \SI{e9}{\Msun}$.
We investigate this difference by calculating a $p$-value for the null hypothesis that both early- and late-type galaxies have the same {\nucfrac}.
We split the data in this mass range into bins as indicated in \Cref{fig:nucfrac_hubble_type}.
The value of {\nucfrac} of the underlying galaxy population is unknown and, hence, we use the joint value of both Hubble types as an estimator.
Using this estimator, for both early- and late-type galaxies we draw $n$ samples from a binomial distribution where $n$ is the number of galaxies per bin.
We repeat this exercise \num{e7} times to estimate a $p$-value for each bin using both Hubble types.
Depending on the mass bin, we find $p$-values between \num{\approx 0.008} and \num{\approx 0.18}.
Assuming that the null hypothesis is true and that the data in each mass bin are independent, we can combine all $p$-values into a single parameter using Fisher's method \citep{Fisher1992}.
This parameter will then follow a $\chi^{2}$-distribution where the number of degrees of freedom equals twice the number of $p$-values (\num{12} in our case).
Based on this parameter, we can determine a final $p$-value, again drawing \num{e7} times from a $\chi^{2}$-distribution.
The final $p$-value is \num{\approx 6e-4}.
Thus, it is very likely that early- and late-type galaxies originate from galaxy populations with different {\nucfrac}.

However, note that there also is some unreliability in the split into early- and late-types:
as already pointed out by \cta{Karachentsev2013}, early- and late-type dwarf galaxies (spheroidal and irregular) are often wrongfully classified due to their low surface brightness.
Therefore, some early-type galaxies might have been classified as an irregular-type galaxy (and vice versa).
Given the total number of galaxies in this mass range, it seems unlikely that this effect can compensate for the observed discrepancy.

In the literature, \citet{Habas2020} uses \num{2210} dwarf galaxies around massive early-type galaxies in the nearby Universe ($z < 0.01$) to find the same trend, albeit as a function of absolute $g$-band magnitude.
In comparison, the difference is greater in their analysis which is likely related to their magnitude limited sample.
Figure~\num{3} of \citet{Neumayer2020} shows \num{1191} galaxies split into early- and late-type galaxies based on $(g - i)_{0}$ colour.
No significant enhancements are seen there, however, the uniformity of our current sample provides more reliable data than the collection of literature values published there.

\subsection{Nucleation fraction as a function of tidal index}
\label{subsec:nucleation_fraction_as_a_function_of_tidal_index}

From the previous sections we have already seen that the nucleation fraction appears to depend on environment.
Early-type galaxy cluster members show an increased nucleation fraction at low galaxy stellar masses compared to the LV.
Here we examine this environmental dependence in more detail.

To investigate the significance of the local environment, we calculate the `tidal index' (denoted as $\Theta$) which is a measure of the local density surrounding a galaxy.
This parameter was first introduced by \citet{Karachentsev1999b} and takes into account the distance to and stellar mass of nearby galaxies.
In contrast to the original definition, we constrain the possibility of a galaxy to become a disturber by requiring that its stellar mass lies above \SI{5e9}{\Msun}.
The `main disturber' for the $i$th galaxy can be calculated via
\begin{equation}
  \label{equ:theta}
  \Theta_{i} = \max_{j \in [1, \, N]} \log_{10} \bigg( \frac{M_{\star, \, j}}{d_{ij}^{3}} \bigg) + \mathcal{C}, \;\; \forall \, M_{\star, \, j} \geq \SI{5e9}{\Msun} \quad ,
\end{equation}
where $N$ is the total number of galaxies in the data set, $d_{ij}$ the 3D distance between the galaxies, and $\mathcal{C} = -11.76$ a constant reference value chosen such that each galaxy can be divided into isolated ($\Theta_{i} < 0$) and clustered ($\Theta_{i} > 0$).
This parameter can be expanded to include also effects of more than one disturbing galaxy \cpa{Karachentsev2013}.
Therefore,
\begin{equation}
  \label{equ:theta_s}
  \Theta_{i, \, s} = \log_{10} \bigg( \sum_{j = 1}^{s} \frac{M_{\star, \, j}} {d_{ij}^{3}} \bigg) + \mathcal{C}, \;\; \forall \, M_{\star, \, j} \geq \SI{5e9}{\Msun}, \;\; s < N \quad,
\end{equation}
where we sum over the $s$-most influential neighbouring galaxies of the $i$th galaxy.
While this is a calculation of the gravitational tidal index for a galaxy, it also is a proxy for other environmental effects, i.e.\ ram pressure stripping or strangulation.

Using \Cref{equ:theta,equ:theta_s}, we calculate $\Theta_{1}$, $\Theta_{5}$, and {\Tten} for each galaxy in the LV and Coma, Fornax, and Virgo galaxy clusters.
While we use 3D positions to calculate the distance between LV galaxies, we use the angular separation for cluster members as reliable relative distances are often not available.
All three parameter values and the ten main disturbers are presented \Cref{tab:data_table_local_volume,tab:data_table_coma,tab:data_table_fornax,tab:data_table_virgo}, but here we will only consider $\Theta_{10}$ as it is less sensitive to uncertainties in stellar mass estimates and distances between galaxies than both $\Theta_{1}$ and $\Theta_{5}$.

\begin{figure}
  \centering
  \includegraphics[width=\columnwidth]{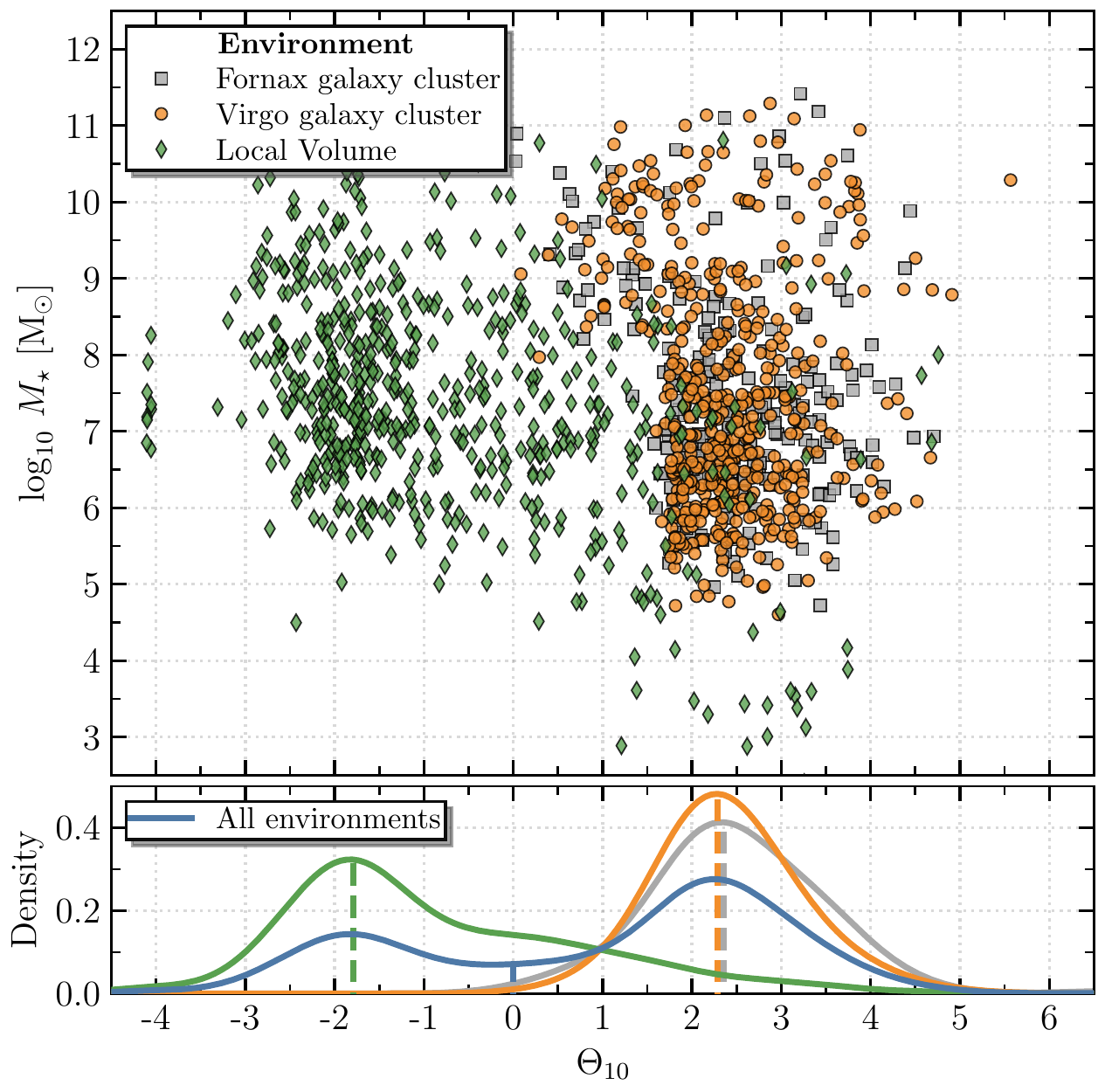}
  \caption{%
    \textit{Top panel}: Logarithmic galaxy stellar mass ($\log_{10} \, M_{\star}$) as a function of tidal index ({\Tten}) for the Local Volume (green diamonds) and the Fornax (gray squares) and Virgo (orange circles) galaxy clusters.
    A Gaussian kernel density estimate with a bandwidth of \num{0.5} is applied to each environment and shown in the \textit{bottom} panel.
    The blue solid line shows the combined data sets.
    Except for the combined data set, where we set $\Theta_{10}^{\mathrm{peak}} = \num{0}$, dashed lines indicate the maxima of the kernel density estimate.
    Their values for each environment are given in \Cref{fig:nucfrac_ti10}.
  }
  \label{fig:2d_mass_ti10}
\end{figure}

In \Cref{fig:2d_mass_ti10} we show the distribution of classified galaxies for the LV, the Fornax and Virgo galaxy clusters, and the combined data set.
The bottom panel shows the kernel density estimate of the {\Tten} distribution.
We use a Gaussian kernel with a bandwidth of \num{0.5}.
The maximum value of the kernel density estimate is shown with dashed lines except for the combined data set where we use a value of zero.
Galaxies with {\Tten} below this peak value (denoted as {\Tpeak}) are assigned to a `loose' environment group; galaxies with $\Theta_{10} > \Theta_{10}^{\mathrm{peak}}$ are assigned to a `dense' environment group.

We calculate {\nucfrac} as a function of {\Tten} and present the results in \S\ref{subsec:ti10_nucleation_fraction}.
As an example, we show the distribution of NGVS members in the Virgo galaxy cluster as a function of stellar mass, tidal index, and angular coordinates in \S\ref{subsec:theta10_bimodial_distribtion}.
However, since it becomes apparent that the stellar mass influences the nucleation fraction as a function of tidal index, we choose a different approach:
we split each data set into two groups based on the peak of the kernel density estimate of {\Tten} for each environment.

It is important to note the differences in {\Tpeak} between environments.
\citet{Karachentsev1999b} consider a galaxy to be rather isolated if its {\Tten} value is smaller than zero.
Therefore, many galaxies in the LV, which are part of the `dense' environment group, can still be considered isolated galaxies.
For both the Fornax and Virgo galaxy clusters the values of {\Tpeak} are significantly larger than zero.

\begin{figure*}
  \centering
  \includegraphics[width=\textwidth]{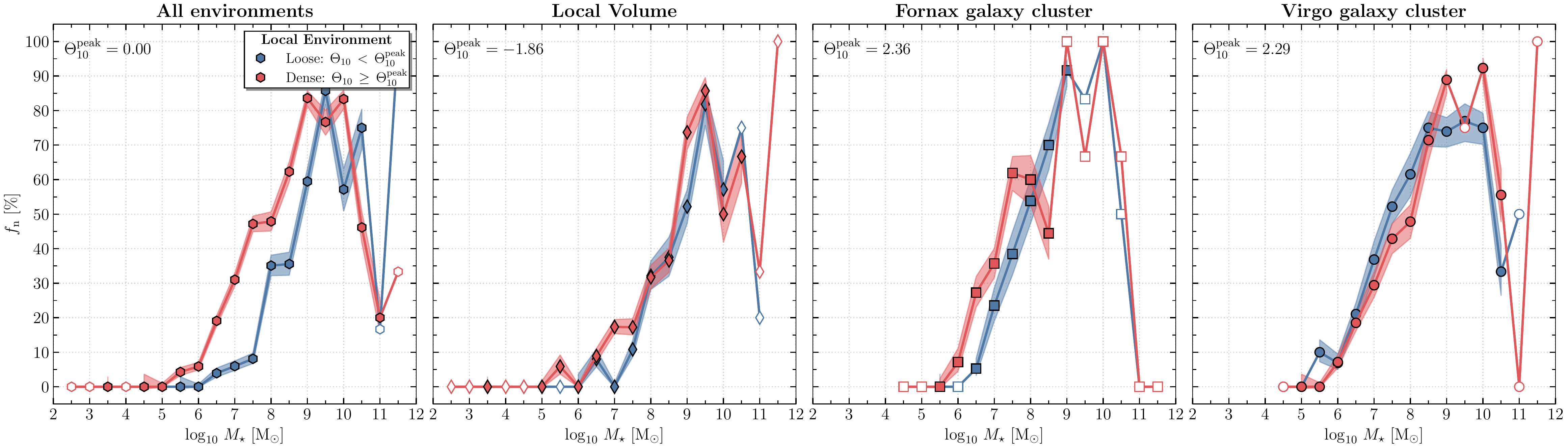}
  \caption{%
    Nucleation fraction ({\nucfrac}) as a function of logarithmic stellar mass ($\log_{10} \, M_{\star}$) for all combined environments (\textit{first}), the Local Volume (\textit{second}), and the Fornax and Virgo galaxy clusters (\textit{third} and \textit{fourth} panels).
    Data for each environment are split based on the peak of the kernel density distribution of {\Tten} (denoted as {\Tpeak} at the top of each panel).
    The combined data set is split at zero.
    Galaxies in a `loose' / `dense' environment are shown in blue / red colour.
  }
  \label{fig:nucfrac_ti10}
\end{figure*}

Based on the peak as a function of {\Tten}, we then determine {\nucfrac} for both subgroups as a function of $M_{\star}$ and present the results in \Cref{fig:nucfrac_ti10}.
The values of {\Tpeak} are indicated at the top of each panel.
We do not perform this analysis for the Coma galaxy cluster as the data sample of \citet{denBrok2014} is incomplete at the low-mass end \citep[see e.g.][]{Madrid2010,Chiboucas2011,Lim2018,Zanatta2021}; this incompleteness heavily influences the {\Tten} values.

First, we combine the LV and the Fornax and Virgo galaxy cluster environments to gain the statistical significance to split the combined data sample at $\Theta_{10} = 0$ (first panel).
For the combined data set we can clearly see a difference in {\nucfrac}, which is expected based on the comparison of the LV with the galaxy clusters in \Cref{fig:nucfrac_stellar_mass}.
For the LV (second panel), we find that galaxies in a `dense' local environment have a slightly higher {\nucfrac} than in a `loose' local environment.
For the Fornax galaxy cluster (third panel), {\nucfrac} is elevated for galaxies in a `dense' local environment until $M_{\star} \approx \SI{e8}{\Msun}$.
On the other hand in the Virgo galaxy cluster, this same trend is not visible, and in fact the trend is reversed, with loose regions having higher nucleation fractions than dense ones.
We test a potential bias of classifiying `non-nucleated high-surface brightness' in \citet{Lisker2007} as unclassified by excluding their entire data set and repeating the analysis.
This limits the data beyond $r \, / \, r_{\mathrm{vir.}} = 0.33$ to \SI{\geq e8}{\Msun}.
Therefore, due to low number statistics, we cannot find significant trends for the Virgo galaxy cluster.

\subsection{Nucleation fraction as a function of galactocentric distance}
\label{subsec:nucleation_fraction_as_a_function_of_galactocentric_distance}

In this section we investigate the nucleation fraction as a function of galactocentric distance.
In the literature it seems to be established that, in a cluster environment, nucleated galaxies are more centrally concentrated than non-nucleated ones \citep[e.g.][]{Binggeli1987,Ferguson1989,Lisker2007,Ordenes-Briceno2018}.
For the LV, the situation was not clear until very recently when \citet{Carlsten2020b} investigated dwarf elliptical satellites of massive LV hosts.
They found that galaxies with a projected distance \SI{\leq 100}{\kilo\pc} have a higher {\nucfrac} than galaxies at larger distances.
The authors point out that this increase could be related to the parent halo mass, but no decisive conclusions could be made.
The aim of this section is, therefore, to analyse the radial dependence of {\nucfrac} using the LV and galaxy clusters.

For the LV, we follow the approach by \citet{Carlsten2021b} and calculate the distance between satellites and their hosts before stacking the results.
We use the \num{30} most massive galaxies ($\num{10.18} \lesssim \log_{10} \, M_{\star} \, / \, \mathrm{M}_{\odot} \lesssim \num{11.33}$) in the LV as hosts and select all galaxies with a 3D distance $r$ of \SI{2}{\mega\pc}.
Each galaxy is only selected once at the distance to the closest of the massive galaxies.
This value equals multiple virial radii of the host galaxies and will be useful for a comparison to the galaxy clusters.
In total, we find \num{148} galaxies within \SI{2}{\mega\pc} around the \num{30} hosts.

In the LV, we need an estimate of $r_{\mathrm{vir.}}$ to put all radial measurements on comparable footing.
In the literature only few virial radii are available and many are estimated via the half-mass radius \citep[e.g.][]{Byun2020,Karachentsev2020b}.
Because the estimates range typically between \SI{200}{\kilo\pc} and \SI{400}{\kilo\pc}, we assume a common value of \SI{300}{\kilo\pc} for the virial radii of all hosts.

For the Fornax galaxy cluster we include the data of \citet{Su2021} from the FDS.
In comparison to \citet{Munoz2015} and the NGFS, the data of the FDS reaches the virial radius of the NGC\,1399 (\SI{\approx 700}{\kilo\pc}; \citealp{Drinkwater2001}).
Similar to the NGFS, the NGVS covers only the central region around M\,87 in the Virgo galaxy cluster.
To study {\nucfrac} at larger distances, we include the data set of \citet{Lisker2007} who uses {\SDSS} imaging to classify galaxies.
As noted by the authors, NSCs in high-surface brightness galaxies may escape detection.
Therefore, we classify all non-nucleated high-surface brightness galaxies of \citet{Lisker2007} as unclassified, thus, not impacting the nucleation fraction estimates.
We take the cluster-centric radii as the angular separation from NGC\,1399 and M\,87, respectively.

\begin{figure*}
  \centering
  \includegraphics[width=0.9\textwidth]{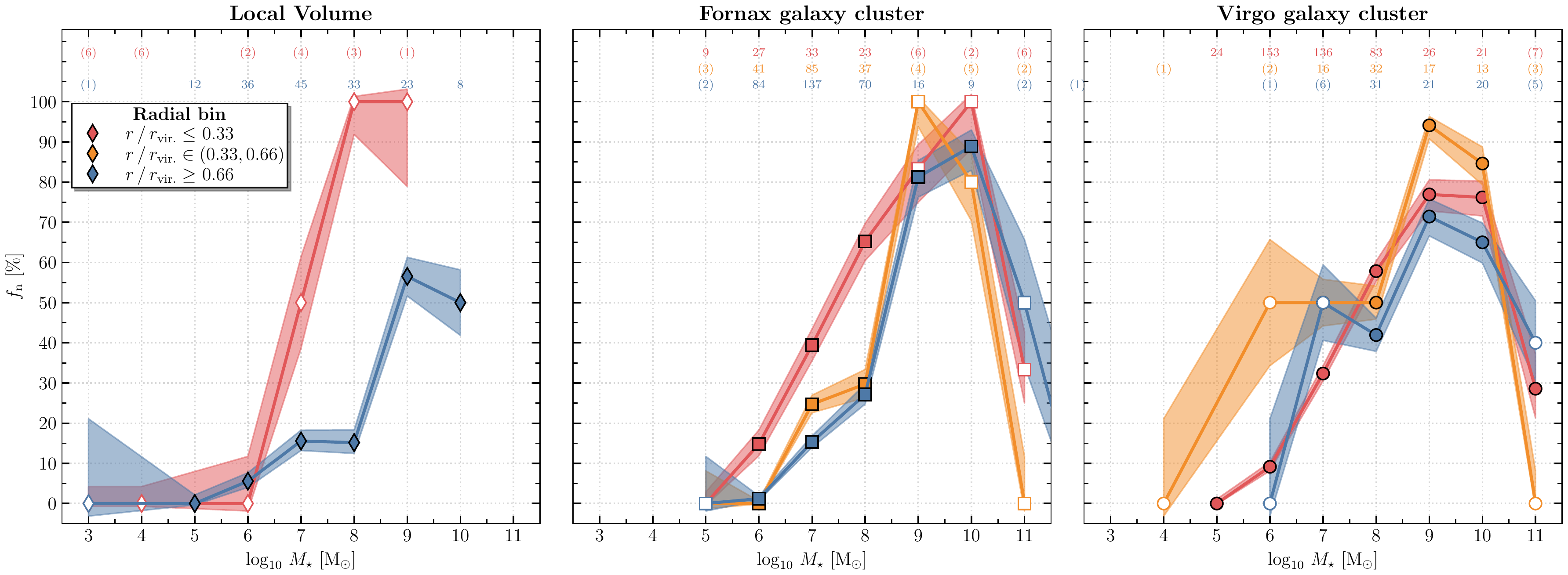}
  \caption{%
    Nucleation fraction ({\nucfrac}) versus stellar mass ($\log_{10} \, M_{\star}$) for the Local Volume (\textit{left panel}), and the Fornax (\textit{middle panel}) and Virgo (\textit{right panel}) galaxy clusters.
    The data are split based on their distance to the central host galaxy.
    With increasing galactocentric distance, the data are shown in red, orange, and blue colour, respectively.
    Numbers at the top of each panel give the number of galaxies per bin.
    For the Local Volume, the satellite populations of \num{30} massive hosts are stacked, assuming a common virial radius of \SI{300}{\kilo\pc}.
  }
  \label{fig:nucfrac_radius}
\end{figure*}

In \Cref{fig:nucfrac_radius} we show the nucleation fraction as a function of stellar mass where the data have been split based on their proximity to their host galaxy.
With increasing distance, red, orange, and blue colour show different populations.
The number of galaxies per bin are indicated at the top.
In contrast to previous figures, we also show uncertainties for all data points (even those with low number statistics).
Although only few galaxies reside in the innermost radial bin ($r \, / \, r_{\mathrm{vir.}} \leq \num{0.33}$), a trend is visible in that {\nucfrac} increases, at the same stellar mass, with decreasing galactocentric distance.
This result is in agreement with \citet{Carlsten2021b} who split their sample at $r = \SI{100}{\kilo\pc}$ ($r \, / \, r_{\mathrm{vir.}} \approx 0.33$ in our figure).
The same trend continues for the Fornax galaxy cluster where the number statistics are better than in the LV.
It is noteworthy that only galaxies within $0.33 r_{\mathrm{vir.}}$ (\SI{\approx 233}{\kilo\pc}) have an elevated {\nucfrac} and that the other two curves closely follow each other.
Our results for the Fornax galaxy cluster are in agreement with \citet{Ferguson1989} and \citet{Munoz2015}.
\citet{Ferguson1989} found that nucleated dwarf ellipticals are concentrated in the centre of both the Fornax and Virgo galaxy clusters, and \citet{Munoz2015} found that the surface number density of nucleated dwarf galaxies increases with decreasing galactocentric radius while the it remains roughly constant for non-nucleated ones.
However, \citet{Carlsten2021b} compared the results of \citet{Munoz2015} and \citet{Ordenes-Briceno2018}, those galaxy samples reside at different galactocentric radii ($r \lesssim 0.25 r_{\mathrm{vir}}$ and $r \, / \, r_{\mathrm{vir}.} \in [0.25, 0.5]$, respectively), and found no significant increase.

For the Virgo galaxy cluster, no clear trend with galactocentric distance is apparent.
This result disagrees with \citet{Binggeli1987}, \citet{Cote2006}, and \citet{Lisker2007} who found that nucleated dwarf ellipticals are more centrally concentrated than non-nucleated ones (i.e.\ an increase in {\nucfrac} with decreasing galactocentric distance).
One explanation, as already pointed out by \citet{Cote2006}, could be a seletion bias of these studies as mainly high-mass ($M_{\star} \geq \SI{e8}{\Msun}$) galaxies were considered.
The difference between {\nucfrac} due to the position within the cluster at high masses seems to become insignificant (\textit{cf}.~\Cref{fig:nucfrac_radius}).

\section{Discussion}
\label{sec:discussion}

Overall, including the LV and cluster samples, we find that {\nucfrac} positively correlates with large-scale environment which is in agreement with the recent literature \citep{Sanchez-Janssen2019a,Carlsten2021b,Zanatta2021}.
However, there are other secondary effects which can influence the nucleation of a galaxy.
In \Cref{subsec:low_mass_galaxies} these effects will be discussed in detail.
In \Cref{subsec:intermediate_high_mass_galaxies} we elaborate further on the decrease of {\nucfrac} at the highest masses ($M_{\star} \geq \SI{e9.5}{\Msun}$) which seems to be independent of environment.

\subsection{Low-mass galaxies with $M_{\star} < \SI{e9}{\Msun}$}
\label{subsec:low_mass_galaxies}

\subsubsection{Stellar mass \& large-scale environment}
\label{subsubsec:stellar_mass_and_large-scale_environment}

Significant differences between {\nucfrac} for the LV and cluster environments are observed in the low-mass regime.
One possible origin for this trend stems from enhanced NSC formation in dense environments.
NSC formation has been studied in various papers and two different formation mechanisms were postulated:
(1) for high-mass galaxies ($M_{\star} \geq \SI{e9}{\Msun}$) in-situ star-formation contributes a significant part to the stellar population(s) of NSCs \citep[see][and references therein]{Neumayer2020}.
(2) For low-mass galaxies ($M_{\star} < \SI{e9}{\Msun}$) globular cluster (GC) migration towards the galactic centre is important and constitutes the largest contributor to NSC populations \citep[e.g.][]{Tremaine1975,Hartmann2011,Turner2012,Antonini2015}.
NSCs in galaxies with stellar masses $M_{\star} \approx \SI{e9}{\Msun}$ likely have contributions from both mechanisms \citep[e.g.][]{Fahrion2021}.
If GC migration is the dominant contributor to NSC masses in low-mass galaxies, we would expect to find an increase in GC abundance in dense environments.
Indeed, \citet{Carlsten2021b} found that, at fixed stellar mass, dwarf ellipticals in the Virgo galaxy cluster host more GCs than in the LV.
Therefore, GC abundance seems to relate to NSC occupation.
Additional arguments supporting the correlation between NSCs and GCs stem from \cta{Sanchez-Janssen2019a} who found that the NSC and GC occupation fractions vary similarly with stellar mass for dwarf early-types in the core of Virgo.
Recently, \citet{Carlsten2021b} showed that this is also true for dwarf early-types in the LV, suggesting that the connection between NSC and GC occupation fraction is independent of environment.

Within each environment we find an increase in {\nucfrac} for galaxies $\lesssim \SI{e9}{\Msun}$ such that at fixed stellar mass, the nucleation fraction is elevated for satellite galaxies residing close to their host galaxies.
Therefore, if GC abundance and NSC occupation correlate with each other, we would expect to find higher GC abundance in galaxies close to their host galaxy.
For galaxy clusters this trend has been observed \citep{Peng2008,Lim2018,Liu2019}, but data is lacking for the LV.

This interpretation seems to fit to our data for the LV and the literature data for the Fornax galaxy cluster, however, for the Virgo galaxy cluster we do not find an increase in {\nucfrac} with galactocentric distance.
Because GC abundance seems to correlate with galactocentric distance, it seems unclear why {\nucfrac} would not.
This could be related to (1) selection biases in removing `non-nucleated high-surface brightness' galaxies from \citet{Lisker2007} or (2) selection biases in the data sample.
However, no significant differences between {\nucfrac} at different radii persists at \SI{\geq e8}{\Msun}, albeit with small number statistics.
If no selection bias is present, one possibility could be that the correlation between {\nucfrac} and the environment weakens once a certain density is reached.
One possibile physical mechanism that could cause this is tidal heating which could prolong (or stop) the inspiral of GCs into a galaxy's centre.
\citet{Carlsten2021a} found that the sizes of their dwarf elliptical galaxies are, on average, \SI{\approx 10}{\percent} smaller than dwarf ellipticals in the core of the Virgo galaxy cluster.
If tidal heating plays a role, we would also expect that {\nucfrac} remains constant (or even decreases) with decreasing galactocentric distance in the Coma galaxy cluster, which is denser than the Virgo galaxy cluster.
\citet{Lim2018} presented an analysis of ultra-diffuse galaxies in combination with dwarf elliptical galaxies in the Coma galaxy cluster to find that {\nucfrac} increases with decreasing galactocentric distance (their Figure \num{8}).
However, their data cover a range of \SI{\approx 3}{\mega\pc} with a bin width of \SI{\approx 0.7}{\mega\pc}.
While a global increase in {\nucfrac} is expected, we can currently not resolve the regions closer to the cluster centre, where we would expect the trend to reverse.
Such an analysis is not possible as the data sample of \citet{denBrok2014} is incomplete at the low-mass end and the data set of \citet{Zanatta2021} 1) lacks photometric parameters to determine stellar masses and 2) could potentially contain a significant number of background objects \citep{Carlsten2021b}.

\subsubsection{Hubble type \& local environment}
\label{subsubsec:hubble_type_and_local_environment}

Because both the analyses of the galaxy clusters and that of \citet{Carlsten2021b} focused on dwarf elliptical galaxies, we have tested whether the morphological type of a galaxy is important for {\nucfrac}.
In \Cref{subsec:nucleation_fraction_as_a_function_of_hubble_type} we showed that, at fixed stellar mass, dwarf early-type galaxies have a higher {\nucfrac} than dwarf late-type galaxies.
The argument presented in the previous section would, therefore, suggest that dwarf elliptical galaxies have a higer GC abundance than dwarf irregular galaxies.
This increase in GC abundance is found in both a dense cluster \citep[e.g.][]{Miller1998,Seth2004,Miller2007,Sanchez-Janssen2012} and field environment \citep[e.g.][]{Georgiev2008}.

\begin{figure}
  \centering
  \includegraphics[width=\columnwidth]{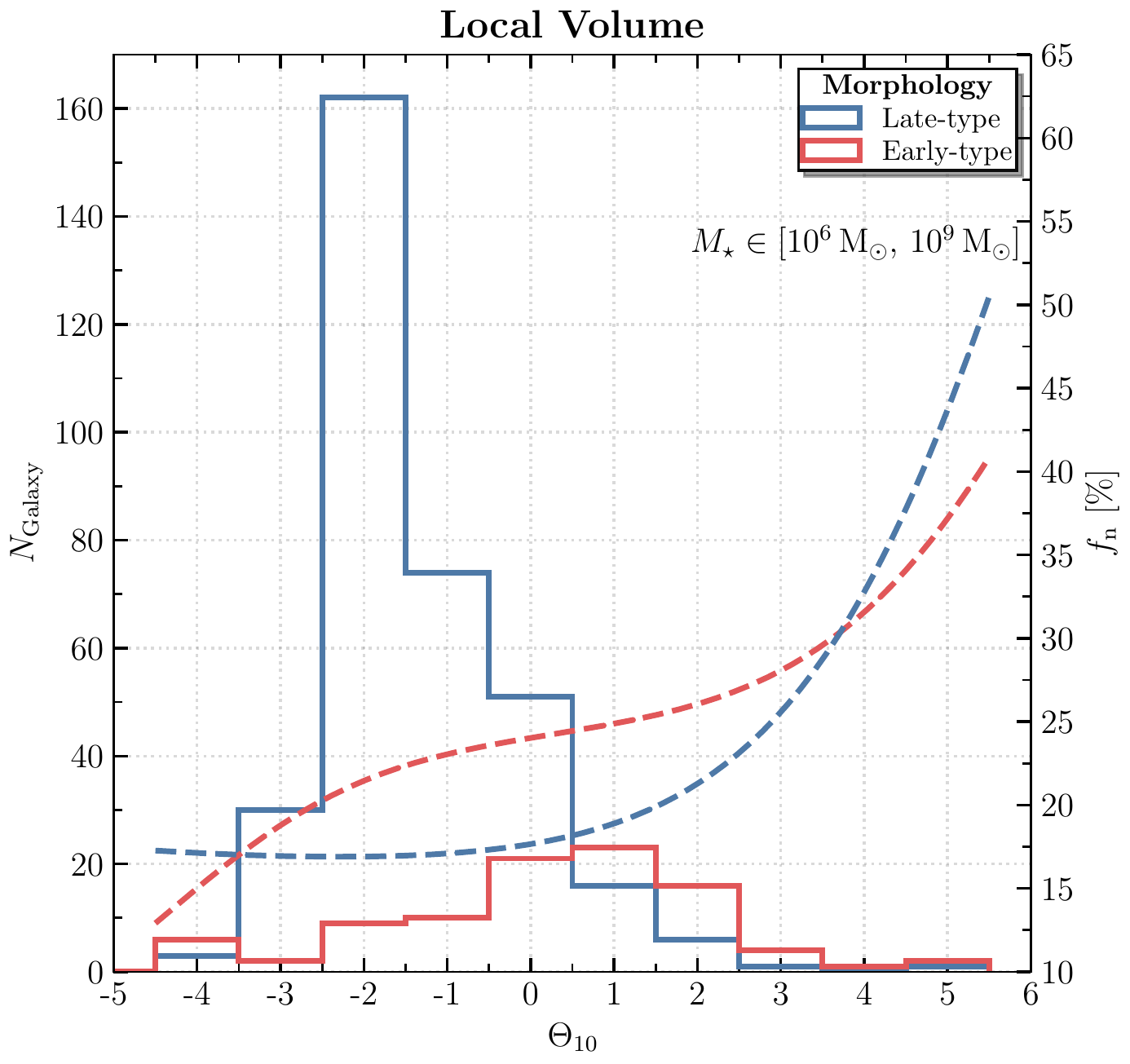}
  \caption{%
    Distribution of classified early- (red) and late-type (blue colour) galaxies in the Local Volume, limited to $M_{\star} \in [\SI{e6}{\Msun}, \, \SI{e9}{\Msun}]$, as a function of tidal index ($\Theta_{10}$) which indicates the local environmental density.
    We use a Gaussian kernel density estimate with a bandwidth of \num{2} to indicate the nucleation fraction ({\nucfrac}; dashed lines).
    The higher nucleation fraction of early-type galaxies relative to late-types suggests galaxy morphology plays a role in setting the nucleation fraction independently of environment.
  }
  \label{fig:nucfrac_hubble_type_tidal_index}
\end{figure}

In \Cref{subsec:nucleation_fraction_as_a_function_of_tidal_index,subsec:nucleation_fraction_as_a_function_of_galactocentric_distance} we also showed that {\nucfrac} in galaxies correlates with galaxy density and inversely correlates with distance from a massive host.
Given that early-type galaxies are typically found in denser environments near massive hosts, it is unclear if the dominant factor in determining {\nucfrac} is from galaxy morphology or environment, or if both contribute independently.
In \Cref{fig:nucfrac_hubble_type_tidal_index} we show the distribution of classified early- and late-type dwarf galaxies in the LV separated by colour.
To increase number statistics, we use the tidal index instead of the galactocentric distance\footnote{Using the galactocentric distance leads to the same conclusions, albeit the trends are not as clear.}.
We limit the galaxy sample to a mass range $M_{\star} \in [\SI{e6}{\Msun}, \, \SI{e9}{\Msun}]$.
We find that, with the exception of the highest $\Theta_{10}$ bins where the number statistics are poor, early-type galaxies are nucleated more often than their dwarf late-type counterparts.
This increase is not related to stellar mass as no significant difference, at fixed tidal index, could be determined between both populations.
Furthermore, dwarf elliptical galaxies reside in denser environments than dwarf irregular galaxies.
This obsevation suggests that the Hubble type, irrespective of stellar mass and environment, correlates with {\nucfrac}.

\subsection{Intermediate- and high-mass galaxies with $M_{\star} \geq \SI{e9}{\Msun}$}
\label{subsec:intermediate_high_mass_galaxies}

At the highest masses we were able to see that {\nucfrac} decreases beyond $M_{\star} \geq \SI{e9.5}{\Msun}$ (\textit{cf}.\ \Cref{fig:nucfrac_stellar_mass,fig:nucfrac_hubble_type}).
It is speculated that this drop is due to the build-up of massive early-type galaxies due to galaxy mergers during which NSCs get disrupted via dynamical heating of the supermassive black holes \citep[e.g.][]{Cote2006}.
The difference between early- and late-type galaxies becomes clearer if we combine the data sets of \citet{Georgiev2014} and the LV for late-type galaxies as no such sharp drop appears (\textit{cf}.~\Cref{fig:nf_g14}).

\begin{figure}
  \centering
  \includegraphics[width=\columnwidth]{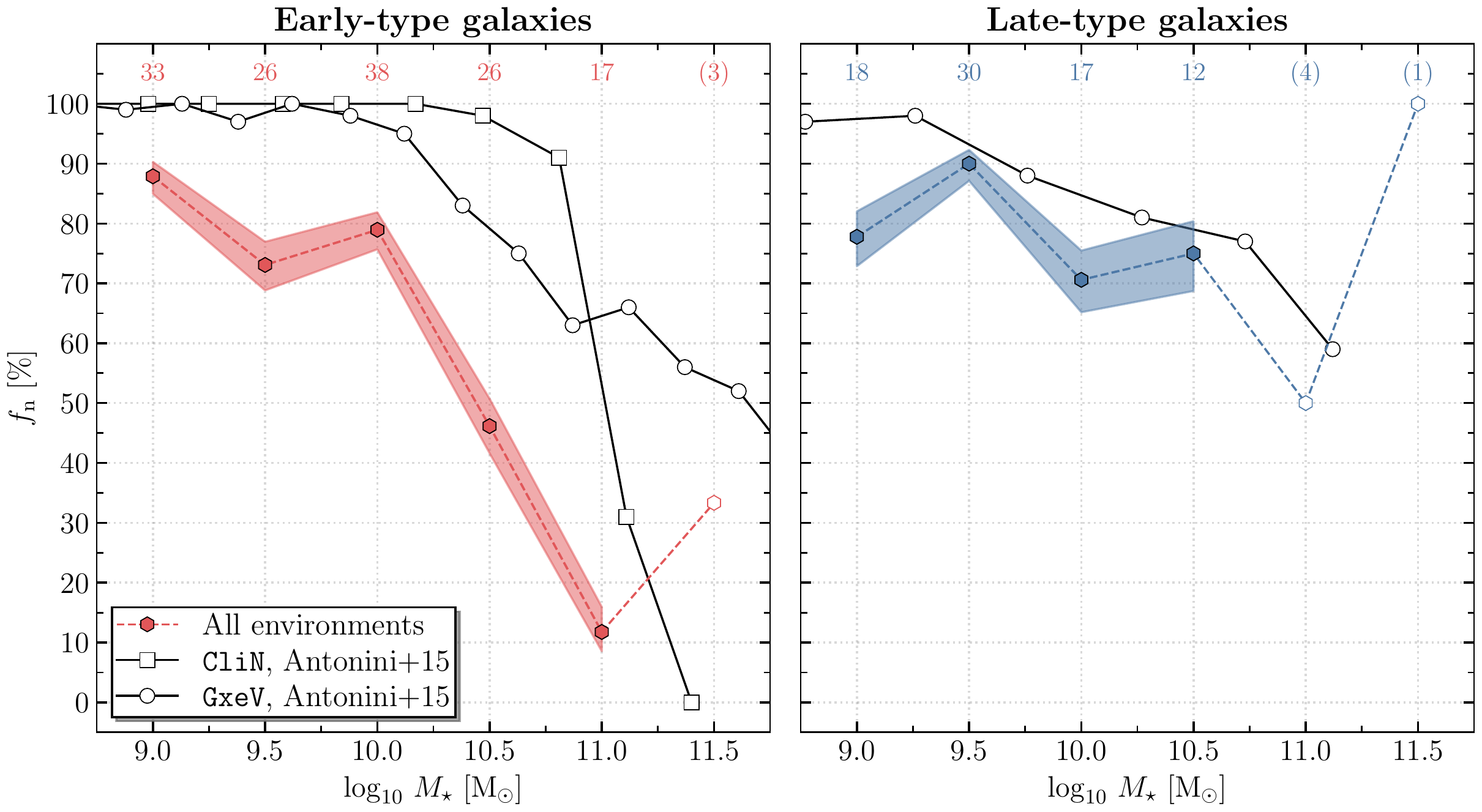}
  \caption{%
    Nucleation fraction ({\nucfrac}) as a function of logarithmic stellar mass ($\log_{10} M_{\star}$) for early-type (\textit{left}) and late-type (\textit{right} panel) galaxies across all environments (colour-coded hexagons) and the two models \texttt{CliN} (squares) and \texttt{GxeV} (circles) of \citet{Antonini2015}.
    The models where dynamical heating from supermassive black holes is disabled show a constant {\nucfrac} at \SI{100}{\percent} (not shown here).
  }
  \label{fig:nucfrac_antonini}
\end{figure}

\citet{Antonini2015} studied the evolution of NSCs and massive black holes in high-mass galaxies.
Using two semi-analytical models named Cluster Inspiral Model (\texttt{CliN}) and Galaxy Formation Model (\texttt{GxeV}), they were able to investigate {\nucfrac}.
We compare their models with observational results for early- and late-type galaxies in the top and bottom panels of \Cref{fig:nucfrac_antonini}, respectively.
We combine early-type galaxies of all environments (LV and Coma, Fornax, and Virgo galaxy clusters) and late-types of the LV and the data sample of \citet{Georgiev2014} to gain statistical significance.

For early-types (left panel) we find that, although their models overestimate {\nucfrac} at all masses, the decline of {\nucfrac} is similar between their models and the observations.
The slope of the \texttt{CliN} model seems to decrease with a steeper slope, but this model only takes into account cluster migration as an NSC formation scenario.
We note that the authors found a better agreement between the \texttt{CliN} model and the data sets of \citet{Erwin2012,Neumayer2012,Turner2012}, and \citet{Scott2013}, however, they assumed $M_{\star} / L_{K} = 0.8$ based on \citet{Bell2001} which results in an overestimation of stellar masses (\textit{cf}.~\Cref{tab:mass_to_light_ratios}).
For the late-type sample we find better agreement between their models and the observations.
This is impressive given that the identification of nucleation in their models is uncertain since they are unable to compare the density profile of the NSC and the host galaxy.

Finally, once \citet{Antonini2015} disabled dynamical heating from the central supermassive black hole, they found that {\nucfrac} stays constant at \SI{100}{\percent} for both morphologies.
Therefore, it seems likely that this effect is a dominant mechanism of NSC destruction.

\section{Conclusions}
\label{sec:conclusions}

In this work we investigated the nucleation fraction ({\nucfrac}) of Local Volume (LV) galaxies as a proxy for field galaxies in a low-density environment.
Based on the `Updated Nearby Galaxy Catalog' \citep{Karachentsev2013} and ancillary data from the HyperLEDA and SIMBAD data bases, we calculated stellar masses for galaxies in a homogeneous way and combine our nuclear classification with the literature to construct {\nucfrac} as a function of galactic stellar mass, Hubble type, and tidal index.
Our catalogue contains \num{601} classified galaxies across eight orders of magnitude in stellar mass.
With the addition of literature data for members of the Coma, Fornax, and Virgo galaxy clusters and a homogeneous calculation of stellar masses for all environments, we obtain the following results:
\begin{enumerate}
  \item[$\bullet$] We identify \num{22} new NSCs in LV galaxies.
  Their photometric and structural parameters will be analysed in a future paper.
  
  \item[$\bullet$] As a function of stellar mass ($M_{\star}$) we find that {\nucfrac} is significantly lower in the LV compared to that in galaxy clusters (\textit{cf}.\ \Cref{fig:nucfrac_stellar_mass}).
  This holds true at masses up to \SI{\approx e9.5}{\Msun} and is most significant between \SI{e7.5}{\Msun} and \SI{e8.5}{\Msun} reaching a difference in {\nucfrac} of \SI{\approx 40}{\percent}.
  These observations confirm previous investigations of more limited studies (e.g.\ \citealp{Sanchez-Janssen2019a,Carlsten2021b,Zanatta2021}).

  \item[$\bullet$] Regardless of environment, nucleation seems to start between \SI{e5}{\Msun} and \SI{e5.5}{\Msun} and seems to peak at \SI{\approx e9.5}{\Msun}.
  Poor statistics at the high-mass end does not allow us to set tight constrains on the decline of nucleation at the high mass end, however, it seems likely that this is the case at \SI{\approx e11.5}{\Msun}.

  \item[$\bullet$] We fit a logistic function to {\nucfrac} of each environment below \SI{e9.5}{\Msun} using a Markov-Chain Monte Carlo analysis.
  We find that the slope and midpoint of the logistic function has higher values for the LV than for both the Fornax and Virgo galaxy clusters, indicating that nucleation starts at lower stellar masses and rises more steeply with mass in dense environments.

  \item[$\bullet$] At fixed stellar mass, LV dwarf early-types have a higher {\nucfrac} than dwarf late-type galaxies.

  \item[$\bullet$] In addition to a large-scale environmental dependence (field versus cluster), {\nucfrac} also correlates with local environment:
  at fixed stellar mass, LV satellite galaxies in close proximity of their host galaxy have higher {\nucfrac} than distant galaxies.
  This trend is also observed for the Fornax galaxy cluster but is not found in Virgo.

  \item[$\bullet$] The galactic stellar mass is the dominant factor that determines the nucleation of galaxies.
  We also find a correlation between {\nucfrac} and both the Hubble type and local environmental density (or galactocentric distance).
  Both effects of Hubble type and local environment seem to contribute independently to {\nucfrac}.

  \item[$\bullet$] Our results further strengthen the evidence that globular cluster (GC) abundance and {\nucfrac} in galaxies below \SI{e9}{\Msun} are correlated.
  It is unclear whether this connection is sufficient to explain {\nucfrac}.
  For the Virgo galaxy cluster, no correlation could be found between the nucleation fraction and the local environment of a galaxy.
  We speculate that, if no selection bias is present, one contributing factor could be the tidal heating of galaxies which could hinder (or completely stop) the inspiral of GCs into a galaxy's centre.
  Furthermore, it is unclear whether GC abundance of dwarf irregulars in the LV correlates with local environmental density.

  \item[$\bullet$] At the high-mass end, {\nucfrac} drops for all environments and nucleation seems to stop at $M_{\star} \approx \SI{e11.5}{\Msun}$.
  A comparison to the numerical study of \citet{Antonini2015} reveals that the likely reason for the decline of {\nucfrac} is dynamical heating from merging supermassive black holes (SMBHs) during galaxy mergers.
\end{enumerate}

Follow-up investigations are clearly required to solve the remaining open questions.
In particular, observational studies should investigate the dependence of {\nucfrac} on the local environment of the host galaxy in the centre of the Coma galaxy cluster.
Theoretical and numerical investigations are also important to constrain the relative influence of the stellar mass, the Hubble type, and the local environmental density on {\nucfrac}.
Additionally, the correlation between GC abundance and environment would benefit from further investigation.

Another interesting connection can be made between NSC and SMBHs at the high-mass end where only few classified galaxies reside.
A sophisticated analysis of NSCs and their {\nucfrac} in high-mass galaxies may help to constrain the decline of nucleation in this mass regime and allow for a more detailed study of the interplay between NSCs and central SMBHs.

\section*{Data availability}

The observational data underlying this article were accessed from the \textit{Hubble Legacy Archive}\footnote{See Footnote 8.} and the \textit{Mikulski Archive for Space Telescopes}\footnote{\url{https://archive.stsci.edu/}}.
The data underlying this article are available in the article and in its online supplementary material.

\section*{Acknowledgements}
\label{sec:acknowledgements}

This research is based on observations with the NASA/ESA Hubble Space Telescope, obtained at the Space Telescope Science Institute, which is operatedby AURA, Inc., under NASA contract NAS5-26555.
This research has made use of the Updated Nearby Galaxy Catalog \citep{Karachentsev2013}; the HyperLEDA data base \citep{Makarov2014}; the SIMBAD data base \citep{Wegner2000}; NASA's Astrophysics Data System (ADS); Astropy \citep{Astropy2013,Astropy2018}; NumPy (\url{https://numpy.org/}); dustmaps \citep{Green2018}; Matplotlib \citep{Hunter2007}; SciPy \citep{Virtanen2020}; AstroQuery \citep{Ginsburg2019}; emcee \citep{Foreman-Mackey2013}; and corner \citep{Foreman-Mackey2016}.

NH thanks Joel Roediger for the communication of star-forming galaxies in the NGVS sample and Mark den Brok for details on his data table for the Coma galaxy cluster.
The authors would like to thank the anonymous referee for a detailed and constructive report.




\bibliographystyle{mnras}
\bibliography{bibliography.bib}

\begin{thebibliography}{}
\makeatletter
\relax
\def\mn@urlcharsother{\let\do\@makeother \do\$\do\&\do\#\do\^\do\_\do\%\do\~}
\def\mn@doi{\begingroup\mn@urlcharsother \@ifnextchar [ {\mn@doi@}
  {\mn@doi@[]}}
\def\mn@doi@[#1]#2{\def\@tempa{#1}\ifx\@tempa\@empty \href
  {http://dx.doi.org/#2} {doi:#2}\else \href {http://dx.doi.org/#2} {#1}\fi
  \endgroup}
\def\mn@eprint#1#2{\mn@eprint@#1:#2::\@nil}
\def\mn@eprint@arXiv#1{\href {http://arxiv.org/abs/#1} {{\tt arXiv:#1}}}
\def\mn@eprint@dblp#1{\href {http://dblp.uni-trier.de/rec/bibtex/#1.xml}
  {dblp:#1}}
\def\mn@eprint@#1:#2:#3:#4\@nil{\def\@tempa {#1}\def\@tempb {#2}\def\@tempc
  {#3}\ifx \@tempc \@empty \let \@tempc \@tempb \let \@tempb \@tempa \fi \ifx
  \@tempb \@empty \def\@tempb {arXiv}\fi \@ifundefined
  {mn@eprint@\@tempb}{\@tempb:\@tempc}{\expandafter \expandafter \csname
  mn@eprint@\@tempb\endcsname \expandafter{\@tempc}}}

\bibitem[\protect\citeauthoryear{Abazajian et~al.,}{Abazajian
  et~al.}{2009}]{Abazajian2009}
Abazajian K.~N.,  et~al., 2009, \mn@doi [ApJS] {10.1088/0067-0049/182/2/543},
  182, 543

\bibitem[\protect\citeauthoryear{Agresti \& Coull}{Agresti \&
  Coull}{1998}]{Agresti1998}
Agresti A.,  Coull B.,  1998, \mn@doi [The American Statistican]
  {10.2307/2685469}, 52, 119

\bibitem[\protect\citeauthoryear{Antonini, Barausse  \& Silk}{Antonini
  et~al.}{2015}]{Antonini2015}
Antonini F.,  Barausse E.,   Silk J.,  2015, \mn@doi [ApJ]
  {10.1088/0004-637X/812/1/72}, 812, 24

\bibitem[\protect\citeauthoryear{Balcells, Graham  \& Peletier}{Balcells
  et~al.}{2007}]{Balcells2007}
Balcells M.,  Graham A.,   Peletier R.,  2007, \mn@doi [ApJ] {10.1086/519752},
  665, 1084

\bibitem[\protect\citeauthoryear{Baldassare, Gallo, Miller, Plotkin, Treu,
  Valluri  \& Woo}{Baldassare et~al.}{2014}]{Baldassare2014}
Baldassare V.,  Gallo E.,  Miller B.,  Plotkin R.,  Treu T.,  Valluri M.,   Woo
  J.-H.,  2014, \mn@doi [ApJ] {10.1088/0004-637x/791/2/133}, 791, 133

\bibitem[\protect\citeauthoryear{Barth, Strigari, Bentz, Greene  \& Ho}{Barth
  et~al.}{2008}]{Barth2009}
Barth A.,  Strigari L.,  Bentz M.,  Greene J.,   Ho L.,  2008, \mn@doi [ApJ]
  {10.1088/0004-637x/690/1/1031}, 690, 1031

\bibitem[\protect\citeauthoryear{Baumgardt \& Hilker}{Baumgardt \&
  Hilker}{2018}]{Baumgardt2018}
Baumgardt H.,  Hilker M.,  2018, \mn@doi [MNRAS] {10.1093/mnras/sty1057}, 478,
  1520

\bibitem[\protect\citeauthoryear{Beifiori, Sarzi, Corsini, Dalla~Bont{\'{a}},
  Pizzella, Coccato  \& Bertola}{Beifiori et~al.}{2009}]{Beifiori2009}
Beifiori A.,  Sarzi M.,  Corsini E.,  Dalla~Bont{\'{a}} E.,  Pizzella A.,
  Coccato L.,   Bertola F.,  2009, \mn@doi [ApJ] {10.1088/0004-637X/692/1/856},
  692, 856

\bibitem[\protect\citeauthoryear{Bell \& de Jong}{Bell \&
  de~Jong}{2001}]{Bell2001}
Bell E.,  de Jong R.,  2001, \mn@doi [ApJ] {10.1086/319728}, 550, 212

\bibitem[\protect\citeauthoryear{Bell, McIntosh, Katz  \& Weinberg}{Bell
  et~al.}{2003}]{Bell2003}
Bell E.,  McIntosh D.,  Katz N.,   Weinberg M.,  2003, \mn@doi [ApJS]
  {10.1086/378847}, 149, 289

\bibitem[\protect\citeauthoryear{Bellazzini et~al.,}{Bellazzini
  et~al.}{2008}]{Bellazzini2008}
Bellazzini M.,  et~al., 2008, \mn@doi [AJ] {10.1088/0004-6256/136/3/1147}, 136,
  1147

\bibitem[\protect\citeauthoryear{Bellazzini, Annibali, Tosi, Mucciarelli,
  Cignoni, Beccari, Nipoti  \& Pascale}{Bellazzini
  et~al.}{2020}]{Bellazzini2020}
Bellazzini M.,  Annibali F.,  Tosi M.,  Mucciarelli A.,  Cignoni M.,  Beccari
  G.,  Nipoti C.,   Pascale R.,  2020, \mn@doi [A\&A]
  {10.1051/0004-6361/201937284}, 634, 10

\bibitem[\protect\citeauthoryear{Binggeli, Sandage  \& Tammann}{Binggeli
  et~al.}{1985}]{Binggeli1985}
Binggeli B.,  Sandage A.,   Tammann G.,  1985, \mn@doi [AJ] {10.1086/113874},
  90, 1681

\bibitem[\protect\citeauthoryear{Binggeli, Tammann  \& Sandage}{Binggeli
  et~al.}{1987}]{Binggeli1987}
Binggeli B.,  Tammann G.,   Sandage A.,  1987, \mn@doi [AJ] {10.1086/114467},
  94, 251

\bibitem[\protect\citeauthoryear{Blakeslee et~al.,}{Blakeslee
  et~al.}{2009}]{Blakeslee2009}
Blakeslee J.,  et~al., 2009, \mn@doi [ApJ] {10.1088/0004-637X/694/1/556}, 694,
  556

\bibitem[\protect\citeauthoryear{B{\"o}ker, van~den Marel  \& Vacca}{B{\"o}ker
  et~al.}{1999a}]{Boeker1999}
B{\"o}ker T.,  van~den Marel R.,   Vacca W.,  1999a, \mn@doi [ApJ]
  {10.1086/300985}, 118, 831

\bibitem[\protect\citeauthoryear{B{\"o}ker et~al.,}{B{\"o}ker
  et~al.}{1999b}]{Boeker1999b}
B{\"o}ker T.,  et~al., 1999b, \mn@doi [ApJS] {10.1086/313253}, 124, 95

\bibitem[\protect\citeauthoryear{B{\"o}ker, van~den Marel, Mazzuca, Rix,
  Rudnick, Ho  \& Shields}{B{\"o}ker et~al.}{2001}]{Boeker2001}
B{\"o}ker T.,  van~den Marel R.,  Mazzuca L.,  Rix H.-W.,  Rudnick G.,  Ho L.,
   Shields J.,  2001, \mn@doi [ApJ] {10.1086/319415}, 121, 1473

\bibitem[\protect\citeauthoryear{B{\"o}ker, Laine, van~der Marel, Sarzi, Rix,
  Ho  \& Shields}{B{\"o}ker et~al.}{2002}]{Boeker2002}
B{\"o}ker T.,  Laine S.,  van~der Marel R.,  Sarzi M.,  Rix H.-W.,  Ho L.,
  Shields J.,  2002, \mn@doi [ApJ] {10.1086/339025}, 123, 1389

\bibitem[\protect\citeauthoryear{B{\"o}ker, Lisenfeld  \& Schinnerer}{B{\"o}ker
  et~al.}{2003}]{Boeker2003}
B{\"o}ker T.,  Lisenfeld U.,   Schinnerer E.,  2003, \mn@doi [ApJ]
  {10.1051/0004-6361:20030755}, 406, 87

\bibitem[\protect\citeauthoryear{B{\"o}ker, Laine, van~der Marel, Sarzi, Rix,
  Ho  \& Shields}{B{\"o}ker et~al.}{2004}]{Boeker2004b}
B{\"o}ker T.,  Laine S.,  van~der Marel R.,  Sarzi M.,  Rix H.-W.,  Ho L.,
  Shields J.,  2004, \mn@doi [ApJ] {10.1086/380231}, 127, 105

\bibitem[\protect\citeauthoryear{Bruzual \& Charlot}{Bruzual \&
  Charlot}{2003}]{Bruzual2003}
Bruzual G.,  Charlot S.,  2003, \mn@doi [MNRAS]
  {10.1046/j.1365-8711.2003.06897.x}, 344, 1000

\bibitem[\protect\citeauthoryear{Butler \& Mart\'inez-Delgado}{Butler \&
  Mart\'inez-Delgado}{2005}]{Butler2005}
Butler D.,  Mart\'inez-Delgado D.,  2005, \mn@doi [AJ] {10.1086/429524}, 129,
  2217

\bibitem[\protect\citeauthoryear{Byun et~al.,}{Byun et~al.}{2020}]{Byun2020}
Byun W.,  et~al., 2020, \mn@doi [ApJ] {10.3847/1538-4357/ab6f6e}, 891, 13

\bibitem[\protect\citeauthoryear{Calzetti et~al.,}{Calzetti
  et~al.}{2015}]{Calzetti2015}
Calzetti D.,  et~al., 2015, \mn@doi [ApJ] {10.1088/0004-637X/811/2/75}, 811, 26

\bibitem[\protect\citeauthoryear{Carlsten, Greene, Peter, Beaton  \&
  Greco}{Carlsten et~al.}{2020a}]{Carlsten2020b}
Carlsten S.,  Greene J.,  Peter A.,  Beaton R.,   Greco J.,  2020a, preprint
  (\mn@eprint {arXiv} {2006.02443})

\bibitem[\protect\citeauthoryear{Carlsten, Greco, Beaton  \& Greene}{Carlsten
  et~al.}{2020b}]{Carlsten2020a}
Carlsten S.,  Greco J.,  Beaton R.,   Greene J.,  2020b, \mn@doi [ApJ]
  {10.3847/1538-4357/ab7758}, 891, 37

\bibitem[\protect\citeauthoryear{Carlsten, Greene, Greco, Beaton  \&
  Kado-Fong}{Carlsten et~al.}{2021b}]{Carlsten2021a}
Carlsten S.~G.,  Greene J.~E.,  Greco J.~P.,  Beaton R.~L.,   Kado-Fong E.,
  2021b, preprint (\mn@eprint {arXiv} {2105.03435})

\bibitem[\protect\citeauthoryear{Carlsten, Greene, Beaton  \& Greco}{Carlsten
  et~al.}{2021a}]{Carlsten2021b}
Carlsten S.~G.,  Greene J.~E.,  Beaton R.~L.,   Greco J.~P.,  2021a, preprint
  (\mn@eprint {arXiv} {2105.03440})

\bibitem[\protect\citeauthoryear{Carollo, Stiavelli, de Zeeuw  \& Mack}{Carollo
  et~al.}{1997}]{Carollo1997}
Carollo C.,  Stiavelli M.,  de Zeeuw P.,   Mack J.,  1997, \mn@doi [AJ]
  {10.1086/118654}, 114, 2366

\bibitem[\protect\citeauthoryear{Carollo, Stiavelli  \& Mack}{Carollo
  et~al.}{1998}]{Carollo1998}
Carollo C.,  Stiavelli M.,   Mack J.,  1998, \mn@doi [AJ] {10.1086/300407},
  116, 68

\bibitem[\protect\citeauthoryear{Carollo, Stiavelli, Seigar, de Zeeuw  \&
  Dejonghe}{Carollo et~al.}{2002}]{Carollo2002}
Carollo C.,  Stiavelli M.,  Seigar M.,  de Zeeuw P.,   Dejonghe H.,  2002,
  \mn@doi [AJ] {10.1086/324725}, 123, 159

\bibitem[\protect\citeauthoryear{Carretta et~al.,}{Carretta
  et~al.}{2010}]{Carretta2010}
Carretta E.,  et~al., 2010, \mn@doi [A\&A] {10.1051/0004-6361/201014924}, 520,
  A95

\bibitem[\protect\citeauthoryear{Carson, Barth, Seth, den Brok, Cappellari,
  Greene, Ho  \& Neumayer}{Carson et~al.}{2015}]{Carson2015}
Carson D.,  Barth A.,  Seth A.,  den Brok M.,  Cappellari M.,  Greene J.,  Ho
  L.,   Neumayer N.,  2015, \mn@doi [AJ] {10.1088/0004-6256/149/5/170}, 149,
  170

\bibitem[\protect\citeauthoryear{Chiboucas et~al.,}{Chiboucas
  et~al.}{2011}]{Chiboucas2011}
Chiboucas K.,  et~al., 2011, \mn@doi [ApJ] {10.1088/0004-637X/737/2/86}, 737,
  26

\bibitem[\protect\citeauthoryear{Cohen et~al.,}{Cohen et~al.}{2018}]{Cohen2018}
Cohen Y.,  et~al., 2018, \mn@doi [ApJ] {10.3847/1538-4357/aae7c8}, 868, 14

\bibitem[\protect\citeauthoryear{Cole et~al.,}{Cole et~al.}{2017}]{Cole2017}
Cole A.,  et~al., 2017, \mn@doi [ApJ] {10.3847/1538-4357/aa5df6}, 837, 12

\bibitem[\protect\citeauthoryear{Collaboration}{Collaboration}{2013}]{Astropy2013}
Collaboration A.,  2013, \mn@doi [A\&A] {10.1051/0004-6361/201322068}, 558, 9

\bibitem[\protect\citeauthoryear{Collaboration}{Collaboration}{2018}]{Astropy2018}
Collaboration A.,  2018, \mn@doi [AJ] {10.3847/1538-3881/aabc4f}, 156, 19

\bibitem[\protect\citeauthoryear{Conroy, Gunn  \& White}{Conroy
  et~al.}{2009}]{Conroy2009}
Conroy C.,  Gunn J.,   White M.,  2009, \mn@doi [ApJ]
  {10.1088/0004-637X/699/1/486}, 699, 486

\bibitem[\protect\citeauthoryear{Contenta et~al.,}{Contenta
  et~al.}{2018}]{Contenta2018}
Contenta F.,  et~al., 2018, \mn@doi [MNRAS] {10.1093/mnras/sty424}, 476, 3124

\bibitem[\protect\citeauthoryear{C{\^o}t\'e et~al.,}{C{\^o}t\'e
  et~al.}{2006}]{Cote2006}
C{\^o}t\'e P.,  et~al., 2006, \mn@doi [ApJS] {10.1086/504042}, 165, 57

\bibitem[\protect\citeauthoryear{Das, Sengupta, Ramya  \& Misra}{Das
  et~al.}{2012}]{Das2012}
Das M.,  Sengupta C.,  Ramya S.,   Misra K.,  2012, \mn@doi [MNRAS]
  {10.1111/j.1365-2966.2012.21120.x}, 423, 3274

\bibitem[\protect\citeauthoryear{Davidge \& Courteau}{Davidge \&
  Courteau}{2002}]{Davidge2002}
Davidge T.,  Courteau S.,  2002, \mn@doi [AJ] {10.1086/338901}, 123, 1438

\bibitem[\protect\citeauthoryear{Desroches \& Ho}{Desroches \&
  Ho}{2008}]{Desroches2009}
Desroches L.-B.,  Ho L.,  2008, \mn@doi [AJ] {10.1088/0004-637x/690/1/267},
  690, 267

\bibitem[\protect\citeauthoryear{Drinkwater, Gregg  \& Colless}{Drinkwater
  et~al.}{2001}]{Drinkwater2001}
Drinkwater M.,  Gregg M.,   Colless M.,  2001, \mn@doi [ApJ] {10.1086/319113},
  548, L139

\bibitem[\protect\citeauthoryear{Du, Cheng, Zheng  \& Wu}{Du
  et~al.}{2020}]{Du2020b}
Du W.,  Cheng C.,  Zheng Z.,   Wu H.,  2020, \mn@doi [AJ]
  {10.3847/1538-3881/ab6efb}, 138, 17

\bibitem[\protect\citeauthoryear{Eigenthaler et~al.,}{Eigenthaler
  et~al.}{2018}]{Eigenthaler2018}
Eigenthaler P.,  et~al., 2018, \mn@doi [ApJ] {10.3847/1538-4357/aaab60}, 855,
  19

\bibitem[\protect\citeauthoryear{Emsellem, Dejonghe  \& Bacon}{Emsellem
  et~al.}{1999}]{Emsellem1999}
Emsellem E.,  Dejonghe H.,   Bacon R.,  1999, \mn@doi [MNRAS]
  {10.1046/j.1365-8711.1999.02210.x}, 303, 495

\bibitem[\protect\citeauthoryear{Erwin \& Gadotti}{Erwin \&
  Gadotti}{2012}]{Erwin2012}
Erwin P.,  Gadotti D.,  2012, \mn@doi [Advances in Astronomy]
  {10.1155/2012/946368}, 2012, 11

\bibitem[\protect\citeauthoryear{Fahrion et~al.,}{Fahrion
  et~al.}{2020}]{Fahrion2020}
Fahrion K.,  et~al., 2020, \mn@doi [A\&A] {10.1051/0004-6361/201937120}, 634,
  13

\bibitem[\protect\citeauthoryear{Fahrion et~al.,}{Fahrion
  et~al.}{2021}]{Fahrion2021}
Fahrion K.,  et~al., 2021, preprint (\mn@eprint {arXiv} {2104.06412})

\bibitem[\protect\citeauthoryear{Ferguson \& Sandage}{Ferguson \&
  Sandage}{1989}]{Ferguson1989}
Ferguson H.,  Sandage A.,  1989, \mn@doi [ApJL] {10.1086/185577}, 346, 4

\bibitem[\protect\citeauthoryear{Ferrarese et~al.,}{Ferrarese
  et~al.}{2000}]{Ferrarese2000b}
Ferrarese L.,  et~al., 2000, \mn@doi [ApJ] {10.1086/308309}, 529, 745

\bibitem[\protect\citeauthoryear{Ferrarese et~al.,}{Ferrarese
  et~al.}{2006}]{Ferrarese2006}
Ferrarese L.,  et~al., 2006, \mn@doi [ApJ] {10.1086/505388}, 644, L21

\bibitem[\protect\citeauthoryear{Ferrarese et~al.,}{Ferrarese
  et~al.}{2020}]{Ferrarese2020}
Ferrarese L.,  et~al., 2020, \mn@doi [ApJ] {10.3847/1538-4357/ab339f}, 890, 108

\bibitem[\protect\citeauthoryear{Filippenko \& Ho}{Filippenko \&
  Ho}{2003}]{Filippenko2003}
Filippenko A.,  Ho L.,  2003, \mn@doi [ApJ] {10.1086/375361}, 588, L13

\bibitem[\protect\citeauthoryear{Filippenko \& Sargent}{Filippenko \&
  Sargent}{1989}]{Filippenko1989}
Filippenko A.,  Sargent W.,  1989, \mn@doi [ApJ] {10.1086/185472}, 342, L11

\bibitem[\protect\citeauthoryear{Fisher}{Fisher}{1992}]{Fisher1992}
Fisher R.~A.,  1992, Statistical Methods for Research Workers.
Springer New York, New York, NY, pp 66--70,
  \mn@doi{10.1007/978-1-4612-4380-9_6}, \url
  {https://doi.org/10.1007/978-1-4612-4380-9_6}

\bibitem[\protect\citeauthoryear{Fitzpatrick}{Fitzpatrick}{1999}]{Fitzpatrick1999}
Fitzpatrick E.,  1999, \mn@doi [PASP] {10.1086/316293}, 111, 63

\bibitem[\protect\citeauthoryear{Foreman-Mackey}{Foreman-Mackey}{2016}]{Foreman-Mackey2016}
Foreman-Mackey D.,  2016, \mn@doi [The Journal of Open Source Software]
  {10.21105/joss.00024}, 1, 24

\bibitem[\protect\citeauthoryear{Foreman-Mackey, Hogg, Lang  \&
  Goodman}{Foreman-Mackey et~al.}{2013}]{Foreman-Mackey2013}
Foreman-Mackey D.,  Hogg D.,  Lang D.,   Goodman J.,  2013, \mn@doi [PASP]
  {10.1086/670067}, 125, 306

\bibitem[\protect\citeauthoryear{Ganda, Peletier, Balcells  \&
  Falc{\'o}n-Barroso}{Ganda et~al.}{2009}]{Ganda2009}
Ganda K.,  Peletier R.,  Balcells M.,   Falc{\'o}n-Barroso J.,  2009, \mn@doi
  [MNRAS] {10.1111/j.1365-2966.2009.14658.x}, 395, 1669

\bibitem[\protect\citeauthoryear{Georgiev \& B{\"o}ker}{Georgiev \&
  B{\"o}ker}{2014}]{Georgiev2014}
Georgiev I.,  B{\"o}ker T.,  2014, \mn@doi [MNRAS] {10.1093/mnras/stu797}, 441,
  3570

\bibitem[\protect\citeauthoryear{Georgiev, Goudfrooij, Puzia  \&
  Hilker}{Georgiev et~al.}{2008}]{Georgiev2008}
Georgiev I.,  Goudfrooij P.,  Puzia T.,   Hilker M.,  2008, \mn@doi [AJ]
  {10.1088/0004-6256/135/5/1858}, 135, 1858

\bibitem[\protect\citeauthoryear{Georgiev, Puzia, Hilker  \&
  Goudfrooij}{Georgiev et~al.}{2009a}]{Georgiev2009b}
Georgiev I.,  Puzia T.,  Hilker M.,   Goudfrooij P.,  2009a, \mn@doi [MNRAS]
  {10.1111/j.1365-2966.2008.14104.x}, 392, 879

\bibitem[\protect\citeauthoryear{Georgiev, Puzia, Hilker, Goudfrooij  \&
  Baumgardt}{Georgiev et~al.}{2009b}]{Georgiev2009}
Georgiev I.,  Puzia T.,  Hilker M.,  Goudfrooij P.,   Baumgardt H.,  2009b,
  \mn@doi [MNRAS] {10.1111/j.1365-2966.2009.14776.x}, 396, 1075

\bibitem[\protect\citeauthoryear{Ginsburg et~al.,}{Ginsburg
  et~al.}{2019}]{Ginsburg2019}
Ginsburg A.,  et~al., 2019, \mn@doi [AJ] {10.3847/1538-3881/aafc33}, 157, 7

\bibitem[\protect\citeauthoryear{Gnerucci, Marconi, Capetti, Axon  \&
  Robinson}{Gnerucci et~al.}{2013}]{Gnerucci2013}
Gnerucci A.,  Marconi A.,  Capetti A.,  Axon D.,   Robinson A.,  2013, \mn@doi
  [A\&A] {10.1051/0004-6361/201118709}, 549, A139

\bibitem[\protect\citeauthoryear{Gonz{\'a}lez~Delgado, P{\'e}rez, Cid~Fernandes
   \& Schmitt}{Gonz{\'a}lez~Delgado et~al.}{2008}]{Gonzalez_Delgado2008}
Gonz{\'a}lez~Delgado R.,  P{\'e}rez E.,  Cid~Fernandes R.,   Schmitt H.,  2008,
  \mn@doi [AJ] {10.1088/0004-6256/135/3/747}, 135, 747

\bibitem[\protect\citeauthoryear{Gordon, Hanson, Clayton, Rieke  \&
  Misselt}{Gordon et~al.}{1999}]{Gordon1999}
Gordon K.,  Hanson M.,  Clayton G.,  Rieke H.,   Misselt K.,  1999, \mn@doi
  [ApJ] {10.1086/307350}, 519, 165

\bibitem[\protect\citeauthoryear{Graham}{Graham}{2008}]{Graham2008}
Graham A.,  2008, \mn@doi [Publications of The Astronomical Society of
  Australia] {10.1071/as08013}, 25, 167

\bibitem[\protect\citeauthoryear{Graham}{Graham}{2012}]{Graham2012}
Graham A.,  2012, \mn@doi [MNRAS] {10.1111/j.1365-2966.2012.20734.x}, 422, 1586

\bibitem[\protect\citeauthoryear{Graham \& Driver}{Graham \&
  Driver}{2007}]{Graham2007b}
Graham A.,  Driver S.,  2007, \mn@doi [ApJ] {10.1086/509758}, 655, 77

\bibitem[\protect\citeauthoryear{Graham \& Scott}{Graham \&
  Scott}{2013}]{Graham2013}
Graham A.,  Scott N.,  2013, \mn@doi [ApJ] {10.1088/0004-637x/764/2/151}, 764,
  151

\bibitem[\protect\citeauthoryear{Graham \& Spitler}{Graham \&
  Spitler}{2009}]{Graham2009}
Graham A.,  Spitler L.,  2009, \mn@doi [MNRAS]
  {10.1111/j.1365-2966.2009.15118.x}, 397, 2148

\bibitem[\protect\citeauthoryear{Green}{Green}{2018}]{Green2018}
Green G.,  2018, \mn@doi [The Journal of Open Source Software]
  {10.21105/joss.00695}, 3, 695

\bibitem[\protect\citeauthoryear{Habas et~al.,}{Habas et~al.}{2020}]{Habas2020}
Habas R.,  et~al., 2020, \mn@doi [MNRAS] {10.1093/mnras/stz3045}, 491, 1901

\bibitem[\protect\citeauthoryear{Hartmann, Debattista, Seth, Cappellari  \&
  Quinn}{Hartmann et~al.}{2011}]{Hartmann2011}
Hartmann M.,  Debattista V.,  Seth A.,  Cappellari M.,   Quinn T.,  2011,
  \mn@doi [MNRAS] {10.1111/j.1365-2966.2011.19659.x}, 418, 2697

\bibitem[\protect\citeauthoryear{Ho, Filippenko  \& Sargant}{Ho
  et~al.}{1997}]{Ho1997}
Ho L.,  Filippenko A.,   Sargant W.,  1997, \mn@doi [ApJS] {10.1086/313041},
  112, 315

\bibitem[\protect\citeauthoryear{Hopkins, Cox, Dutta, Hernquist, Kormendy  \&
  Lauer}{Hopkins et~al.}{2009}]{Hopkins2009}
Hopkins P.,  Cox T.,  Dutta S.,  Hernquist L.,  Kormendy J.,   Lauer T.,  2009,
  \mn@doi [ApJS] {10.1088/0067-0049/181/1/135}, 181, 135

\bibitem[\protect\citeauthoryear{Hunter}{Hunter}{2007}]{Hunter2007}
Hunter J.,  2007, \mn@doi [Computing in Science \& Engineering]
  {10.1109/MCSE.2007.55}, 9, 90

\bibitem[\protect\citeauthoryear{Huxor et~al.,}{Huxor et~al.}{2014}]{Huxor2014}
Huxor A.,  et~al., 2014, \mn@doi [MNRAS] {10.1093/mnras/stu771}, 442, 2165

\bibitem[\protect\citeauthoryear{Into \& Portinari}{Into \&
  Portinari}{2013}]{Into2013}
Into T.,  Portinari L.,  2013, \mn@doi [MNRAS] {10.1093/mnras/stt071}, 430,
  2715

\bibitem[\protect\citeauthoryear{Iodice et~al.,}{Iodice
  et~al.}{2016}]{Iodice2016}
Iodice E.,  et~al., 2016, \mn@doi [ApJ] {10.3847/0004-637X/820/1/42}, 820, 17

\bibitem[\protect\citeauthoryear{Jarrett, Chester, Cutri, Schneider  \&
  Huchra}{Jarrett et~al.}{2003}]{Jarrett2003}
Jarrett T.,  Chester T.,  Cutri R.,  Schneider S.,   Huchra J.,  2003, \mn@doi
  [AJ] {10.1086/345794}, 125, 525

\bibitem[\protect\citeauthoryear{Jones et~al.,}{Jones et~al.}{1996}]{Jones1996}
Jones D.,  et~al., 1996, \mn@doi [ApJ] {10.1086/177547}, 466, 742

\bibitem[\protect\citeauthoryear{Kacharov \& Makarov}{Kacharov \&
  Makarov}{1999}]{Karachentsev1999b}
Kacharov I.,  Makarov D.,  1999, in Barnes J.,  Sanders D.,  eds, Galaxy
  Interactions at Low and High Redshift. International Astronomical Union, pp
  107--116

\bibitem[\protect\citeauthoryear{Kacharov, Neumayer, Seth, Cappellari,
  McDermid, Walcher  \& B{\"o}ker}{Kacharov et~al.}{2018}]{Kacharov2018}
Kacharov N.,  Neumayer N.,  Seth A.,  Cappellari M.,  McDermid R.,  Walcher C.,
    B{\"o}ker T.,  2018, \mn@doi [MNRAS] {10.1093/mnras/sty1985}, 480, 1973

\bibitem[\protect\citeauthoryear{Karachentsev \& Karachentseva}{Karachentsev \&
  Karachentseva}{2002}]{Karachentsev2002}
Karachentsev I.,  Karachentseva V.,  2002, in Green R.,  Khachikian E.,
  Sanders D.,  eds, Proceedings of IAU Colloquium 184. ASP Conference
  Proceedings, pp 325--334

\bibitem[\protect\citeauthoryear{Karachentsev, Karachentseva, Huchtmeier  \&
  Makarov}{Karachentsev et~al.}{2004}]{Karachentsev2004}
Karachentsev I.,  Karachentseva V.,  Huchtmeier W.,   Makarov D.,  2004,
  \mn@doi [AJ] {10.1086/382905}, 127, 2031

\bibitem[\protect\citeauthoryear{Karachentsev, Makarov  \&
  Kaisina}{Karachentsev et~al.}{2013}]{Karachentsev2013}
Karachentsev I.,  Makarov D.,   Kaisina E.,  2013, \mn@doi [AJ]
  {10.1088/0004-6256/145/4/101}, 145, 101

\bibitem[\protect\citeauthoryear{Karachentsev, Makarova, Tully, Anand, Rizzi,
  Shaya  \& Afansiev}{Karachentsev et~al.}{2020a}]{Karachentsev2020}
Karachentsev I.,  Makarova L.,  Tully R.,  Anand G.,  Rizzi L.,  Shaya E.,
  Afansiev V.,  2020a, \mn@doi [A\&A] {10.1051/0004-6361/202037993}, 638, 6

\bibitem[\protect\citeauthoryear{Karachentsev, Makarova, Brent, Anand, Rizzi
  \& Shaya}{Karachentsev et~al.}{2020b}]{Karachentsev2020b}
Karachentsev I.~D.,  Makarova L.~N.,  Brent T.~R.,  Anand G.~S.,  Rizzi L.,
  Shaya E.~J.,  2020b, \mn@doi [A\&A] {10.1051/0004-6361/202038928}, 643, 5

\bibitem[\protect\citeauthoryear{Kim et~al.,}{Kim et~al.}{2014}]{Kim2014}
Kim S.,  et~al., 2014, \mn@doi [ApJS] {10.1088/0067-0049/215/2/22}, 215, 29

\bibitem[\protect\citeauthoryear{Kim, Jerjen, Milone, Mackey  \& Da~Costa}{Kim
  et~al.}{2015}]{Kim2015}
Kim D.,  Jerjen H.,  Milone A.,  Mackey D.,   Da~Costa G.,  2015, \mn@doi [ApJ]
  {10.1088/0004-637X/803/2/63}, 803, 9

\bibitem[\protect\citeauthoryear{Kormendy \& Bender}{Kormendy \&
  Bender}{1999}]{Kormendy1999}
Kormendy J.,  Bender R.,  1999, \mn@doi [ApJ] {10.1086/307665}, 522, 772

\bibitem[\protect\citeauthoryear{Kormendy \& Ho}{Kormendy \&
  Ho}{2013}]{Kormendy2013}
Kormendy J.,  Ho L.,  2013, \mn@doi [ARA\&A]
  {10.1146/annurev-astro-082708-101811}, 51, 511

\bibitem[\protect\citeauthoryear{Kormendy et~al.,}{Kormendy
  et~al.}{1996}]{Kormendy1996}
Kormendy J.,  et~al., 1996, \mn@doi [ApJ] {10.1086/310399}, 473, L91

\bibitem[\protect\citeauthoryear{Kormendy, Drory, Bender  \& Cornell}{Kormendy
  et~al.}{2010}]{Kormendy2010}
Kormendy J.,  Drory N.,  Bender R.,   Cornell M.,  2010, \mn@doi [ApJ]
  {10.1088/0004-637x/723/1/54}, 723, 54

\bibitem[\protect\citeauthoryear{Kraan-Korteweg \& Tammann}{Kraan-Korteweg \&
  Tammann}{1979}]{Kraan-Korteweg1979}
Kraan-Korteweg R.,  Tammann G.,  1979, \mn@doi [Astronomische Nachrichten]
  {10.1002/asna.19793000403}, 300, 181

\bibitem[\protect\citeauthoryear{Krist}{Krist}{1993}]{Krist1993}
Krist J.,  1993, in Hanisch R.,  Brissenden R.,   Barnes J.,  eds, Astronomical
  Data Analysis Software and Systems II. A.S.P. Conference Series, pp 536--539,
  \url {https://ui.adsabs.harvard.edu/abs/1993ASPC...52..536K}

\bibitem[\protect\citeauthoryear{Krist}{Krist}{1995}]{Krist1995}
Krist J.,  1995, in Shaw R.,  Payne H.,   Hayes J.,  eds, Astronomical Data
  Analysis Software and Systems IV. A.S.P. Conference Series, pp 349--352, \url
  {https://ui.adsabs.harvard.edu/abs/1995ASPC...77..349K}

\bibitem[\protect\citeauthoryear{Lauer et~al.,}{Lauer et~al.}{2005}]{Lauer2005}
Lauer T.,  et~al., 2005, \mn@doi [AJ] {10.1086/429565}, 129, 2139

\bibitem[\protect\citeauthoryear{Lim, Peng, C{\^{o}}t{\"{e}}, Sales, den Brok,
  Blakeslee  \& Guhathakurta}{Lim et~al.}{2018}]{Lim2018}
Lim S.,  Peng E.,  C{\^{o}}t{\"{e}} P.,  Sales L.,  den Brok M.,  Blakeslee J.,
    Guhathakurta P.,  2018, \mn@doi [ApJ] {10.3847/1538-4357/aacb81}, 862, 11

\bibitem[\protect\citeauthoryear{Lisker, Grebel, Binggeli  \& Glatt}{Lisker
  et~al.}{2007}]{Lisker2007}
Lisker T.,  Grebel E.~K.,  Binggeli B.,   Glatt K.,  2007, \mn@doi [ApJ]
  {10.1086/513090}, 660, 1186

\bibitem[\protect\citeauthoryear{Liu, Peng, Jord{\'{a}}, Blakeslee,
  C{\^{o}}t{\'{e}}, Ferrarese  \& Puzia}{Liu et~al.}{2019}]{Liu2019}
Liu Y.,  Peng E.~W.,  Jord{\'{a}} A.,  Blakeslee J.~P.,  C{\^{o}}t{\'{e}} P.,
  Ferrarese L.,   Puzia T.~H.,  2019, \mn@doi [ApJ] {10.3847/1538-4357/ab12d9},
  875, 13

\bibitem[\protect\citeauthoryear{Luo et~al.,}{Luo et~al.}{2012}]{Luo2012}
Luo B.,  et~al., 2012, \mn@doi [ApJ] {10.1088/0004-637x/749/2/130}, 749, 130

\bibitem[\protect\citeauthoryear{Madrid et~al.,}{Madrid
  et~al.}{2010}]{Madrid2010}
Madrid J.,  et~al., 2010, \mn@doi [ApJ] {10.1088/0004-637X/722/2/1707}, 722,
  1707

\bibitem[\protect\citeauthoryear{Makarov, Terekhova, Courtois  \&
  Vauglin}{Makarov et~al.}{2014}]{Makarov2014}
Makarov D.,  Terekhova N.,  Courtois H.,   Vauglin I.,  2014, \mn@doi [A\&A]
  {10.1051/0004-6361/201423496}, 570, A13

\bibitem[\protect\citeauthoryear{Martocchia, Dalessandro, Salaris, Larsen  \&
  Rejkuba}{Martocchia et~al.}{2020}]{Martocchia2020}
Martocchia S.,  Dalessandro E.,  Salaris M.,  Larsen S.,   Rejkuba M.,  2020,
  \mn@doi [MNRAS] {10.1093/mnras/staa1482}, 495, 4518

\bibitem[\protect\citeauthoryear{Matthews \& Gallagher}{Matthews \&
  Gallagher}{2002}]{Matthews2002}
Matthews L.,  Gallagher J.,  2002, \mn@doi [ApJS] {10.1086/340647}, 141, 429

\bibitem[\protect\citeauthoryear{Matthews et~al.,}{Matthews
  et~al.}{1999}]{Matthews1999}
Matthews L.,  et~al., 1999, \mn@doi [AJ] {10.1086/300909}, 118, 208

\bibitem[\protect\citeauthoryear{McConnachie et~al.,}{McConnachie
  et~al.}{2018}]{McConnachie2018}
McConnachie A.,  et~al., 2018, \mn@doi [ApJ] {10.3847/1538-4357/aae8e7}, 868,
  36

\bibitem[\protect\citeauthoryear{McGaugh \& Schombert}{McGaugh \&
  Schombert}{2014}]{McGaugh2014}
McGaugh S.,  Schombert J.,  2014, \mn@doi [AJ] {10.1088/0004-6256/148/5/77},
  148, 12

\bibitem[\protect\citeauthoryear{McMillan}{McMillan}{2011}]{McMillan2011}
McMillan P.,  2011, \mn@doi [MNRAS]
  {https://academic.oup.com/mnras/article/414/3/2446/1042117}, 414, 2446

\bibitem[\protect\citeauthoryear{Mei et~al.,}{Mei et~al.}{2007}]{Mei2007}
Mei S.,  et~al., 2007, \mn@doi [ApJ] {10.1086/509598}, 655, 144

\bibitem[\protect\citeauthoryear{Miller \& Lotz}{Miller \&
  Lotz}{2007}]{Miller2007}
Miller B.~W.,  Lotz J.~M.,  2007, \mn@doi [ApJ] {10.1086/522323}, 670, 1074

\bibitem[\protect\citeauthoryear{Miller, Lotz, Ferguson, Stiavelli  \&
  Whitmore}{Miller et~al.}{1998}]{Miller1998}
Miller B.~W.,  Lotz J.~M.,  Ferguson H.~C.,  Stiavelli M.,   Whitmore B.~C.,
  1998, \mn@doi [ApJ] {10.1086/311739}, 508, L133

\bibitem[\protect\citeauthoryear{Milosavljevi{\'{c}}}{Milosavljevi{\'{c}}}{2004}]{Milosavljevic2004}
Milosavljevi{\'{c}} M.,  2004, \mn@doi [ApJ] {10.1086/420696}, 605, L13

\bibitem[\protect\citeauthoryear{Milosavljevi{\'{c}} \&
  Bromm}{Milosavljevi{\'{c}} \& Bromm}{2014}]{Milosavljevic2014}
Milosavljevi{\'{c}} M.,  Bromm V.,  2014, \mn@doi [MNRAS]
  {10.1093/mnras/stu285}, 440, 50

\bibitem[\protect\citeauthoryear{Misgeld \& Hilker}{Misgeld \&
  Hilker}{2011}]{Misgeld2011}
Misgeld I.,  Hilker M.,  2011, \mn@doi [MNRAS]
  {10.1111/j.1365-2966.2011.18669.x}, 414, 3699

\bibitem[\protect\citeauthoryear{Mitzkus, Cappellari  \& Walcher}{Mitzkus
  et~al.}{2016}]{Mitzkus2017}
Mitzkus M.,  Cappellari M.,   Walcher J.,  2016, \mn@doi [MNRAS]
  {10.1093/mnras/stw2677}, 464, 4789

\bibitem[\protect\citeauthoryear{Monaco, Saviane, Perina, Bellazzini, Buzzoni,
  Federici, Fusi~Pecci  \& Galleti}{Monaco et~al.}{2009}]{Monaco2009}
Monaco L.,  Saviane I.,  Perina S.,  Bellazzini M.,  Buzzoni A.,  Federici L.,
  Fusi~Pecci F.,   Galleti S.,  2009, \mn@doi [A\&A]
  {10.1051/0004-6361/200912412}, 502, L9

\bibitem[\protect\citeauthoryear{Mu{\~{n}}oz et~al.,}{Mu{\~{n}}oz
  et~al.}{2015}]{Munoz2015}
Mu{\~{n}}oz R.,  et~al., 2015, \mn@doi [ApJL] {10.1088/2041-8205/813/1/L15},
  813, L15

\bibitem[\protect\citeauthoryear{Neumayer \& Walcher}{Neumayer \&
  Walcher}{2012}]{Neumayer2012}
Neumayer N.,  Walcher J.,  2012, \mn@doi [Advances in Astronomy]
  {10.1155/2012/709038}, 2012, 13

\bibitem[\protect\citeauthoryear{Neumayer, Walcher, Andersen, S{\'{a}}nchez,
  B{\"{o}}ker  \& Rix}{Neumayer et~al.}{2011}]{Neumayer2011}
Neumayer N.,  Walcher C.,  Andersen D.,  S{\'{a}}nchez F.,  B{\"{o}}ker T.,
  Rix H.-W.,  2011, \mn@doi [MNRAS] {10.1111/j.1365-2966.2011.18266.x}, 413,
  1875

\bibitem[\protect\citeauthoryear{Neumayer, Seth  \& B{\"{o}}ker}{Neumayer
  et~al.}{2020}]{Neumayer2020}
Neumayer N.,  Seth A.,   B{\"{o}}ker T.,  2020, \mn@doi [A\&ARv]
  {10.1007/s00159-020-00125-0}, 28, 75

\bibitem[\protect\citeauthoryear{Nguyen et~al.,}{Nguyen
  et~al.}{2017}]{Nguyen2017}
Nguyen D.,  et~al., 2017, \mn@doi [ApJ] {10.3847/1538-4357/aa5cb4}, 836, 237

\bibitem[\protect\citeauthoryear{Nguyen et~al.,}{Nguyen
  et~al.}{2018}]{Nguyen2018}
Nguyen D.,  et~al., 2018, \mn@doi [ApJ] {10.3847/1538-4357/aabe28}, 858, 118

\bibitem[\protect\citeauthoryear{Ordenes-Brice{\~{n}}o
  et~al.,}{Ordenes-Brice{\~{n}}o et~al.}{2018}]{Ordenes-Briceno2018}
Ordenes-Brice{\~{n}}o Y.,  et~al., 2018, \mn@doi [ApJ]
  {10.3847/1538-4357/aac1b8}, 860, 20

\bibitem[\protect\citeauthoryear{Pechetti, Seth, Neumayer, Georgiev, Kacharov
  \& den Brok}{Pechetti et~al.}{2020}]{Pechetti2019}
Pechetti R.,  Seth A.,  Neumayer N.,  Georgiev I.,  Kacharov N.,   den Brok M.,
   2020, \mn@doi [ApJ] {10.3847/1538-4357/abaaa7}, 900, 19

\bibitem[\protect\citeauthoryear{Peng, Ho, Impey  \& Rix}{Peng
  et~al.}{2002}]{Peng2002}
Peng C.,  Ho L.,  Impey C.,   Rix H.-W.,  2002, \mn@doi [AJ] {10.1086/340952},
  124, 266

\bibitem[\protect\citeauthoryear{Peng et~al.,}{Peng et~al.}{2008}]{Peng2008}
Peng E.,  et~al., 2008, \mn@doi [ApJ] {10.1086/587951}, 681, 197

\bibitem[\protect\citeauthoryear{Peng, Ho, Impey  \& Rix}{Peng
  et~al.}{2010}]{Peng2010}
Peng C.,  Ho L.,  Impey C.,   Rix H.-W.,  2010, \mn@doi [AJ]
  {10.1088/0004-6256/139/6/2097}, 139, 2097

\bibitem[\protect\citeauthoryear{Phillips, Illingworth, MacKenty  \&
  Franx}{Phillips et~al.}{1996}]{Phillips1996}
Phillips A.,  Illingworth G.,  MacKenty J.,   Franx M.,  1996, \mn@doi [AJ]
  {10.1086/117896}, 111, 1566

\bibitem[\protect\citeauthoryear{Piqueras~L{\'{o}}pez, Davies, Colina  \&
  Orban~de Xivry}{Piqueras~L{\'{o}}pez et~al.}{2012}]{Piqueras_Lopez2012}
Piqueras~L{\'{o}}pez J.,  Davies R.,  Colina L.,   Orban~de Xivry G.,  2012,
  \mn@doi [ApJ] {10.1088/0004-637X/752/1/47}, 752, 13

\bibitem[\protect\citeauthoryear{Portinari, Sommer-Larsen  \&
  Tantalo}{Portinari et~al.}{2004}]{Portinari2004}
Portinari L.,  Sommer-Larsen J.,   Tantalo R.,  2004, \mn@doi [MNRAS]
  {10.1111/j.1365-2966.2004.07207.x}, 347, 691

\bibitem[\protect\citeauthoryear{Puzia \& Sharina}{Puzia \&
  Sharina}{2008}]{Puzia2007}
Puzia T.,  Sharina M.,  2008, \mn@doi [ApJ] {10.1086/525038}, 674, 909

\bibitem[\protect\citeauthoryear{Ravindranath, Ho, Peng, Filippenko  \&
  Sargent}{Ravindranath et~al.}{2001}]{Ravindranath2001}
Ravindranath S.,  Ho L.,  Peng C.,  Filippenko A.,   Sargent W.,  2001, \mn@doi
  [AJ] {10.1086/321175}, 122, 653

\bibitem[\protect\citeauthoryear{Roediger \& Courteau}{Roediger \&
  Courteau}{2015}]{Roediger2015}
Roediger J.,  Courteau S.,  2015, \mn@doi [MNRAS] {10.1093/mnras/stv1499}, 452,
  3209

\bibitem[\protect\citeauthoryear{Rossa, van~der Marel, B{\"o}ker, Gerssen, Ho,
  Rix, Shields  \& Walcher}{Rossa et~al.}{2006}]{Rossa2006}
Rossa J.,  van~der Marel R.,  B{\"o}ker T.,  Gerssen J.,  Ho L.,  Rix H.-W.,
  Shields J.,   Walcher C.,  2006, \mn@doi [AJ] {10.1086/505968}, 132, 1074

\bibitem[\protect\citeauthoryear{S{\'{a}}nchez-Janssen \&
  Aguerri}{S{\'{a}}nchez-Janssen \& Aguerri}{2012}]{Sanchez-Janssen2012}
S{\'{a}}nchez-Janssen R.,  Aguerri J. A.~L.,  2012, \mn@doi [MNRAS]
  {10.1111/j.1365-2966.2012.21301.x}, 424, 2614

\bibitem[\protect\citeauthoryear{S{\'a}nchez-Janssen
  et~al.,}{S{\'a}nchez-Janssen et~al.}{2019}]{Sanchez-Janssen2019a}
S{\'a}nchez-Janssen R.,  et~al., 2019, \mn@doi [ApJ]
  {10.3847/1538-4357/aaf4fd}, 878, 18

\bibitem[\protect\citeauthoryear{Sarzi, Rix, Shield, Ho, Barth, Rudnick,
  Filippenko  \& Sargent}{Sarzi et~al.}{2005}]{Sarzi2005}
Sarzi M.,  Rix H.-W.,  Shield J.,  Ho L.,  Barth A.,  Rudnick G.,  Filippenko
  A.,   Sargent W.,  2005, \mn@doi [ApJ] {10.1086/428637}, 628, 169

\bibitem[\protect\citeauthoryear{Scarlata et~al.,}{Scarlata
  et~al.}{2004}]{Scarlata2004}
Scarlata C.,  et~al., 2004, \mn@doi [AJ] {10.1086/423036}, 128, 1124

\bibitem[\protect\citeauthoryear{Schlafly \& Finkbeiner}{Schlafly \&
  Finkbeiner}{2011}]{Schlafly2011}
Schlafly E.,  Finkbeiner D.,  2011, \mn@doi [ApJ]
  {10.1088/0004-637X/737/2/103}, 737, 13

\bibitem[\protect\citeauthoryear{Schlegel, Finkbeiner  \& Davis}{Schlegel
  et~al.}{1998}]{Schlegel1998}
Schlegel D.,  Finkbeiner D.,   Davis M.,  1998, \mn@doi [ApJ] {10.1086/305772},
  500, 525

\bibitem[\protect\citeauthoryear{Scott \& Graham}{Scott \&
  Graham}{2013}]{Scott2013}
Scott N.,  Graham A.,  2013, \mn@doi [ApJ] {10.1088/0004-637X/763/2/76}, 763,
  76

\bibitem[\protect\citeauthoryear{Scott, Graham  \& Schombert}{Scott
  et~al.}{2013}]{Scott2013b}
Scott N.,  Graham A.,   Schombert J.,  2013, \mn@doi [ApJ]
  {10.1088/0004-637X/768/1/76}, 768, 76

\bibitem[\protect\citeauthoryear{S{\'{e}}rsic}{S{\'{e}}rsic}{1968}]{Sersic1968}
S{\'{e}}rsic J.,  1968, Observatorio Astronomico, p.~142

\bibitem[\protect\citeauthoryear{Seth, Olsen, Miller, Lotz  \& Telford}{Seth
  et~al.}{2004}]{Seth2004}
Seth A.~C.,  Olsen K.,  Miller B.,  Lotz J.,   Telford R.,  2004, \mn@doi [ApJ]
  {10.1086/381070}, 127, 798

\bibitem[\protect\citeauthoryear{Seth, Dalcanton, Hodge  \& Debattista}{Seth
  et~al.}{2006}]{Seth2006}
Seth A.,  Dalcanton J.,  Hodge P.,   Debattista V.,  2006, \mn@doi [AJ]
  {10.1086/508994}, 132, 2539

\bibitem[\protect\citeauthoryear{Seth et~al.,}{Seth et~al.}{2010}]{Seth2010}
Seth A.,  et~al., 2010, \mn@doi [ApJ] {10.1088/0004-637X/714/1/713}, 714, 713

\bibitem[\protect\citeauthoryear{Smith, Crowther, Calzetti  \& Sidoli}{Smith
  et~al.}{2016}]{Smith2016}
Smith L.,  Crowther P.,  Calzetti D.,   Sidoli F.,  2016, \mn@doi [ApJ]
  {10.3847/0004-637X/823/1/38}, 823, 10

\bibitem[\protect\citeauthoryear{Su et~al.,}{Su et~al.}{2021}]{Su2021}
Su A.~H.,  et~al., 2021, \mn@doi [A\&A] {10.1051/0004-6361/202039633}, 647, 36

\bibitem[\protect\citeauthoryear{Tremaine, Ostriker  \& Spitzer}{Tremaine
  et~al.}{1975}]{Tremaine1975}
Tremaine S.,  Ostriker J.,   Spitzer L.,  1975, \mn@doi [ApJ] {10.1086/153422},
  196, 407

\bibitem[\protect\citeauthoryear{Tully}{Tully}{1988}]{Tully1988}
Tully R.,  1988, Nearby galaxies catalog.
Cambridge University Press, \url
  {https://ui.adsabs.harvard.edu/abs/1988ngc..book.....T}

\bibitem[\protect\citeauthoryear{Turner, C{\^{o}}t{\'{e}}, Ferrarese,
  Jord{\'{a}}n, Blakeslee, Mei, Peng  \& West}{Turner
  et~al.}{2012}]{Turner2012}
Turner M.,  C{\^{o}}t{\'{e}} P.,  Ferrarese L.,  Jord{\'{a}}n A.,  Blakeslee
  J.,  Mei S.,  Peng E.,   West M.,  2012, \mn@doi [ApJS]
  {10.1088/0067-0049/203/1/5}, 203, 33

\bibitem[\protect\citeauthoryear{Vasiliev \& Belokurov}{Vasiliev \&
  Belokurov}{2020}]{Vasiliev2020}
Vasiliev E.,  Belokurov V.,  2020, \mn@doi [MNRAS] {10.1093/mnras/staa2114},
  497, 4162

\bibitem[\protect\citeauthoryear{Veljanoski et~al.,}{Veljanoski
  et~al.}{2013}]{Veljanoski2013}
Veljanoski J.,  et~al., 2013, \mn@doi [MNRAS] {10.1093/mnras/stt1557}, 435,
  3654

\bibitem[\protect\citeauthoryear{Venhola et~al.,}{Venhola
  et~al.}{2018}]{Venhola2018}
Venhola A.,  et~al., 2018, \mn@doi [A\&A] {10.1051/0004-6361/201833933}, 620,
  31

\bibitem[\protect\citeauthoryear{Virtanen et~al.,}{Virtanen
  et~al.}{2020}]{Virtanen2020}
Virtanen P.,  et~al., 2020, \mn@doi [Nature Methods]
  {10.1038/s41592-019-0686-2}, \href {https://rdcu.be/b08Wh} {17, 261}

\bibitem[\protect\citeauthoryear{Walcher et~al.,}{Walcher
  et~al.}{2005}]{Walcher2005}
Walcher C.,  et~al., 2005, \mn@doi [ApJ] {10.1086/425977}, 618, 237

\bibitem[\protect\citeauthoryear{Wegner et~al.,}{Wegner
  et~al.}{2000}]{Wegner2000}
Wegner M.,  et~al., 2000, \mn@doi [A\&AS] {10.1051/aas:2000332}, 143, 9

\bibitem[\protect\citeauthoryear{Yong, Da~Costa  \& Norris}{Yong
  et~al.}{2016}]{Yong2016}
Yong D.,  Da~Costa G.,   Norris J.,  2016, \mn@doi [MNRAS]
  {10.1093/mnras/stw1091}, 460, 1846

\bibitem[\protect\citeauthoryear{Yuan, Smith, Xue, Li, Liu, Wang, Li  \&
  Chang}{Yuan et~al.}{2019}]{Yuan2019}
Yuan Z.,  Smith M.,  Xue X.-X.,  Li J.,  Liu C.,  Wang Y.,  Li L.,   Chang J.,
  2019, \mn@doi [ApJ] {10.3847/1538-4357/ab2e09}, 881, 10

\bibitem[\protect\citeauthoryear{Zanatta, S{\'{a}}nchez-Janssen, Chies-Santos,
  de Souza  \& Blakeslee}{Zanatta et~al.}{2021}]{Zanatta2021}
Zanatta E.,  S{\'{a}}nchez-Janssen R.,  Chies-Santos A.,  de Souza R.,
  Blakeslee J.,  2021, preprint (\mn@eprint {arXiv} {2103.02123})

\bibitem[\protect\citeauthoryear{Zibetti, Charlot  \& Rix}{Zibetti
  et~al.}{2009}]{Zibetti2009}
Zibetti S.,  Charlot S.,   Rix H.-W.,  2009, \mn@doi [MNRAS]
  {10.1111/j.1365-2966.2009.15528.x}, 400, 1181

\bibitem[\protect\citeauthoryear{den Brok et~al.,}{den Brok
  et~al.}{2014}]{denBrok2014}
den Brok M.,  et~al., 2014, \mn@doi [MNRAS] {10.1093/mnras/stu1906}, 445, 2385

\makeatother
\end{thebibliography}




\appendix

\section{Selection effects}
\label{sec:selection_effects}

\subsection{Biases due to incompleteness}
\label{subsec:missing_galaxies}

The UNGC is incomplete at the low stellar mass end \cpa{Karachentsev2013}.
We test if our sample is biased by comparing our results to the analysis of \citet{Carlsten2020a} and \citet{Habas2020} who investigate dwarf galaxy populations around high-mass field galaxies.

\citet{Carlsten2020a} detected \num{155} dwarf elliptical galaxy candidates around ten primary host galaxies in the Local Volume from which \num{93} are considered new detections and are \emph{not} part of the most recent iteration of the UNGC.
Although we do not calculate stellar masses for their galaxy sample as accurate distance estimates are unavailable for \SI{66.5}{\percent} (103 out of 155 galaxies; \citealp{Carlsten2020b}), they noted that their galaxy candidates are similar to known satellites in the Local Group, i.e.\ $M_{\star} \lesssim \SI{e7}{\Msun}$.
The authors argued that their sample has a completeness of \SI{> 90}{\percent} and, based on visual inspection of exposures, found \num{19} nucleated and \num{90} non-nucleated galaxies ($f_{\mathrm{n}} \approx \SI{19.3}{\percent}$).

\citet{Habas2020} identified \num{2210} dwarf galaxies from which \SI{75}{\percent} were assigned an early-type morphology.
Similar to \citet{Carlsten2020a}, they argued that, based on scaling relations, their sample is comparable to dwarf galaxies in the Local Group and the Fornax and Virgo galaxy clusters (again, $M_{\star} \lesssim \SI{e7}{\Msun}$).
The authors found $f_{\mathrm{n}} \approx \SI{28.2}{\percent}$ for dwarf early-types and $f_{\mathrm{n}} \approx \SI{9.4}{\percent}$ for irregular-type galaxies.
Both values are in excellent agreement with our results as most of the early- and irregular-type galaxies in the Local Volume are system of similar mass compared to their samples (see also \Cref{fig:nucfrac_hubble_type}).
Thus, our data are unbiased for low-mass early- and irregular-type galaxies.

\begin{figure}
  \centering
  \includegraphics[width=\columnwidth]{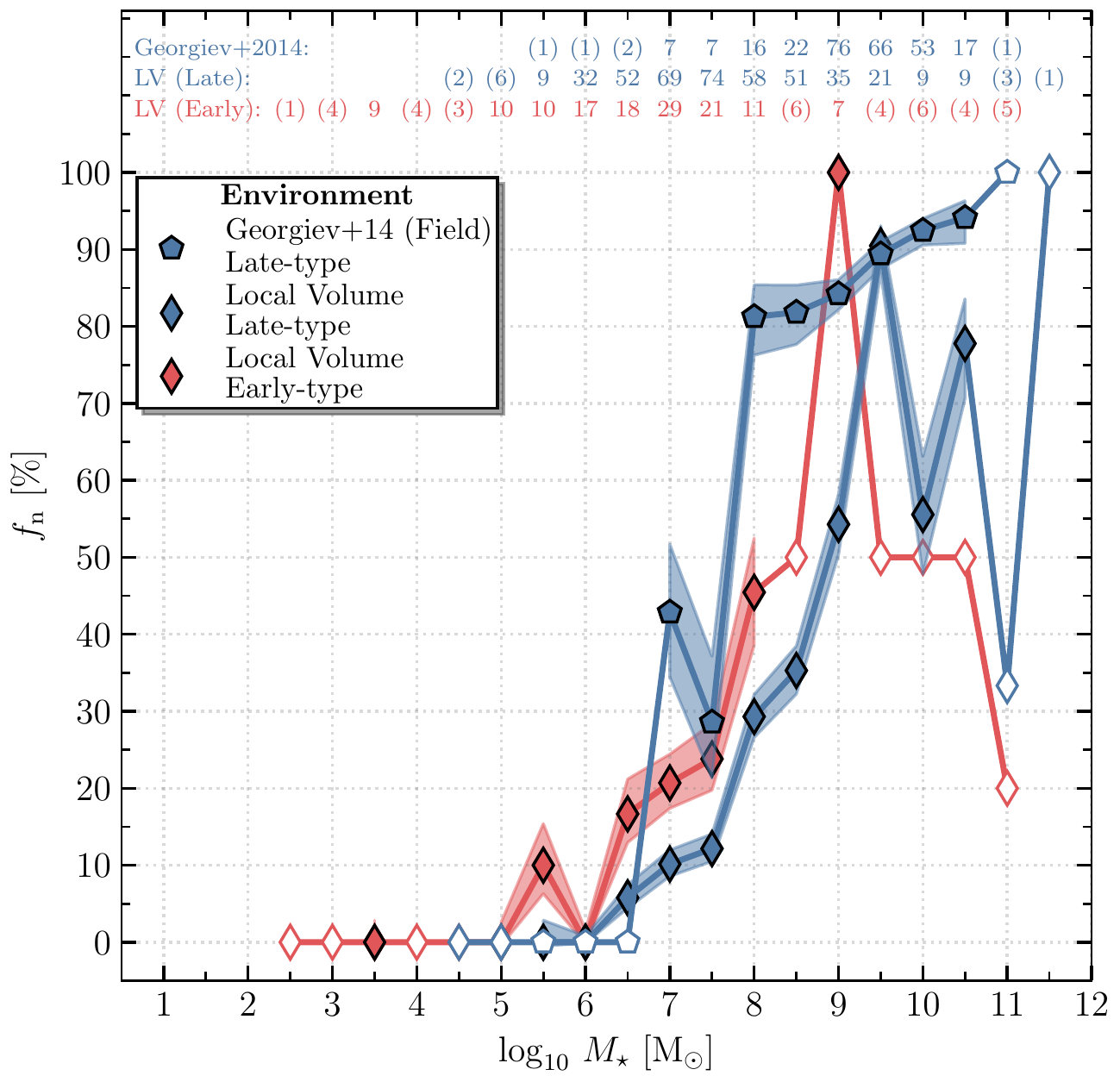}
  \caption{%
    Nucleation fraction ({\nucfrac}) as a function of logarithmic stellar mass ($\log_{10} \, M_{\star}$) for the Local Volume (diamonds) and the late-type sample of \citet{Georgiev2014} (\textbf{pentagons}).
    The presentation is the same as in \Cref{fig:nucfrac_hubble_type}.
  }
  \label{fig:nf_g14}
\end{figure}

Next, we compare our results to the study of \citet{Georgiev2014} who analysed NSCs in \num{323} late-type (Hubble type \num{> 3.5}) galaxies which do not lie edge-on (inclination $\lesssim \SI{88}{\degree}$) and have a distance estimate \SI{\leq 40}{\Mpc}.
A total of \num{88} galaxies (\SI{\approx 27.2}{\percent}) are contained in the UNGC and are removed from their sample.
We also exclude \num{4} / \num{25} galaxies which belong to the Fornax / Virgo galaxy cluster (\SI{\approx 1.2}{\percent} / \SI{\approx 7.7}{\percent}).
All other \num{206} galaxies are part of a heterogeneous sample from a mix of local environments (galaxy group, cluster of galaxies) whose completeness study goes beyond the scope of this paper.
In \Cref{fig:nf_g14} we compare {\nucfrac} from their sample (blue plus-signs) to our results (diamonds; same as in \Cref{fig:nucfrac_hubble_type}).
We find that their {\nucfrac} is larger than our results only using the Local Volume.
Since we do not have detailed information about the direct environment of these galaxies, we cannot draw any conclusions why this difference occurs.
Note that if you combine the data sets of \citet{Georgiev2014} and the UNGC, the difference between {\nucfrac} of field galaxies and e.g.\ the Virgo galaxy cluster shrinks between \SI{e8}{\Msun} and \SI{e8.5}{\Msun}, while beyond \SI{e9.5}{\Msun} it still persists.
Thus, although adding late-type field galaxies may increase {\nucfrac} above \SI{e8}{\Msun}, it is insufficient to explain the observed differences between galaxy environments.

\begin{figure}
  \centering
  \includegraphics[width=\columnwidth]{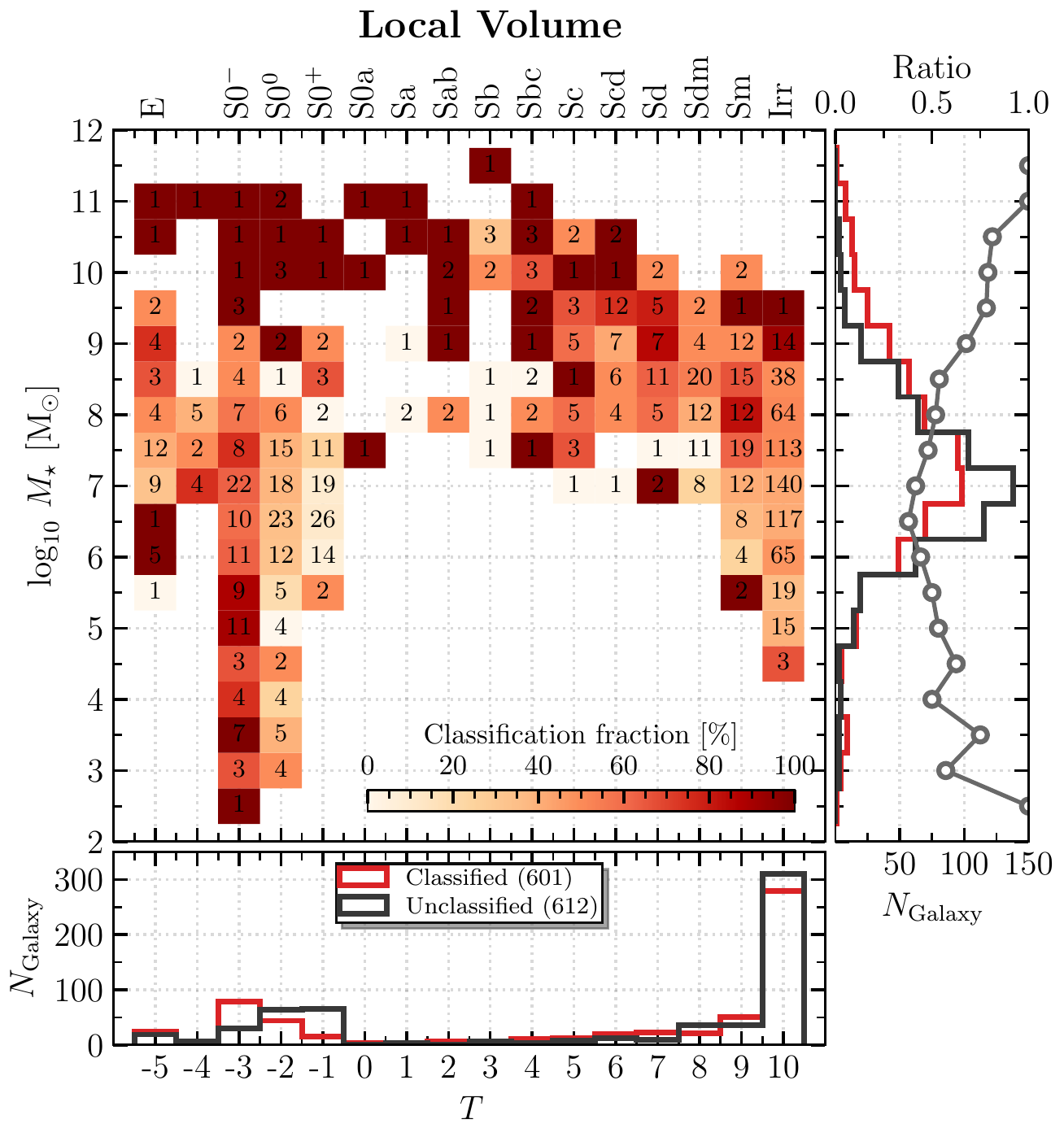}
  \caption{%
    Two-dimensional histogram of logarithmic stellar mass ($\log_{10} \, M_{\star}$) and Hubble type ($T$).
    Bins are colour-coded based on the fraction of classified galaxies:
    dark red colour corresponds to a high completeness / classification fraction.
    The projections along both axes are indicated with histograms where galaxies are split into classified (red) and unclassified (black).
    Additionally, we show the ratio between classified and unclassified galaxies as a function of stellar mass with dark gray circles in the \textit{right} panel.
  }
  \label{fig:completeness_local_volume}
\end{figure}

We also investigate whether the distribution of classified galaxies represents the overall distribution of Local Volume galaxies.
In \Cref{fig:completeness_local_volume} we show the two-dimensional histogram of Local Volume galaxies where bins are colour-coded based on the fraction of classified galaxies.
A high classification fraction (dark red colour) corresponds to a good representation of the global galaxy population.
In addition, we show the projected histograms of galactic stellar mass and Hubble type where we show the distributions of classified (red) and unclassified (black) galaxies.
We find that the classification fraction of our sample is \SI{\geq 50}{\percent} for most bins, independent of Hubble type and stellar mass.
Note that only S0$^{+}$ galaxies in the mass range \SIrange{e6}{e8}{\Msun} are poorly represented.
However, only \num{81} galaxies are contained in these bins and that is very unlikely that any of the above results are affected.

We repeat this exercise for both the Fornax and Virgo galaxy clusters by obtaining a list of their members from HyperLEDA.
For the Coma galaxy cluster no such list is available.
As described in \S\ref{subsec:galaxy_parameters}, galactic parameters are taken from SIMBAD and HyperLEDA and stellar masses are calculated as described in \S\ref{subsec:stellar_mass_estimates}.
Due to redshift-dependent distance estimates of many galaxies in the Virgo Cluster Catalog which lead to an overestimation of stellar masses, we limit the data set to galaxies which are also part of the Extended Virgo Cluster Catalog.
This catalogue contains Virgo Cluster Catalog members whose radial velocity does not exceed \SI{3000}{\kilo\metre\per\second} \citep{Kim2014}, leading to a removal of more than \num{1000} galaxies.
We only consider nuclear classifications of \citet{Cote2006}, \citet{Lisker2007}, \citet{Turner2012}, \citet{Munoz2015}, \citet{Sanchez-Janssen2019a}, and \citet{Su2021}.
Additionally, we add the classification of \citet{Georgiev2014} to the data sets of the Fornax and Virgo galaxy clusters (\num{4} and \num{25} galaxies, respectively).
As a result, we cannot exclude the possibility that a subset of unclassified galaxies have a classification in the literature and trying to obtain one is beyond the scope of this paper.

For the Virgo galaxy cluster we find that only a small subset of all known cluster members are classified by \citet{Cote2006}, \citet{Lisker2007}, and \citet{Sanchez-Janssen2019a} (\textit{cf}.~\Cref{fig:completeness_virgo}).
While the overall population of galaxies is poorly represented, the completeness of elliptical galaxies is \SI{> 50}{\percent} with $M_{\star}$ < \SI{e8}{\Msun} and $M_{\star} \geq \SI{e10}{\Msun}$.
However, keep note that the analysis of \citet{Sanchez-Janssen2019a} solely focused on the core of Virgo and is complete down to $M_{\star} \approx \SI{e7}{\Msun}$ \citep{Ferrarese2020}.
Despite the addition of the late-type sample of \citet{Georgiev2014}, late-type spiral galaxies are not well represented.
Since {\nucfrac} is similar for elliptical galaxies between the Local Volume and the Virgo galaxy cluster (i.e.~\Cref{fig:nucfrac_hubble_type}) and no also no significant differences can be found in the literature \citep[Figure~\num{3} in][]{Neumayer2020}, we suspect that this holds true for late-type galaxies as well and would therefore expect to find many (potentially) unidentified NSCs.

For the Fornax galaxy cluster we find that the classification fraction is \SI{> 50}{\percent} for most bins for early-type galaxies (\textit{cf}.~\Cref{fig:completeness_fornax}).
Similar to the Virgo galaxy cluster, no late-type galaxies are covered by \citet{Turner2012}, \citet{Munoz2015}, and \citet{Su2021}.
Furthermore, only four late-type galaxies of \citet{Georgiev2014} are Fornax cluster members, leaving a large parameter space unexplored.

In total, the completeness of the Local Volume is \SI{49.3}{\percent} (601 classified and 619 unclassified galaxies) whereas it is \SI{90.7}{\percent} (737 classified and 76 unclassified galaxies) for the Fornax and \SI{31.9}{\percent} (776 classified and 1658 unclassified galaxies) for the Virgo galaxy cluster.

Combining these arguments, we conclude that {\nucfrac} for the Fornax and Virgo galaxy clusters are representative of early-type galaxies\footnote{For the Virgo galaxy cluster this is the case for $M_{\star} \geq \SI{e8}{\Msun}$.}.
If we assume that {\nucfrac} evolves similar to what we observe in the Local Volume, it would be representative of all cluster members.
Based on a classification fraction of \SI{> 50}{\percent} for galaxies across all Hubble types and stellar masses and no visible differences in the distributions of classified and unclassified galaxies, we conclude that our results well represent the Local Volume, and therefore, the local field galaxy population.

\subsection{Completeness fraction for the Fornax and Virgo galaxy clusters}
\label{subsec:missing_galaxies2}

\begin{figure}
  \centering
  \includegraphics[width=\columnwidth]{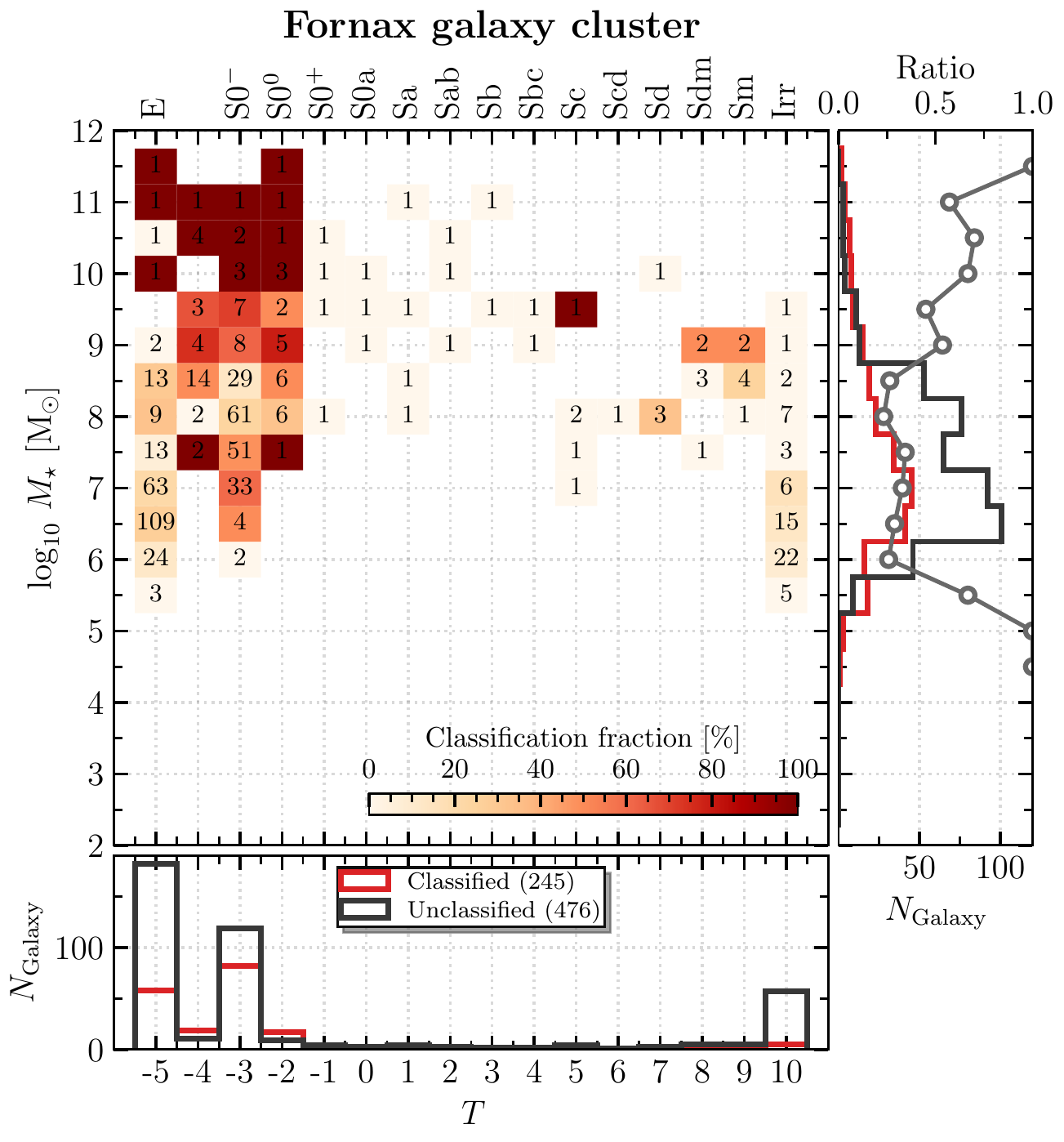}
  \caption{%
    Same as in \Cref{fig:completeness_local_volume} but for the Fornax galaxy cluster.
  }
  \label{fig:completeness_fornax}
\end{figure}

\begin{figure}
  \centering
  \includegraphics[width=\columnwidth]{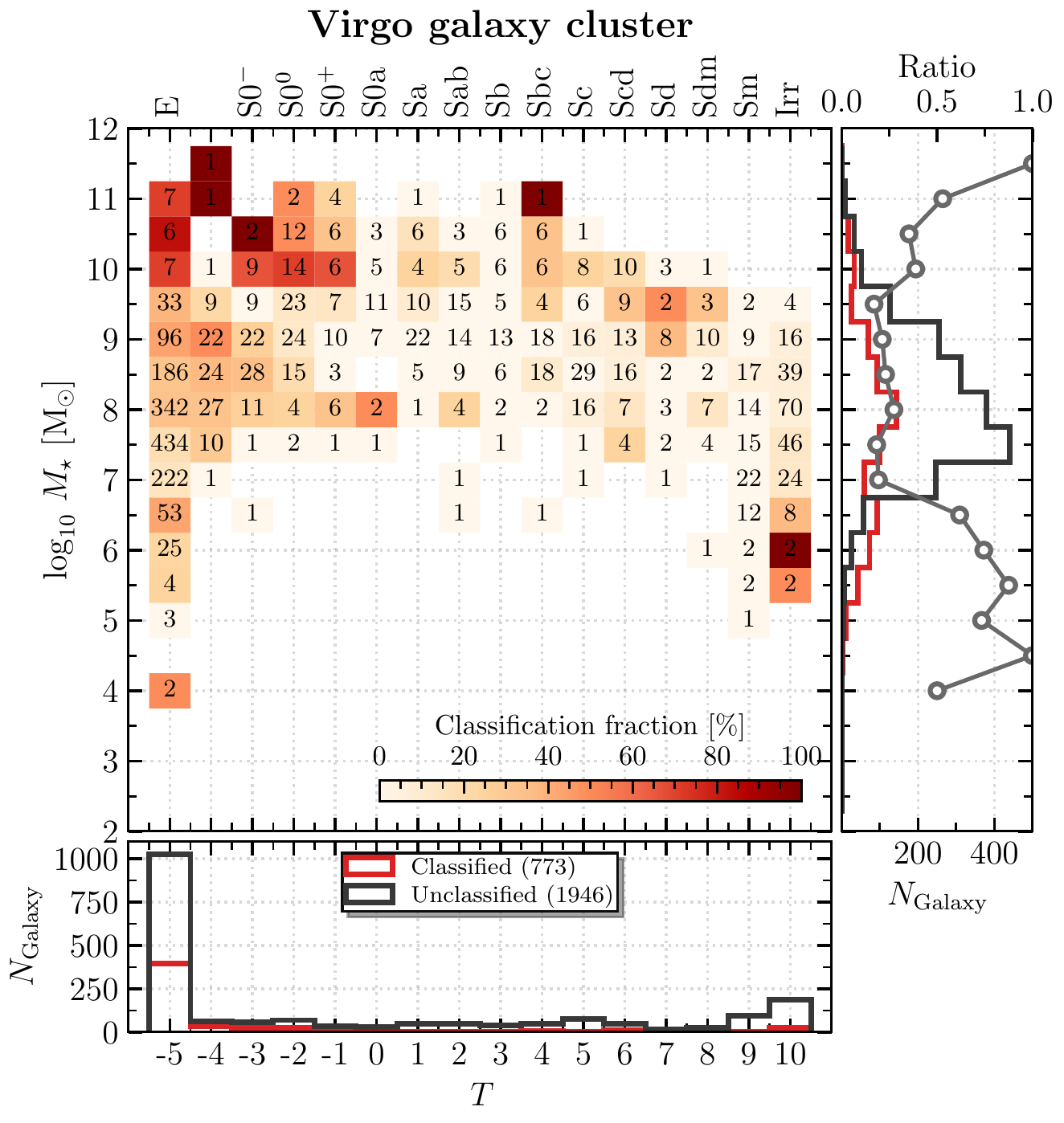}
  \caption{%
    Same as in \Cref{fig:completeness_local_volume} but for the Virgo galaxy cluster.
  }
  \label{fig:completeness_virgo}
\end{figure}

In \Cref{fig:completeness_fornax} and \Cref{fig:completeness_virgo} we show the two-dimensional histograms of the Fornax and Virgo galaxy clusters, respectively, that the show completeness fraction of classified galaxies.
More information about the data sets is presented in the main body of the paper.

\subsection{Bias due to non-detections of nuclei}
\label{subsec:bias_due_to_non-detections_of_nuclei}

Due to the compact nature of nuclear star clusters and the resolution limits of {\HST}, we might wrongfully classify a nucleated galaxy as non-nucleated by flagging a nuclear star cluster as a central star.
We investigate the false-negative classification bias to investigating the resolution limits of different camers of {\HST} which have been used to classify galaxies.

For this investigations we create {\HST} PSFs at the centre of the ACS (WFC, HRC), WFPC2 (WFC, PC), and WFC3 (UVIS, NIR) detectors in the $F814W$ band\footnote{For the WFC3 NIR detector, we use the $F160W$ detector.}.
We use this filter because, as pointed out in \Cref{subsubsec:galaxy_classification_local_volume_galaxies}, our classification process focuses on red / NIR {\HST} exposures to minimise the influence of Galactic and internal dust obscuration.
The light profile of each PSF is then fit with a two-dimensional Gaussian profile.
The width of the PSF is calculated as the geometric mean of the width of the Gaussian profile.

\begin{figure}
  \centering
  \includegraphics[width=0.8\columnwidth]{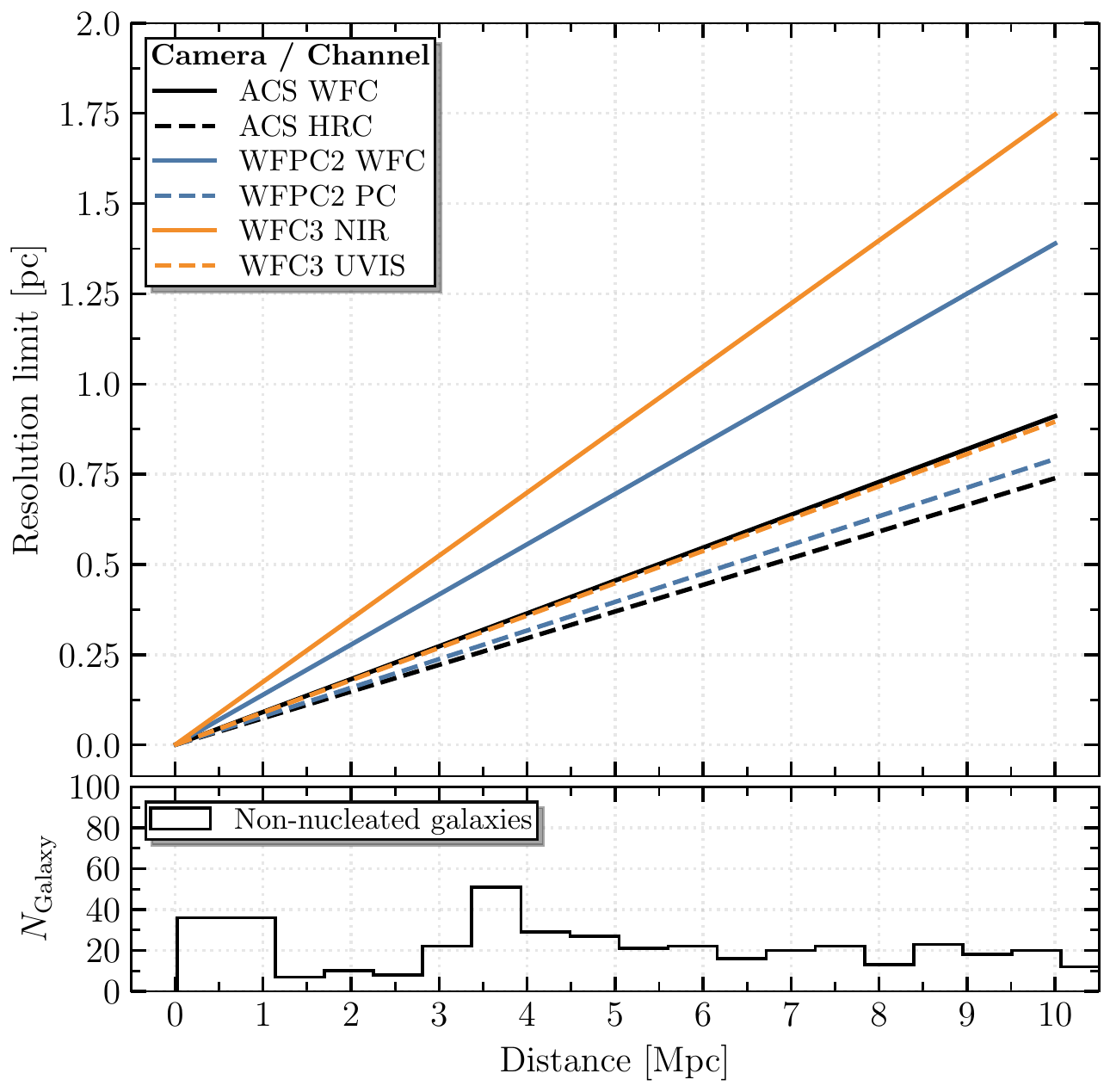}
  \caption{%
    \textit{Top} panel:
    Resolution limit of different {\HST} detectors as a function of galaxy distance.
    A nuclear star cluster is flagged as a star in our analysis if twice its effective radius is below the resolution limit.
    \textit{Bottom} panel:
    Distribution of galaxies classified as non-nucleated.
  }
  \label{fig:limiting_resolution}
\end{figure}

Similar to \citet{Georgiev2014}, we assume that we are able to distinguish a star from an extended source if the difference between the instrument PSF and the extended source is greater than \SI{20}{\percent}.
As argued by the authors, a value of \SI{20}{\percent} is a conservative limit.
In \Cref{fig:limiting_resolution} we show this limit as a function of distance for each detector.
Focus should be put on the ACS WFC (solid black) and WFPC2 WFC (solid blue) lines because these detectors have been used most often in our analysis pipeline.
Note that the resolution limit on the y-axis corresponds to twice the effective radius of a potential nuclear star cluster.
In addition, we show in the bottom panel the distribution of galaxies classified as non-nucleated.
In our analysis pipeline an NSC is wrongfully classified as a star if twice its effective radius is below the resolution limit.

\section{Additions regarding the tidal index}
\label{sec:additions_regarding_the_tidal_index}

\subsection{Bimodial distribution of the tidal index for the Virgo galaxy cluster}
\label{subsec:theta10_bimodial_distribtion}

\begin{figure}
  \centering
  \includegraphics[width=\columnwidth]{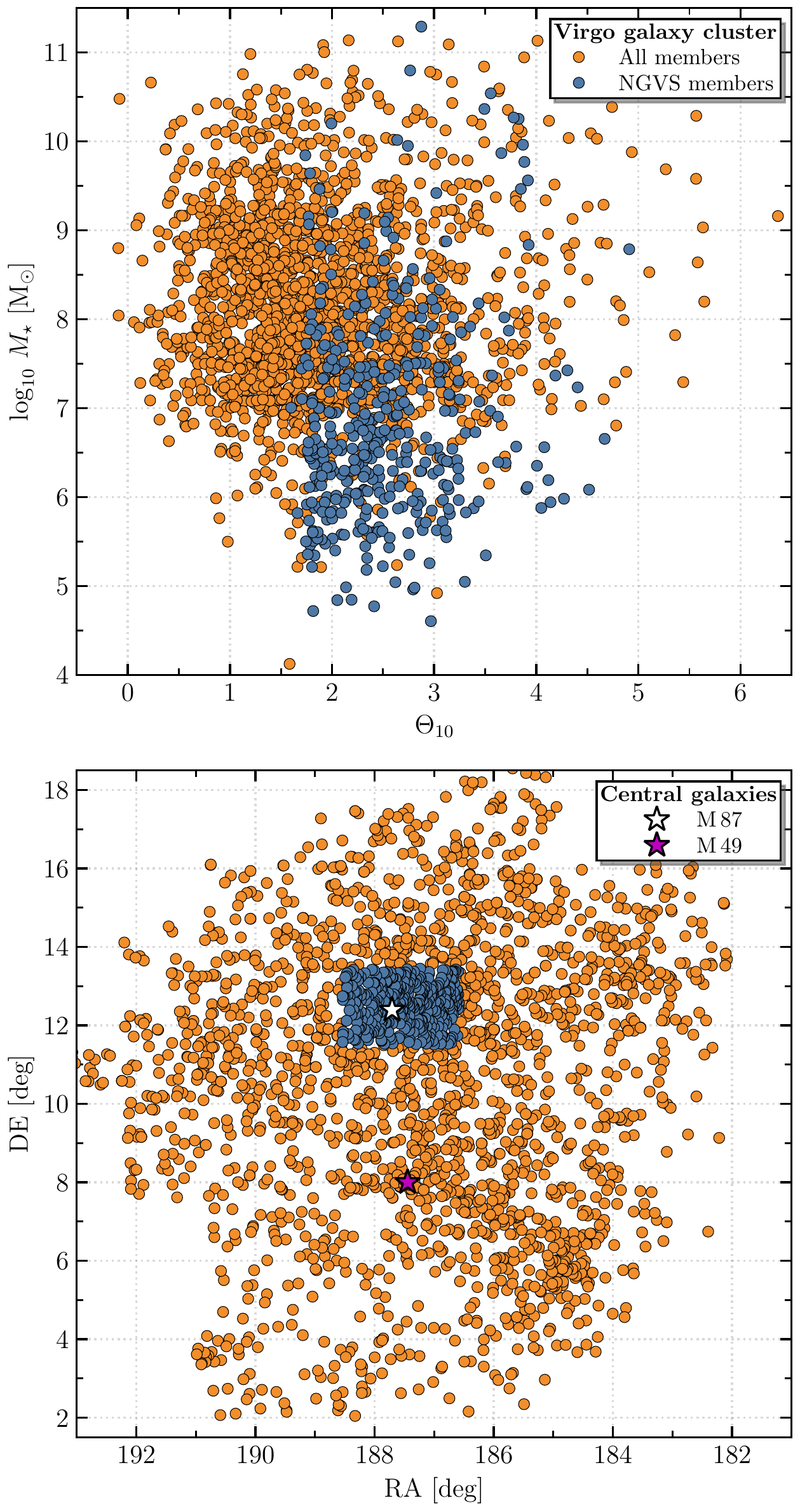}
  \caption{%
    \textit{Top panel}: Logarithmic stellar mass ($\log_{10} \, M_{\star}$) as a function of tidal index ($\Theta_{10}$).
    A small tidal index corresponds to a `loose' local environment.
    NGVS members are shown in blue colour.
    \textit{Bottom panel}: Spatial distribution of Virgo galaxy cluster members.
    A white and magenta star indicate the position of M$\,$49 and M$\,$87, respectively.
  }
  \label{fig:theta10_virgo}
\end{figure}

In the top panel of \Cref{fig:theta10_virgo} we show the logarithmic stellar mass $\log_{10} \, M_{\star}$ as a function of the tidal index $\Theta_{10}$ for thre Virgo galaxy cluster.
In the bottom panel we show the spatial distribution.
The data of the NGVS are shown in blue colour.
For comparison, we show the central M$\,$49 and M$\,$87 with a white and magenta star, respectively.

\subsection{Nucleation fraction}
\label{subsec:ti10_nucleation_fraction}

\begin{figure*}
  \centering
  \includegraphics[width=\textwidth]{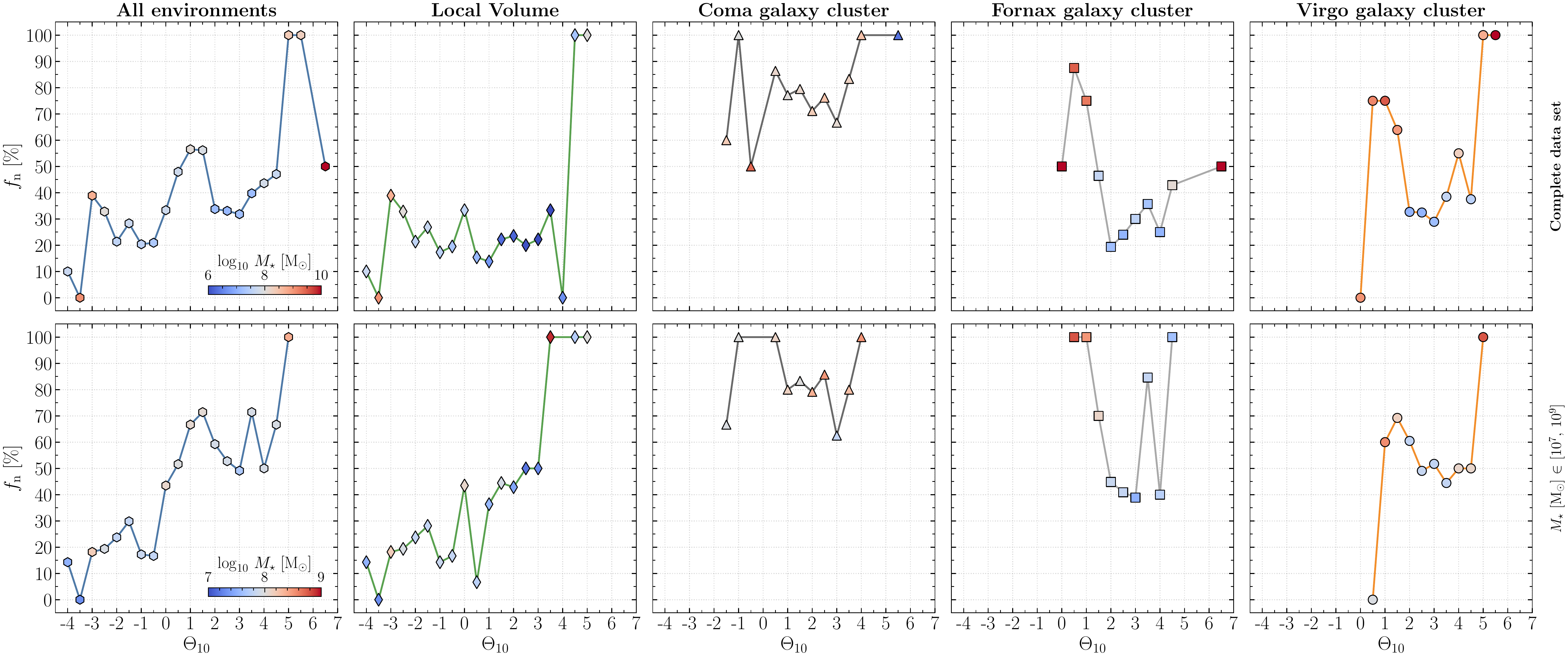}
  \caption{%
    Nucleation fraction ({\nucfrac}) as a function of tidal index ({\Tten}).
    The stellar mass per data point is given by a colourmap.
    This colourmap is different for each row and is given in the left panels, respectively.
    From \textit{left} to \textit{right}: the combined data set from all environments, the Local Volume, the Coma, Fornax, and Virgo galaxy clusters.
    \textit{Top}: The complete data set for each environment.
    \textit{Bottom}: The data are restricted to the mass range $M_{\star} \in [\SI{e7}{\Msun}, \, \SI{e9}{\Msun}]$.
  }
  \label{fig:nucfrac_ti10_2}
\end{figure*}

In \Cref{fig:nucfrac_ti10_2} we construct the nucleation fraction {\nucfrac} as a function of tidal index {\Tten} for all environments combined (first), the Local Volume (second), and the Coma (third), Fornax (fourth), and Virgo (fifth panel) galaxy clusters.
The top row shows the whole data set whereas the data are limited to $M_{\star} \in [\SI{e7}{\Msun}, \, \SI{e9}{\Msun}]$ in the bottom row.
The colourmaps differ between both rows and are indicated in the left panels, respectively.

You can clearly see that {\nucfrac} depends on {\Tten}, however, a significant part of this correlation is due to $M_{\star}$:
for the combined data set the nucleation fraction decreases with increasing tidal index, but also with decreasing stellar mass.
We tried to limit the impact of the stellar mass in the bottom row where one can see that, for the combined data set, {\nucfrac} positively correlates with {\Tten}.
The two data points at the lowest values of {\Tten} highlight that very massive galaxies are barely influenced by neighbouring galaxies (third and fourth panels).
The effects of {\Tten} and $M_{\star}$ on {\nucfrac} are investigated in greater detail in \Cref{subsec:nucleation_fraction_as_a_function_of_stellar_mass,subsec:nucleation_fraction_as_a_function_of_tidal_index}.
For the Local Volume we see that {\nucfrac} positively correlates with {\Tten}.
For the Coma galaxy cluster the trend is unclear and we suspect that our calculation of {\Tten} is inaccurate as many low-mass systems are missing.
In comparison, for both the Fornax and Virgo galaxy clusters, we see a slightly positive correlation bewteen {\nucfrac} and {\Tten}.

\section{Details on the MCMC analysis}
\label{sec:mcmc_analysis}

We use the Markov-Chain Monte Carlo (MCMC) capabilities of the \texttt{emcee} package \citep{Foreman-Mackey2013} to find the best fitting parameter values of our nucleation fraction model for all environments.
Because we only model the low-mass galaxy populations, we limit the data set to $M_{\star} \leq \SI{e9.5}{\Msun}$.
For all fits we use uniformative and uniform priors:
$\kappa \in [0.0, \, 3.0]$, $M_{50} \in [2.0, 12.0]$.
The initial parameters for $\kappa$ and $M_{50}$ are $1.0$ and $8.0$.
The logarithmic likelihood function is
\begin{equation}
  \label{equ:log_likelihood}
  \log \mathcal{L}(\kappa, \, M_{50}) = -\frac{1}{2} \sum_{i} \bigg[ \bigg( \frac{y_{i} - \mathrm{L}_{i}(\kappa, \, M_{50})}{\sigma_{i}} \bigg)^{2}  + \log 2 \pi \sigma_{i}^{2} \bigg] \quad ,
\end{equation}
where $y_{i} \in \{0, \, 1\}$ is the nucleation of the $i$th galaxy and $\mathrm{L}_{i}$ is the Logistic function (\textit{cf}.\ \Cref{equ:model}).

\begin{figure}
  \centering
  \includegraphics[width=0.85\columnwidth]{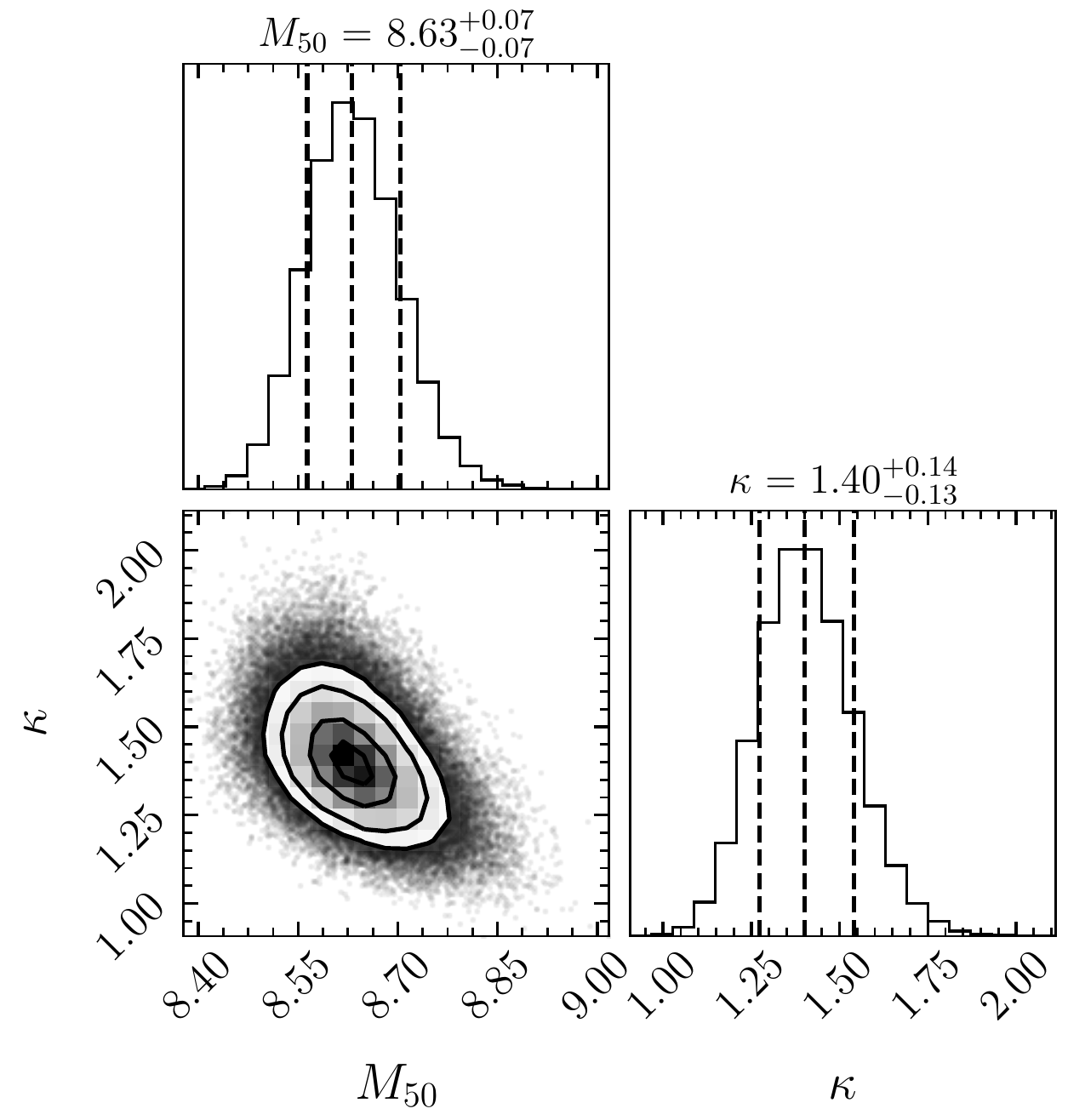}
  \caption{%
    Corner plot of the Markov-Chain Monte Carlo analysis for the data set of the Local Volume.
    Only galaxies with $M_{\star} \leq \SI{e9.5}{\Msun}$ are considered for the fit.
    The median and $1\sigma$ interval are indicated with dashed lines.
  }
  \label{fig:lv_corner}
\end{figure}

\begin{figure}
  \centering
  \includegraphics[width=\columnwidth]{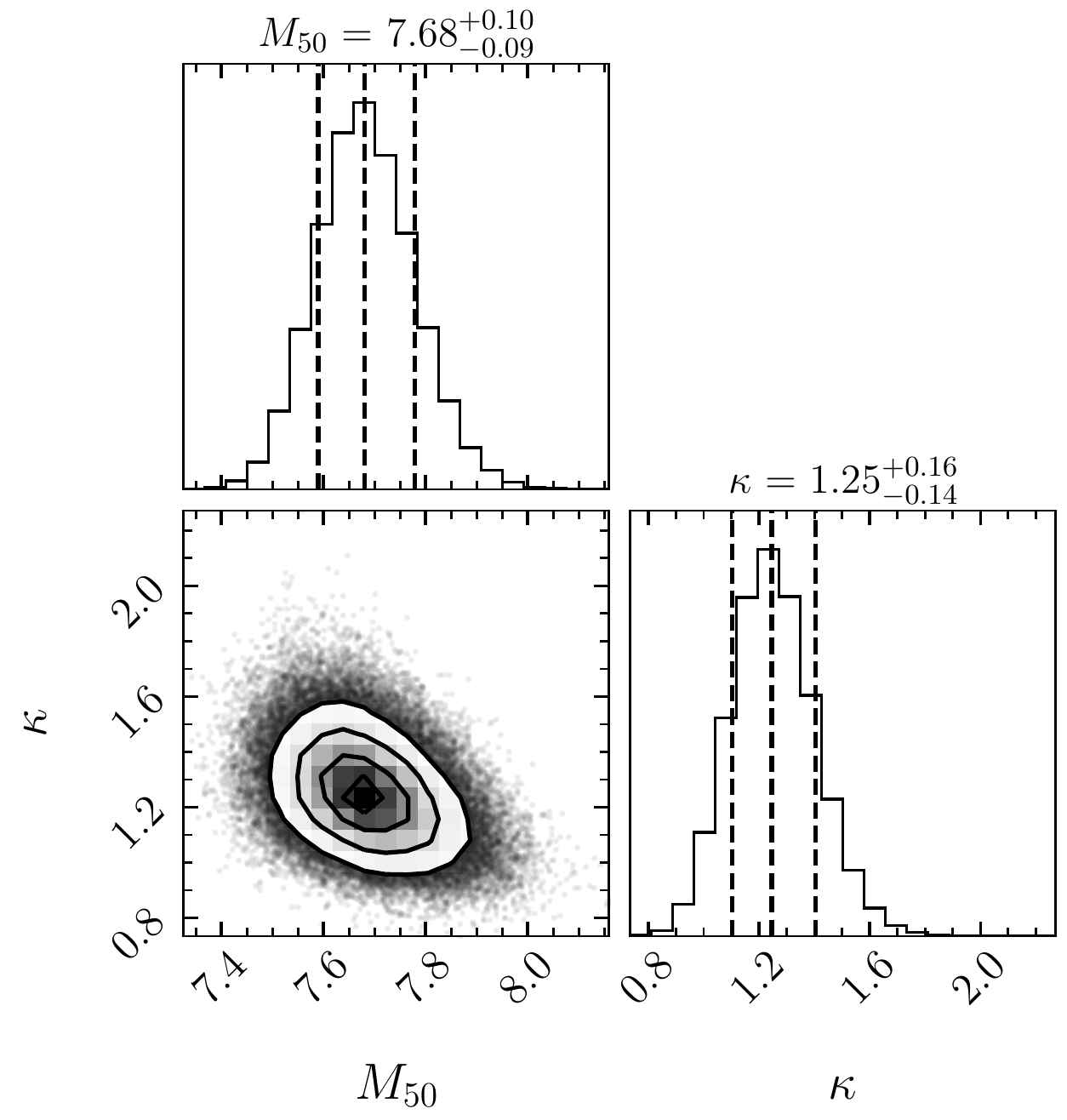}
  \caption{%
    Same as in \Cref{fig:lv_corner} but for the Fornax galaxy cluster.
  }
  \label{fig:f_corner}
\end{figure}

\begin{figure}
  \centering
  \includegraphics[width=\columnwidth]{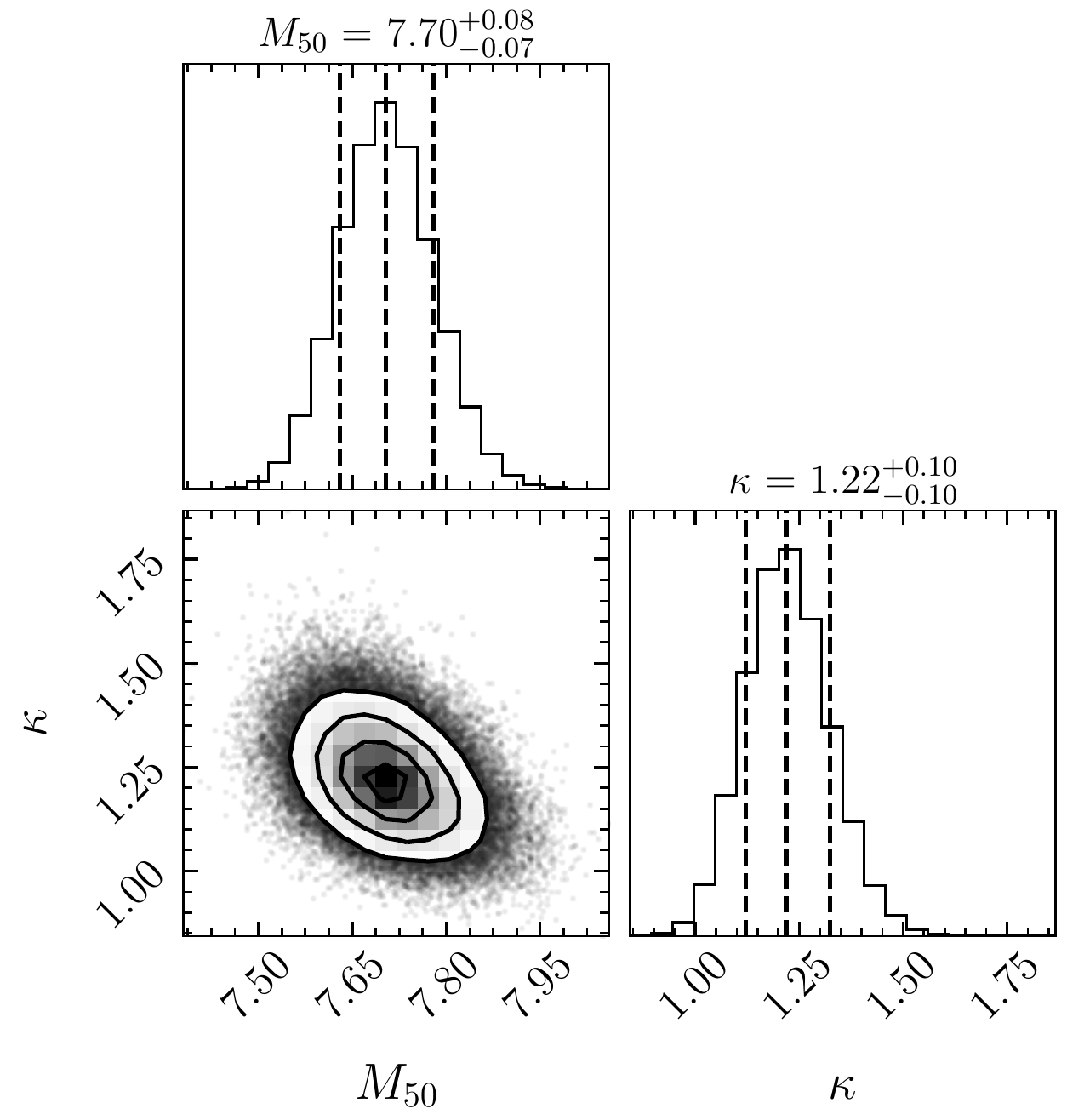}
  \caption{%
    Same as in \Cref{fig:lv_corner} but for the Virgo galaxy cluster.
  }
  \label{fig:v_corner}
\end{figure}

\begin{figure}
  \centering
  \includegraphics[width=\columnwidth]{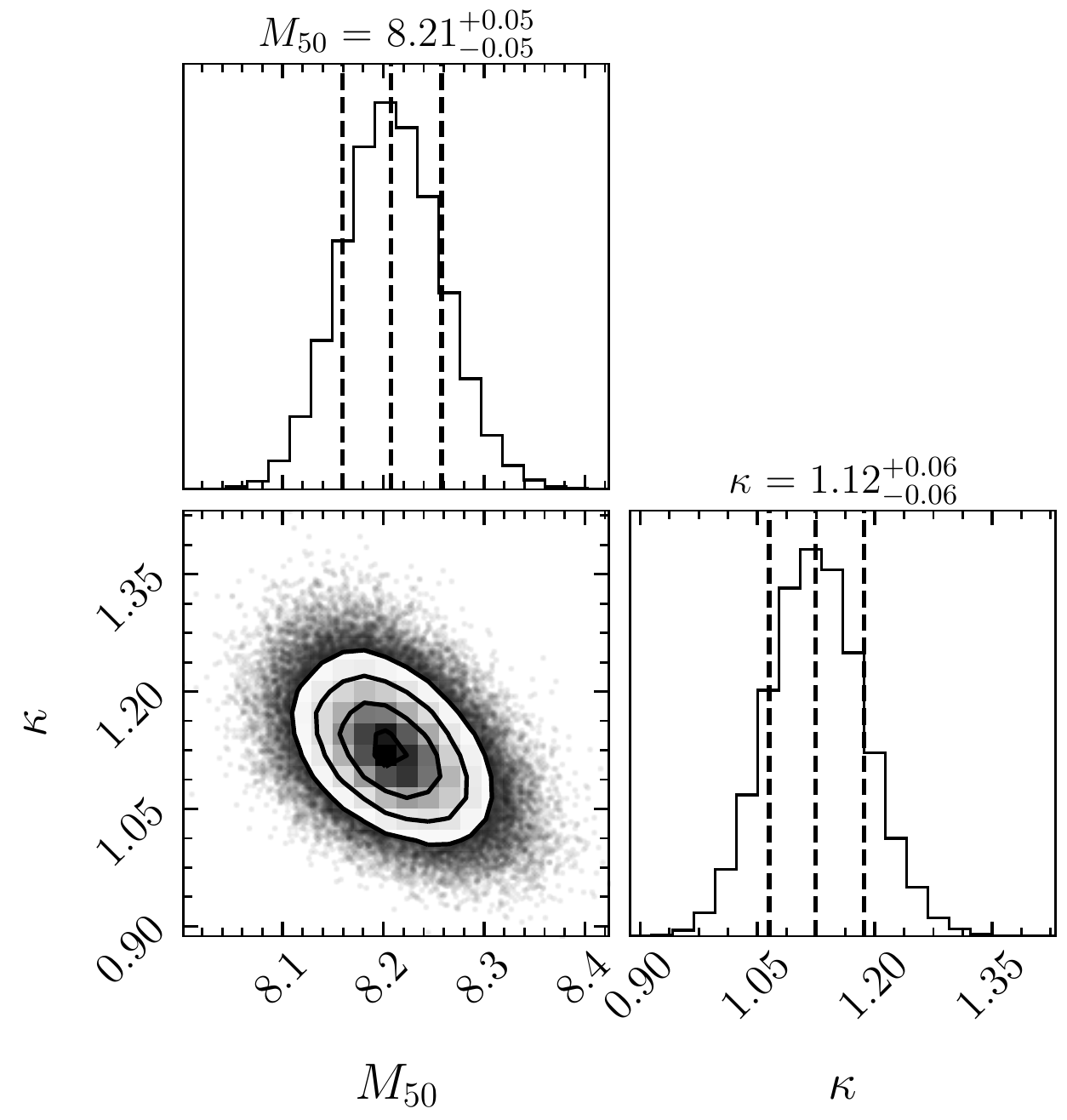}
  \caption{%
    Same as in \Cref{fig:lv_corner} but for the combined data set.
  }
  \label{fig:a_corner}
\end{figure}

We run the MCMC for \num{5000} iterations using \num{500} walkers.
The autocorrelation time varies between \num{20} and \num{40} steps, and the mean acceptance rate is \SI{\approx 70}{\percent}.
We discard the first \num{500} steps from each run and use a thinning factor of \num{25}.
The final posteriors for the Local Volume, the Fornax and Virgo galaxy clusters, and the combined data set are shown in \Cref{fig:lv_corner,fig:f_corner,fig:v_corner,fig:a_corner}, respectively.
All values reported in \Cref{tab:fit_results} are the median values of each distribution in the posteriors.
The uncertainties give the $1\sigma$ interval around the median.

\section{Data tables}
\label{sec:data_tables}

\Cref{tab:data_table_local_volume} presents the first five entries in the data table for the Local Volume.
The data tables for the Coma, Fornax, and Virgo galaxy clusters share the same format.
All data tables are available online in \texttt{.fits.gz} format.
\begin{table*}
  \scriptsize
  \caption{
    List of galaxies in the Local Volume based on the Updated Nearvy Galaxy Catalog \citep{Karachentsev2013} including basic photometric parameters, the tidal index which characterises the local density of a galaxy, a stellar mass estimate, and a nuclear classification.
    Magnitudes are given in the VEGAmag system, except for the $g$- and $r$-bands which are given in the ABmag system.
    If no reference is provided, the nuclear classification is based on available Hubble Legacy Archive data.
    The full data table is available online.
  }
  \begin{center}
    \begin{threeparttable}
      \begin{tabular}{%
        l
        S[table-format=3.5]
        S[table-format=2.5]
        S[table-format=2.2]@{\,\( \pm \)\,}
        S[table-format=1.2]
        l
        l
        S[table-format=2.1]@{\,\( \pm \)\,}
        S[table-format=1.1]
        S[table-format=-1.3]
        S[table-format=-1.3]
        S[table-format=-1.3]
         l
        S[table-format=2.2]@{\,\( \pm \)\,}
        S[table-format=1.2]
        l
        l
      }
      \toprule
      {Name} & {RA} & {DE} & \multicolumn{2}{c}{$d$} & {$d_{\mathrm{M}}$\tnote{(a)}} & {\ldots} & \multicolumn{2}{c}{$T$} & {$\Theta_{1}$} & {$\Theta_{5}$} & {$\Theta_{10}$} & {MD} & \multicolumn{2}{c}{$\log_{10} M_{\star}$} & {Nuc.\tnote{(b)}} & {Ref.\tnote{(c)}} \\
      \cmidrule(lr){2-3}
      \cmidrule(lr){4-5}
      \cmidrule(lr){14-15}
      {} & {[deg]} & {[deg]} & \multicolumn{2}{c}{[mag]} & {} & {} & \multicolumn{2}{c}{} & {} & {} & {} & {} & \multicolumn{2}{c}{[M$_{\odot}$]} & {} & {} \\
      \midrule

      {A0554+07}      & 89.40292  & 7.49194  & 29.95 & {-{-}} & TF     & {\ldots} & 10.0 & 1.0 & -3.65 & -3.12 & -2.94 & {\ldots} & 6.70 & 0.40 & {?} & {-{-}} \\
      {A0952+69}      & 149.37083 & 69.27222 & 27.97 & 0.17   & TRGB   & {\ldots} & 10.0 & 1.0 & 0.40  & 0.40  & 0.40  & {\ldots} & 6.48 & 0.39 & {0} & {-{-}} \\
      {AF7448\_001}   & 344.89708 & 16.76972 & 28.49 & {-{-}} & TF     & {\ldots} & 10.0 & 1.0 & -2.82 & -2.37 & -2.27 & {\ldots} & 6.39 & 0.43 & {0} & {-{-}} \\
      {AGC\,102728}   & 0.08917   & 31.02194 & 29.74 & 0.16   & TRGB   & {\ldots} & 10.0 & 1.0 & -3.52 & -2.91 & -2.72 & {\ldots} & 5.73 & 0.54 & {0} & {-{-}} \\
      {AGC\,112454}   & 23.24875  & 14.37333 & 30.04 & {-{-}} & mem    & {\ldots} & 10.0 & 2.0 & 0.05  & 0.05  & 0.05  & {\ldots} & 6.73 & 0.71 & {?} & {-{-}} \\
      {\ldots}                                                                                                                                                 \\

      \bottomrule
      \end{tabular}
      \begin{tablenotes}
         \item[(a)] Methods, see \citet{Karachentsev2013}. In addition, we use distance estimates from HyperLEDA and SIMBAD.
         \item[(b)] Nuclear classification: `1' nucleated; `0' non-nucleated; `?' unknown.
         \item[(c)] References: (1) \citet{Baldassare2014}; (2) \citet{Barth2009}; (3) \citet{Baumgardt2018}; (4) \citet{Beifiori2009}; (5) \citet{Bellazzini2008}; (6) \citet{Bellazzini2020}; (7) \citet{Boeker1999}; (8) \citet{Boeker1999b}; (9) \citet{Boeker2001}; (10) \citet{Boeker2002}; (11) \citet{Boeker2003}; (12) \citet{Boeker2004b}; (13) \citet{Butler2005}; (14) \citet{Calzetti2015}; (15) \citet{Carlsten2020a}; (16) \citet{Carlsten2021b}; (17) \citet{Carollo2002}; (18) \citet{Carretta2010}; (19) \citet{Carson2015}; (20) \citet{Cohen2018}; (21) \citet{Cole2017}; (22) \citet{Contenta2018}; (23) \citet{Das2012}; (24) \citet{Davidge2002}; (25) \citet{Desroches2009}; (26) \citet{Emsellem1999}; (27) \citet{Fahrion2020}; (28) \citet{Ferrarese2006}; (29) \citet{Filippenko1989}; (30) \citet{Filippenko2003}; (31) \citet{Ganda2009}; (32) \citet{Georgiev2008}; (33) \citet{Georgiev2009}; (34) \citet{Georgiev2009b}; (35) \citet{Georgiev2014}; (36) \citet{Gnerucci2013}; (37) \citet{Gonzalez_Delgado2008}; (38) \citet{Gordon1999}; (39) \citet{Graham2007b}; (40) \citet{Graham2008}; (41) \citet{Graham2009}; (42) \citet{Graham2012}; (43) \citet{Graham2013}; (44) \citet{Hartmann2011}; (45) \citet{Ho1997}; (46) \citet{Hopkins2009}; (47) \citet{Jones1996}; (48) \citet{Kacharov2018}; (49) \citet{Karachentsev2004}; (50) \citet{Karachentsev2020}; (51) \citet{Kormendy1996}; (52) \citet{Kormendy1999}; (53) \citet{Kormendy2010}; (54) \citet{Kormendy2013}; (55) \citet{Lauer2005}; (56) \citet{Luo2012}; (57) \citet{Martocchia2020}; (58) \citet{Matthews1999}; (59) \citet{Matthews2002}; (60) \citet{Milosavljevic2004}; (61) \citet{Milosavljevic2014}; (62) \citet{Misgeld2011}; (63) \citet{Mitzkus2017}; (64) \citet{Monaco2009}; (65) \citet{Neumayer2012}; (66) \citet{Nguyen2017}; (67) \citet{Nguyen2018}; (68) \citet{Pechetti2019}; (69) \citet{Peng2002}; (70) \citet{Phillips1996}; (71) \citet{Piqueras_Lopez2012}; (72) \citet{Puzia2007}; (73) \citet{Ravindranath2001}; (74) \citet{Rossa2006}; (75) \citet{Sanchez-Janssen2019a}; (76) \citet{Sarzi2005}; (77) \citet{Scott2013b}; (78) \citet{Seth2006}; (79) \citet{Seth2010}; (80) \citet{Smith2016}; (81) \citet{Veljanoski2013}; (82) \citet{Walcher2005}; (83) \citet{Yong2016}; (84) \citet{Yuan2019}
      \end{tablenotes}
    \end{threeparttable}
  \end{center}
  \label{tab:data_table_local_volume}
\end{table*}
\begin{table*}
  \scriptsize
  \caption{%
    List of galaxies for the Coma galaxy cluster adopted from \citet{denBrok2014}.
    The authors classify a galaxy as nucleated (`1') if a fit with a central Gaussian source yielded a higher evidence than a fit without one.
    The full data table is available online.
  }
  \begin{center}
    \begin{threeparttable}
      \begin{tabular}{%
        l
        S[table-format=3.5]
        S[table-format=2.5]
        S[table-format=2.2]@{\,\( \pm \)\,}
        S[table-format=1.2]
        l
        l
        S[table-format=2.1]
        S[table-format=-1.3]
        S[table-format=-1.3]
        S[table-format=-1.3]
         l
        S[table-format=2.2]@{\,\( \pm \)\,}
        S[table-format=1.2]
        l
      }
      \toprule
      {Name} & {RA} & {DE} & \multicolumn{2}{c}{$d$} & {$d_{\mathrm{M}}$\tnote{(a)}} & {\ldots} & \multicolumn{1}{c}{$T$} & {$\Theta_{1}$} & {$\Theta_{5}$} & {$\Theta_{10}$} & {MD} & \multicolumn{2}{c}{$\log_{10} M_{\star}$} & {Nuc.} \\
      \cmidrule(lr){2-3}
      \cmidrule(lr){4-5}
      \cmidrule(lr){13-14}
      {} & {[deg]} & {[deg]} & \multicolumn{2}{c}{[mag]} & {} & {} & \multicolumn{1}{c}{} & {} & {} & {} & {} & \multicolumn{2}{c}{[M$_{\odot}$]} & {} \\
      \midrule

      {COMAi125713.240p272437.} & 195.0773 & 28.0972 & 35.10 & 0.55 & \citet{Ferrarese2000b} & {\ldots} & {-{-}} & 1.27 & 1.32 & 1.35 & {\ldots} & 8.73   & 0.11   & {1} \\
      {COMAi125828.358p271315.} & 194.8829 & 27.8613 & 35.22 & 0.15 & HyperLEDA              & {\ldots} & {-{-}} & 1.72 & 2.00 & 2.01 & {\ldots} & 8.59   & 0.12   & {1} \\
      {COMAi125844.089p274647.} & 195.1717 & 28.0451 & 35.10 & 0.55 & \citet{Ferrarese2000b} & {\ldots} & {-{-}} & 0.31 & 0.81 & 0.86 & {\ldots} & 7.83   & 0.09   & {1} \\
      {COMAi125855.735p274528.} & 195.1435 & 27.9347 & 35.10 & 0.55 & \citet{Ferrarese2000b} & {\ldots} & {-{-}} & 1.52 & 1.91 & 1.92 & {\ldots} & {-{-}} & {-{-}} & {0} \\
      {COMAi125924.938p275320.} & 194.9978 & 27.9406 & 35.10 & 0.55 & \citet{Ferrarese2000b} & {\ldots} & {-{-}} & 3.34 & 3.35 & 3.35 & {\ldots} & 8.00   & 0.15   & {1} \\
      {\ldots}                                                                                                                                                                       \\

      \bottomrule
      \end{tabular}
      \begin{tablenotes}
         \item[(a)] Either based on HyperLEDA, SIMBAD, or \citet{Ferrarese2000b}.
         \item[(b)] Nuclear classification: `1' nucleated; `0' non-nucleated.
      \end{tablenotes}
    \end{threeparttable}
  \end{center}
  \label{tab:data_table_coma}
\end{table*}
\begin{table*}
  \scriptsize
  \caption{
    List of galaxies for the Fornax galaxy cluster adopted from the HyperLEDA data base.
    The full data table is available online.
  }
  \begin{center}
    \begin{threeparttable}
      \begin{tabular}{%
        l
        S[table-format=3.5]
        S[table-format=2.5]
        S[table-format=2.2]@{\,\( \pm \)\,}
        S[table-format=1.2]
        l
        l
        S[table-format=2.1]@{\,\( \pm \)\,}
        S[table-format=1.1]
        S[table-format=-1.3]
        S[table-format=-1.3]
        S[table-format=-1.3]
        l
        S[table-format=2.2]@{\,\( \pm \)\,}
        S[table-format=1.2]
        l
        l
        l
      }
      \toprule
      {Name} & {RA} & {DE} & \multicolumn{2}{c}{$d$} & {$d_{\mathrm{M}}$\tnote{(a)}} & {\ldots} & \multicolumn{2}{c}{$T$} & {$\Theta_{1}$} & {$\Theta_{5}$} & {$\Theta_{10}$} & {MD} & \multicolumn{2}{c}{$\log_{10} M_{\star}$} & {\ldots} & {Nuc.\tnote{(b)}} & {Ref.\tnote{(c)}} \\
      \cmidrule(lr){2-3}
      \cmidrule(lr){4-5}
      \cmidrule(lr){14-15}
      {} & {[deg]} & {[deg]} & \multicolumn{2}{c}{[mag]} & {} & {} & \multicolumn{2}{c}{} & {} & {} & {} & {} & \multicolumn{2}{c}{[M$_{\odot}$]} & {} & {} & {}\\
      \midrule

      {ESO\,302-009} & 56.89212 & -38.57648 & 31.38 & 0.64 & HyperLEDA             & {\ldots} &  8.0 & 0.7 & -0.64 & -0.11 & 0.03 & {\ldots} & 9.07 & 0.58 & {\ldots} & {?} & {-{-}} \\
      {ESO\,357-010} & 48.92692 & -33.54269 & 31.51 & 0.15 & \citet{Blakeslee2009} & {\ldots} &  7.9 & 0.7 & -0.45 & -0.07 & 0.04 & {\ldots} & 8.35 & 0.46 & {\ldots} & {?} & {-{-}} \\
      {ESO\,357-012} & 49.22225 & -35.54136 & 31.48 & 0.25 & HyperLEDA             & {\ldots} &  7.0 & 0.4 &  0.41 &  0.76 & 0.79 & {\ldots} & 8.21 & 0.38 & {\ldots} & {1} & 1      \\
      {ESO\,358-011} & 52.53300 & -32.47525 & 31.51 & 0.15 & \citet{Blakeslee2009} & {\ldots} & -1.3 & 4.3 &  0.99 &  1.10 & 1.12 & {\ldots} & 8.10 & 0.38 & {\ldots} & {?} & {-{-}} \\
      {ESO\,358-020} & 53.73871 & -32.63998 & 31.70 & 0.73 & HyperLEDA             & {\ldots} &  9.3 & 2.4 &  0.02 &  0.40 & 0.51 & {\ldots} & 8.67 & 0.41 & {\ldots} & {?} & {-{-}} \\
      {\ldots}                                                                                                                                                                         \\

      \bottomrule
      \end{tabular}
      \begin{tablenotes}
         \item[(a)] Either based on HyperLEDA, SIMBAD, or \citet{Blakeslee2009}.
         \item[(b)] Nuclear classification: `1' nucleated; `0' non-nucleated; `?' unknown.
         \item[(c)] References: (1) \citet{Georgiev2014}; (2) \citet{Munoz2015}; (3) \citet{Su2021}; (4) \citet{Turner2012}; (5) \citet{Venhola2018}.
      \end{tablenotes}
    \end{threeparttable}
  \end{center}
  \label{tab:data_table_fornax}
\end{table*}
\begin{table*}
  \scriptsize
  \caption{
    List of galaxies for the Virgo galaxy cluster adopted from the HyperLEDA data base.
    We removed galaxies which are not part of NGVS sample or the Extended Virgo Cluster Catalog \citep{Kim2014}.
    The full data table is available online.
  }
  \begin{center}
    \begin{threeparttable}
      \begin{tabular}{%
        l
        S[table-format=3.5]
        S[table-format=2.5]
        S[table-format=2.2]@{\,\( \pm \)\,}
        S[table-format=1.2]
        l
        l
        S[table-format=2.1]@{\,\( \pm \)\,}
        S[table-format=1.1]
        S[table-format=-1.3]
        S[table-format=-1.3]
        S[table-format=-1.3]
         l
        S[table-format=2.2]@{\,\( \pm \)\,}
        S[table-format=1.2]
        l
        l
        l
      }
      \toprule
      {Name} & {RA} & {DE} & \multicolumn{2}{c}{$d$} & {$d_{\mathrm{M}}$\tnote{(a)}} & {\ldots} & \multicolumn{2}{c}{$T$} & {$\Theta_{1}$} & {$\Theta_{5}$} & {$\Theta_{10}$} & {MD} & \multicolumn{2}{c}{$\log_{10} M_{\star}$} & {\ldots} & {Nuc.\tnote{(b)}} & {Ref.\tnote{(c)}} \\
      \cmidrule(lr){2-3}
      \cmidrule(lr){4-5}
      \cmidrule(lr){14-15}
      {} & {[deg]} & {[deg]} & \multicolumn{2}{c}{[mag]} & {} & {} & \multicolumn{2}{c}{} & {} & {} & {} & {} & \multicolumn{2}{c}{[M$_{\odot}$]} & {} & {} & {} \\
      \midrule

      {IC\,0767} & 182.76142 & 12.10396 & 32.30 & 0.55 & HyperLEDA       & {\ldots} & -5.0 & 1.0 & 1.28 & 1.41 & 1.42 & {\ldots} &  9.66 & 0.12 & {\ldots} & {?} & {-{-}} \\
      {IC\,0768} & 182.94833 & 12.14361 & 31.09 & 0.01 & \citet{Mei2007} & {\ldots} &  5.9 & 1.0 & 2.18 & 2.20 & 2.21 & {\ldots} &  9.14 & 0.10 & {\ldots} & {?} & {-{-}} \\
      {IC\,0769} & 183.13471 & 12.12372 & 32.86 & 0.19 & HyperLEDA       & {\ldots} &  4.0 & 0.3 & 0.78 & 1.13 & 1.17 & {\ldots} & 10.10 & 0.10 & {\ldots} & {1} & {2}    \\
      {IC\,0771} & 183.80500 & 13.18453 & 31.09 & 0.01 & \citet{Mei2007} & {\ldots} &  5.6 & 1.3 & 2.86 & 2.96 & 2.98 & {\ldots} &  8.55 & 0.19 & {\ldots} & {?} & {-{-}} \\
      {IC\,0773} & 184.53375 &  6.13957 & 31.09 & 0.01 & \citet{Mei2007} & {\ldots} & -0.5 & 1.2 & 2.14 & 2.49 & 2.52 & {\ldots} &  9.21 & 0.09 & {\ldots} & {?} & {-{-}} \\
      {\ldots}                                                                                                                                                 \\

      \bottomrule
      \end{tabular}
      \begin{tablenotes}
         \item[(a)] Either based on HyperLEDA, SIMBAD, or \citet{Mei2007}.
         \item[(b)] Nuclear classification: `1' nucleated; `0' non-nucleated; `?' unknown.
         \item[(c)] References: (1) \citet{Cote2006}; (2) \citet{Georgiev2014}; (3) \citet{Lisker2007}; (4)\citet{Sanchez-Janssen2019a}.
      \end{tablenotes}
    \end{threeparttable}
  \end{center}
  \label{tab:data_table_virgo}
\end{table*}


\bsp	
\label{lastpage}
\end{document}